\documentclass[fleqn,usenatbib, useAMS, a4paper]{mnras}

\usepackage{savesym}
\savesymbol{tablenum}
\usepackage{siunitx}
\restoresymbol{SIX}{tablenum}

\usepackage{newtxtext}
\usepackage[varg,varvw,smallerops]{newtxmath}

\usepackage[T1]{fontenc}
\usepackage{ae,aecompl}

\usepackage{graphicx}	

\usepackage{amsmath}	
\usepackage{amssymb}	
\usepackage{multicol}
\usepackage{enumerate}          
\usepackage{xcolor}

\usepackage[spanish,es-minimal,english]{babel}

\usepackage{booktabs}
\usepackage{array}   
\newcolumntype{L}{>{$}l<{$}} 
\newcolumntype{R}{>{$}r<{$}} 
\newcolumntype{C}{>{$}c<{$}}

\hypersetup{hidelinks=True}

\usepackage{upgreek}
\usepackage{placeins}

\definecolor{NEWcolor}{rgb}{0.7,0.2,0.1}
\newcommand\startNEW{\color{black}}
\newcommand\stopNEW{\color{black}}
\newcommand\NEW[1]{\startNEW #1\stopNEW\relax}

\newcounter{ionstage}
\renewcommand{\ion}[2]{\setcounter{ionstage}{#2}%
  \ensuremath{\mathrm{#1\,\scriptstyle\Roman{ionstage}}}}
  
\newcommand\hii{\ion{H}{2}}
\newcommand\pos{\ensuremath{_{\mathrm{pos}}}}
\newcommand\los{\ensuremath{_{\mathrm{los}}}}
\newcommand\noise{\ensuremath{_{\text{noise}}}}
\newcommand\obs{\ensuremath{_{\mathrm{obs}}}}
\newcommand\model{\ensuremath{_{\mathrm{mod}}}}
\newcommand\halpha{H${\alpha}$}

\newcommand\ha{\ensuremath{\text{H}\upalpha}}
\newcommand\hb{\ensuremath{\text{H}\upbeta}}
\newcommand\Wav[1]{\ensuremath{\lambda #1}}

\newcommand\csound{\ensuremath{c_{\text{s}}}}
\newcommand\FNa{\textsuperscript{a}}
\newcommand\Mach{\ensuremath{\mathcal{M}}}
\newcommand\longsig[1]{\ensuremath{\langle \delta #1^2 \rangle^{1/2} / #1_0}}
\newcommand\shortsig[1]{\ensuremath{\sigma_{#1/#1_0}}}

\newlength\SFwidth
\newcommand\SFtwograph[2]{%
  \includegraphics[width=\SFwidth]{Figures/sf-emcee-#1}
  &  \includegraphics[width=\SFwidth]{Figures/sf-emcee-#2}
}
\newcommand\SFtwocorner[2]{%
  \includegraphics[width=\SFwidth]{Figures/corner-emcee-#1}
  &  \includegraphics[width=\SFwidth]{Figures/corner-emcee-#2}
}

\newcommand\sffigg[2]{%
  \begin{tabular}{@{}ll@{}}
    (a)& (b)\\
    \SFtwograph{#1}{#2}
  \end{tabular}%
}
\newcommand\sffigggg[4]{%
  \begin{tabular}{@{}ll@{}}
    (a)& (b)\\
    \SFtwograph{#1}{#2}\\
    (c)& (d)\\
    \SFtwograph{#3}{#4}\\
  \end{tabular}%
}
  
\newcommand\sfcfigg[2]{%
  \begin{tabular}{@{}ll@{}}
    (a)& (b)\\
    \SFtwocorner{#1}{#2}
  \end{tabular}%
}
\newcommand\sfcfigggg[4]{%
  \begin{tabular}{@{}ll@{}}
    (a)& (b)\\
    \SFtwocorner{#1}{#2}\\
    (c)& (d)\\
    \SFtwocorner{#3}{#4}\\
  \end{tabular}%
}

\title[Turbulence in H II regions]{Turbulence in compact to giant H~II regions}

\author[J. García-Vázquez et al.]{
  J. García-Vázquez$^{1}$\thanks{
    E-mail: \href{mailto:jgarciav1600@alumno.ipn.mx}{jgarciav1600@alumno.ipn.mx}
  },
  William J. Henney$^{2}$
  and H. O. Castañeda$^{1}$\thanks{Deceased.}
\\
$^{1}$Escuela Superior de Física y Matemáticas, Instituto Politécnico Nacional, U.P. Adolfo López Mateos, Zacatenco, Ciudad de México, México C.P. 07738\\
$^{2}$Instituto de Radioastronomía y Astrofísica,
Universidad Nacional Autónoma de México,
Antigua Carretera a Pátzcuaro \#8701,\\
Ex-Hda. San José de la Huerta, 
Morelia, Michoacán, México C.P. 58089\\
}

\date{Accepted XXX. Received YYY; in original form ZZZ}

\pubyear{2023}

\begin{document}
\label{firstpage}
\pagerange{\pageref{firstpage}--\pageref{lastpage}}
\maketitle

\begin{abstract}
  Radial velocity fluctuations on the plane of the sky
  are a powerful tool for studying the turbulent dynamics of emission line regions.
  We conduct a systematic statistical analysis
  of the \halpha{} velocity field for
  a diverse sample of 9 \hii{} regions,
  spanning two orders of magnitude in size and luminosity,
  located in the Milky Way and other Local Group galaxies.
  By fitting a simple model to the second-order spatial structure function
  of velocity fluctuations, we extract three fundamental parameters:
  the velocity dispersion,
  the correlation length,
  and the power law slope.
  We determine credibility limits for these parameters in each region,
  accounting for observational limitations of noise,
  atmospheric seeing, and the finite map size.
  The plane-of-sky velocity dispersion is found to be a better diagnostic
  of turbulent motions than the line width, especially for lower
  luminosity regions where the turbulence is subsonic.
  The correlation length of velocity fluctuations is found to
  be always roughly 2\% of the \hii{} region diameter,
  implying that turbulence is driven on relatively small scales.
  No evidence is found for any steepening of the structure function
  in the transition from subsonic to supersonic turbulence,
  possibly due to the countervailing effect of projection smoothing.
  Ionized density fluctuations are too large to be explained by the action
  of the turbulence in any but the highest luminosity sources.
  A variety of behaviors are seen on scales larger than the correlation length,
  with only a minority of sources showing evidence for homogeneity on the largest scales.
\end{abstract}




\section{Introduction}

Photoionized regions around high-mass stars (\hii{} regions)
show highly vigorous dynamics
as a result of the star formation process and the energy and momentum
injected by the newly-formed stars.
Rather than manifesting a simple pattern such as expansion, infall, or rotation,
the motions are frequently disordered or ``turbulent''
and must be characterised by statistical techniques.

In relatively low luminosity regions, such as the nearby Orion Nebula,
the disordered motions are approximately transonic,
with typical velocities of order \num{5} to \SI{10}{km.s^{-1}}
\citep{castaneda1988, Garcia-Diaz:2008a}.
In larger, higher luminosity regions,
such as 30~Doradus in the Large Magellanic Cloud,
the velocities are significantly supersonic,
of order \SI{30}{km.s^{-1}} \citep{Torres-Flores:2013t, Castro:2018a}.
The same tendency of increasing velocity dispersion (\(\sigma\))
and luminosity (\(L\))
continues up to the scale of entire galaxies
with an approximate relation \(L \propto \sigma^\alpha\) that spans
more than 5 orders of magnitude in luminosity
with a  power-law index \(\alpha = 3\) to~\(7\)
\citep{terlevich1981, Rozas:2006b, Chavez:2014a, Moiseev:2015a}.
However, it is not known whether a single physical mechanism
underlies this relationship at all scales.
The relative importance of gravity and the various stellar feedback mechanisms
(heating, direct radiation pressure, stellar winds, cluster winds)
are unclear in many cases and are frequently disputed \citep{Krumholz:2016a, Melnick:2021x}.

One of the simplest ways to measure the velocity dispersion of an \hii{} region
is to use the Doppler width of a strong emission line, such as the
optical hydrogen recombination line \ha{} \Wav{6563}
\citetext{e.g., \citealp{1986ApJ...300..624R}}.
We will use \(\sigma\los\) to denote
the root-mean-square (RMS) line-of-sight velocity dispersion determined in this way.
Unfortunately, there are many processes that contribute to this width
in addition to the line-of-sight turbulent velocity fluctuations,
such as thermal and fine-structure broadening, instrumental broadening,
dust scattering, 
and large-scale expansion
\citetext{see \citealp{Rozas:2006b} and \citealp{Garcia-Diaz:2008a}
  for detailed discussion}.
All except the last of these can in principal be approximately corrected for,
which is a reasonable strategy to apply in the case of high-luminosity regions
where the turbulent width is expected to be larger than the correction terms.
However, for lower luminosity regions the turbulent velocities are much smaller,
which limits the accuracy of such a correction process
\citetext{see section 3.4 of \citealp{arthur2016turbulence}}. 

An alternative way to study turbulent motions is to measure
the fluctuations on the plane of the sky of the velocity centroids of an emission line
\citep{von1951methode}.
We will denote the RMS magnitude of these fluctuations by \(\sigma\pos\).
In addition to being unaffected by the various nuisance broadening mechanisms
discussed in the previous paragraph,
this also offers the opportunity to study the spatial scale of the fluctuations
by measuring average velocity differences as a function of
the angular separation between two points.
Different mathematical tools can be used to study these spatial fluctuations,
such as the autocorrelation function \citep{lagrois2011}
and \(\Delta\)-variance \citep{Ossenkopf:2006a}.
In this work, we will concentrate on the second order structure function,
\(B(r)\) (see section \ref{sec:second-order-struct}),
which has been used in many previous studies of velocity fluctuations
in \hii{} regions in our own Galaxy
\citep{munch1958internal, castaneda1988, Roy:1985a, 1992ApJ...387..229O, medina2014}
and in external galaxies
\citep{1961MNRAS.122....1F, Medina-Tanco:1997a, lagrois2009multi, lagrois2011, Melnick:2021x}.

In this paper,
we employ archival data from a wide variety of integral field and multi-longslit
spectrographic datasets to analyze spatially resolved \ha{} velocity maps for a sample of
nine \hii{} regions (Figure~\ref{fig:hii-regions}),
ranging in size from \SI{0.5}{pc} to \SI{200}{pc}
and in \ha{} luminosity from \SI{e37}{erg.s^{-1}} to  \SI{e39}{erg.s^{-1}}.
The observations and the physical characteristics of our sample
are described in section~\ref{sec:HIIsample}.
For studying the turbulence we apply a uniform methodology to the analysis of the structure function
by fitting a simple functional form that consists of a power law of slope \(m\)
at small separations,
transitioning to a constant value at separations larger than
an autocorrelation length \(r_0\).
This is described in section~\ref{sec:met}.
We also take into account three observational limitations of the datasets:
the finite angular resolution, \(s_0\) (set by pixel spacing or atmospheric seeing);
the finite map size, \(L_{\text{box}}\);
and a contribution of spatially uncorrelated noise, \(B_0\),
to the observed structure function.
In Appendix~\ref{sec:degr-struct-funct} we use simulated velocity maps to derive
appropriate functional forms for the effects of \(s_0\), \(L_{\text{box}}\) and \(B_0\).
In our fits, we marginalize over these nuisance parameters to obtain robust
credibility limits for the parameters of astrophysical interest:
\(\sigma\pos\), \(r_0\), and \(m\).
The results of these fits are presented in section~\ref{sec:results}.
In section~\ref{sec:discussion} we discuss the relation
of our results to previous studies and the correlations that we find
between the structure function parameters and other properties of
the \hii{} regions in our sample.
Finally, in section~\ref{sec:conclusions} we present our conclusions.


\section{\boldmath Observations of the \hii{} region sample}
\label{sec:HIIsample}

\begin{figure*}
  \centering
  \includegraphics[width=\linewidth]{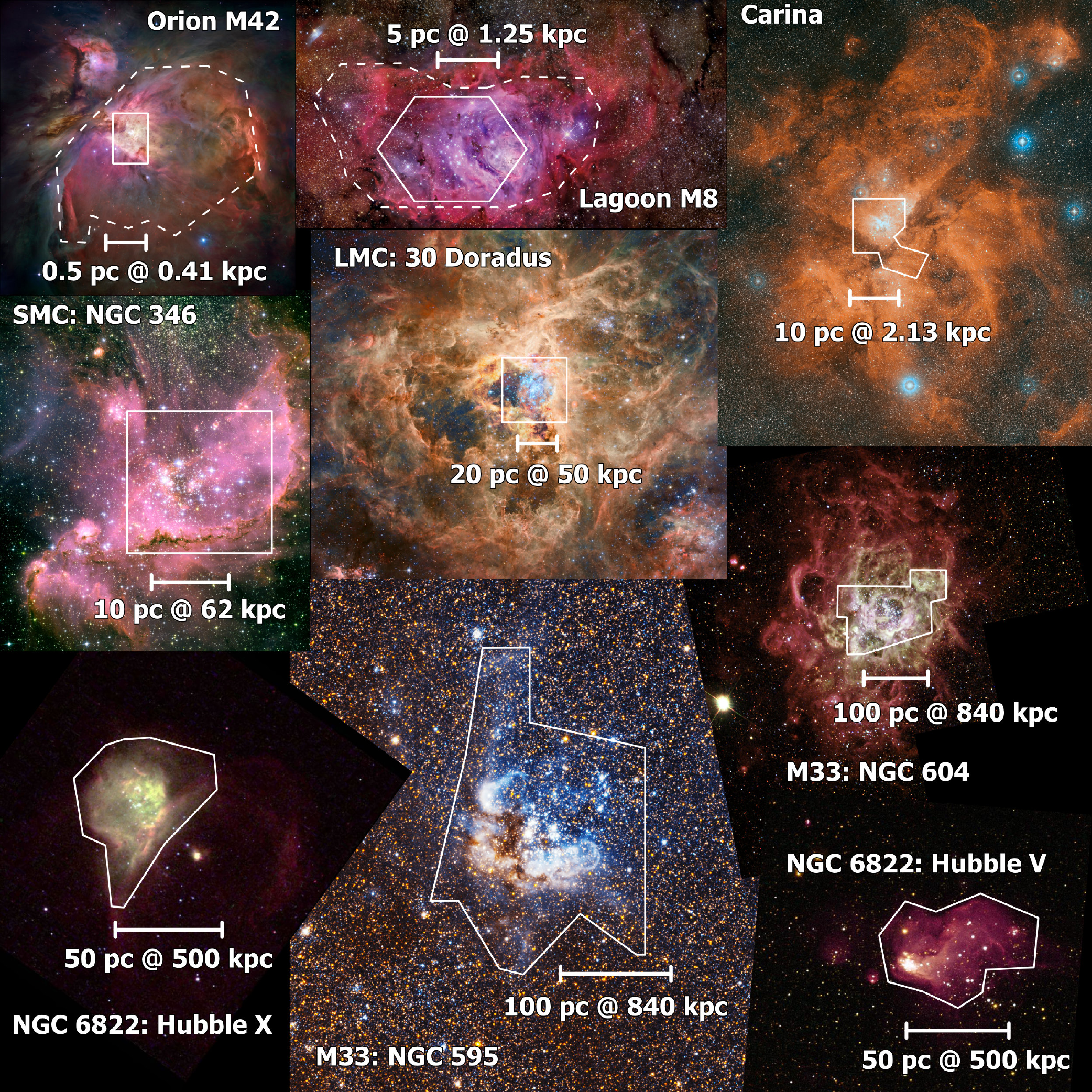}
  \caption{
    The sample of nine \hii{} regions employed in this study,
    arranged from top to bottom in order of distance.
    White shapes show the approximate extents of the
    emission line maps that we use for studying the ionized gas kinematics.
    Scale bars show the equivalent linear length scales at the distance to each region.
    For Orion, two fields are shown: the Extended Orion Nebula,
    sampled at a scale of \SI{0.1}{pc}, and the central Huygens region,
    sampled at a scale of \SI{0.004}{pc}.
    All images are combinations of optical narrow-band and wide-band filters.
    In most cases, red/orange/pink represents \ha{} or other emission lines
    emitted by the ionized gas,
    while blue/white represents continuum starlight from young high-mass stars.
    In some cases, green (for Hubble~X and NGC~604) or blue (for NGC~595)
    represents the [\ion{O}{3}] \Wav{5007} emission line.
    \textit{Image credits as follows.}
    \textbf{Orion Nebula}:
    \textit{HST} Treasury Program on the Orion Nebula Cluster \citep{Robberto:2013a}.
    \textbf{M8 Lagoon}: \href{https://www.cosmotography.com/index.html}{R.~Jay Gabany}.
    \textbf{Carina}: \href{https://www.eso.org/public/images/eso0905b}{
      ESO/Digitized Sky Survey~2, Davide De Martin}.
    \textbf{NGC~346}:
    NASA, ESA and A. Nota (ESA/STScI) \citep{Nota:2006x}.
    \textbf{30 Doradus}:
    \href{http://www.robgendlerastropics.com/Tarantula-HST-ESO.html}{
      Robert Gendler, Roberto Colombari,
      Hubble Legacy Archive, European Southern Observatories}.
    \textbf{Hubble~X}:
    \href{https://hubblesite.org/contents/media/images/2001/01/1012-Image.html}
    {C. R. O'Dell (Vanderbilt University),
      NASA and The Hubble Heritage Team (STScI/AURA)}.
    \textbf{Hubble~V}:
    \href{https://hubblesite.org/contents/media/images/2001/39/1126-Image.html}
    {C. R. O'Dell (Vanderbilt University)
      and L. Bianchi (Johns Hopkins University and Osservatorio Astronomico, Torinese, Italy),
      NASA, ESA, and The Hubble Heritage Team (STScI/AURA)}.
    \textbf{NGC~595}:
    \href{https://esahubble.org/images/heic1901c/}
    {NASA, ESA,
      and M. Durbin, J. Dalcanton, and B. F. Williams (University of Washington)}.
    \textbf{NGC~604}:
    \href{https://hubblesite.org/contents/media/images/2003/30/1423-Image.html}
    {NASA and The Hubble Heritage Team (AURA/STScI)}.
  }
  \label{fig:hii-regions}
\end{figure*}

\begin{figure*}
  \centering
  \includegraphics[width=\linewidth]{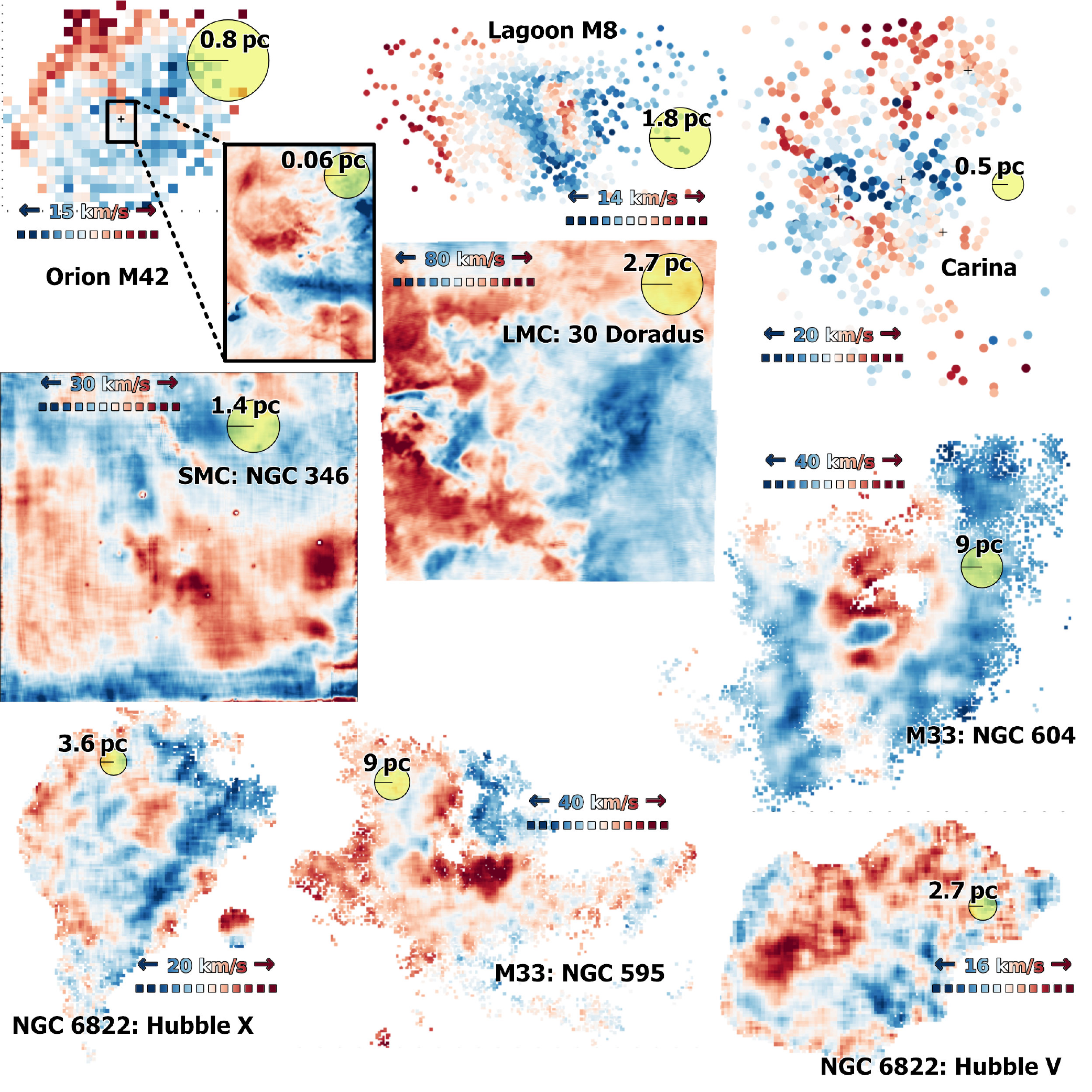}
  \caption{
    Maps of the mean \ha{} velocity for each of the regions in our sample.
    All velocities are relative to the systemic velocity of each region,
    with more negative velocities shown in blue and more positive velocities in red.
    The range of velocities is different for each region, as indicated on the individual maps.
    The correlation length of the velocity fluctuations in each region
    is shown as the radius of a yellow circle and labeled with its value in parsecs. 
  }
  \label{fig:velocity-maps}
\end{figure*}

Previous investigations of the centroid velocity structure function in \hii{} regions
have used a variety of methodologies, which makes it difficult to compare results
between different regions.  \citet{arthur2016turbulence} summarise historical results
for the Orion Nebula in their Table~5.
The most significant differences are seen when different emission line tracers are used,
but even when using the same line, there is some variance between different studies
in the derived values for both the velocity dispersion \(\sigma\pos\) and power-law slope \(m\).
It is therefore worthwhile to employ a uniform approach across a variety of different regions.
To that end, we have selected 9 \hii{} regions,
covering a broad range in size and luminosity,
for which good quality mapping of the \ha{} line exists in the literature
or in data archives.
Figure~\ref{fig:hii-regions} shows optical images of
each region in our sample
and Table~\ref{tab:regions-properties} lists their most important physical parameters.
The derived centroid \ha{} velocity maps are shown in Figure~\ref{fig:velocity-maps}.
Our sample includes three Milky Way regions
(at approximate distance of \num{0.5} to \SI{2}{kpc}),
two regions in the Magellanic Clouds (\num{50} to \SI{60}{kpc}),
and four regions in more distant galaxies of the Local Group
(\num{500} to \SI{800}{kpc}).
Further details of the observations are given in the following section.

\subsection{Spectroscopic datasets}
\label{sec:spectr-datas}

\newcommand\xx{\ensuremath{\boldsymbol{x}}}
Wherever possible, we calculate the centroid velocity \(V_c\)
directly as the normalized first velocity moment of the spectral line intensity profile
for each plane-of-sky position \(\xx\) on the nebula:
\(V_{c}(\xx) = M_1(\xx) / M_0(\xx)\).
The \(k\)th unnormalized moment \(M_k\) of the continuum-subtracted
spectral intensity profile \(I(v)\)
is defined as
\begin{equation}
  \label{eq:kth-moment}
  M_k = \int_{\Delta v} I(v) \, v^k \, dv
\end{equation}
where the Doppler velocity \(v\) is calculated according to
the \texttt{VOPT} convention
\citetext{Eq.~[32] of \citealt{Greisen:2006a}}:
\begin{equation}
  \label{eq:optical-velocity}
  v \equiv c\, (\lambda - \lambda_0) / \lambda_0 ,
\end{equation}
where \(\lambda\) is the observed wavelength,
\(\lambda_0\) is the rest wavelength, and \(c\) is the speed of light.
The spectral window \(\Delta v\) is chosen to be just large enough to include
the entire \ha{} emission line.
So long as the signal-to-noise is high enough,
this mean velocity can be calculated to a much higher precision
than the nominal velocity resolution of the spectrograph. 
In a few cases, as noted below, we are working with already reduced data,
which are provided in the form of Gaussian fits to the line profiles.
In the cases where multiple Gaussian components have been fitted,
we take the flux-weighted mean velocity of these components to be the centroid.

The observed RMS spectral line width \(\sigma'\) can likewise be found from
the second velocity moment as
\begin{equation}
  \label{eq:observed-sigma}
  \sigma'(\xx)^2  = [M_2(\xx) / M_0(\xx)] - V_{c}(\xx)^2 .
\end{equation}
However, many different broadening mechanisms contribute to this width,
including the finite spectral resolution \(\sigma_{\mathrm{ins}}\),
fine-structure splitting \(\sigma_{\mathrm{fs}}\),
and thermal Doppler broadening \(\sigma_{\mathrm{t}}\).
In the approximation that all these broadening mechanisms are independent
and described by Gaussian profiles, then the dispersion of
line-of-sight velocities due to bulk motions \(\sigma\los\) can be found
by subtracting these contributions in quadrature from the observed width:
\begin{equation}
  \label{eq:sigma-los}
  \sigma\los^2 = \sigma'^2 - \sigma_{\mathrm{ins}}^2 - \sigma_{\mathrm{fs}}^2 - \sigma_{\mathrm{t}}^2 .
\end{equation}
See sec~5.1 of \citet{Garcia-Diaz:2008a} for further details.

\subsubsection{Longslit echelle spectroscopy}
\label{sec:longsl-echelle-spect}

For the Orion Nebula, we use data obtained with the echelle spectrograph attached to the \SI{4}{m} telescope at Kitt Peak National Observatory (KPNO) initially published in
\citet{Doi:2004a}.
These observations cover a \(\SI{3}{arcmin} \times \SI{5}{arcmin}\) region of the
central (Huygens) part of the Orion Nebula.
The data consist of 96 North--South orientated \SI{300}{arcsec} slits,
uniformly spaced at \SI{2}{arcsec} intervals with a width of \SI{0.8}{arcsec}
with a velocity resolution of \SI{8}{km.s^{-1}}. 
For the centroid velocities, we use the intensity-weighted 
mean velocities calculated by \citet{Garcia-Diaz:2008a},
who used additional observations with East--West oriented slits
to refine the inter-slit velocity calibration. 
Unlike in the previous analysis of this dataset by \citet{arthur2016turbulence},
we do not mask out regions affected by known high-velocity outflows.
This decision was made for consistency with the analysis of more distant regions,
where such a masking out is not possible.

\subsubsection{Fabry-Pérot observations}
\label{sec:fabry-perot-etalaon}

We use the velocity field presented in \citet{1987A&A...176..347H} to analyze the Extended Orion Nebula \citetext{EON henceforth;  \citealp{2008Sci...319..309G}}.
The EON is an extended region in southwest direction from the Huygens region which presents a diminution in the brightness.
The observational data was obtained using a Fabry-Pérot interferometer with a étalon separation of 0.5 mm on the 106 cm-Cassegrain telescope at Observatorium Hoher List. 
A number of fifteen interferograms have been taken with different pointing directions of the telescope's optical axis in the nebula with an exposure time between 10 and 40 minutes. 
The exposures are overlapped and fall into one square of a grid with a width of \SI{1}{arcmin} centered at \(\uptheta^{1}\)Ori~C.

\subsubsection{FLAMES multi-fiber spectroscopy}
\label{sec:flames-multi-fiber}

Archival data from \citet{Damiani:2016a} and \citet{Damiani:2017b} are used
to study the Carina Nebula and Lagoon Nebula, respectively.
These were obtained as a by-product of a study of young stars in their respective regions
as part of the Gaia-ESO Spectroscopic Survey \citep{Gilmore:2012v, Randich:2013m}
using VLT/FLAMES with the GIRAFFE and UVES spectrographs \citep{2002Msngr.110....1P}.
The spectra are from multiple discrete fiber positions in each nebula
(866 positions in Carina; 1089 in the Lagoon),
visible as colored disks in Figure~\ref{fig:velocity-maps}.
The angular separation between fibers varies across the maps with
average nearest-neighbor distance of \SI{21 \pm 17}{arcsec}.
The spectral resolution ranges from  \SI{6}{km.s^{-1}} (UVES) to \SI{16}{km.s^{-1}} (GIRAFFE).
We do not have access to the individual spectra, but instead use the
Gaussian fits to the line profiles from \citep{Damiani:2016a, Damiani:2017b},
obtained from data tables downloaded from the CDS Vizier service.\footnote{%
  \url{https://doi.org//10.26093/cds/vizier.35910074} and
  \url{https://doi.org//10.26093/cds/vizier.36040135}.}
For the Lagoon, we use single-Gaussian fits, while for Carina we take the
flux-weighted mean of the blue and red component of two-Gaussian fits.

\subsubsection{MUSE integral field spectroscopy}
\label{sec:muse-integral-field}

For the Magellanic Cloud regions 30~Doradus and NGC~346 we use data obtained
with the MUSE spectrograph \citep{Bacon:2010a, Bacon:2014a} on the VLT.\@
Each exposure consists of a \(300 \times 300\) pixel spectral image with a plate scale
of \SI{0.2}{arcsec.pixel^{-1}} and a spectral resolution of \SI{110}{km.s^{-1}}
at \ha{}.
For 30~Doradus, four separate exposures are mosaicked to give a square field of
size \(\SI{2}{arcmin}\) \citep{Castro:2018a}. 
We use centroid velocities of single Gaussian fits to the \ha{} line.\footnote{%
  Data kindly provided by Norberto Castro Rodríguez, priv.~comm.
}
For NGC~346, we use a single field that was observed in 2016 as part of
ESO observing program 098.D-0211(A) (P.I. W.-R.~Hamann).
We obtained the pipeline-reduced data cubes from the ESO data archive\footnote{%
  \url{http://archive.eso.org/wdb/wdb/eso/eso_archive_main/query?prog_id=098.D-0211(A)}
}
and extracted the \ha{} profile using tools in the MPDAF python library.\footnote{
  \url{https://mpdaf.readthedocs.io/en/latest/}
} 

\subsubsection{TAURUS-II Fabry-Perot interferometer}
\label{sec:taurus-ii-fabry}

The archival data for the extragalactic giant regions (GEHRs), NGC 604, NGC 595, Hubble X and Hubble V was obtained with the Fabry Perot TAURUS-II instrument
\citep{Gordon:2000v}
on the \SI{4.2}{m} William Herschel Telescope (WHT) and was retrieved from the La Palma archive\footnote{\url{http://casu.ast.cam.ac.uk/casuadc/ingarch}}.
The IPCS-II detector was used with a etalon of \SI{125}{\mu m}.
Each data cube is \(256 \times 256 \times 100\) with a spatial scale of \SI{0.26}{arcsec.pixel^{-1}} with a total integration time of \SI{3600}{s} (\SI{36}{s} per frame).
For more details on the observational setup see table~1 of \citet{sabalisck1995supersonic}.
The reduction and analysis was done with the packages TAUCAL y TAUFITS \citep{1992ASPC...25..445L} and the emission spectrum was adjusted with a single Gaussian fit.
The NGC~604 observations were previously reported and analyzed in
\citet{sabalisck1995supersonic}, \citet{Medina-Tanco:1997a} and \citet{Melnick:2021x}.

\subsection{Physical properties of the sample regions}
\label{sec:regions-milky-way}

\begin{table}
  \begin{center}\caption{Summary of properties of our \hii{} region sample. Diameters are taken from Table 2 of \citet{1984ApJ...287..116K}.
      References for other properties are given in Appendix~\ref{sec:notes-individual-hii}.
      Mean non-thermal linewidth values (\(\langle \sigma\los \rangle\)) shown here are corrected considering equation \eqref{eq:sigma-los} which takes into account \(\sigma_{\mathrm{fs}}\). }
\begin{tabular}{cCCCC}\toprule
\hii{}    &  \text{Distance,}\ d & \text{Diameter,}\ D_{\hii} & \log L(\ha) &  \langle \sigma\los \rangle \\
  Region    &  [\si{kpc}]          &  [\si{pc}]    &  [\si{erg.s^{-1}}]            &    [\si{km.s^{-1}}]  \\ 
\midrule
Orion     & 0.44 \pm 0.02  & 5 \pm 0.5    &    37.18       &    9.9 \pm 1.2\FNa        \\
Lagoon    & 1.25 \pm 0.12   & 25 \pm 2.5      &    37.47    &   11.2 \pm 1.6    \\
Carina    & 2.35 \pm 0.05   & 200 \pm 20      &    39.01    &   18.6 \pm 3.3\FNa      \\
30 Dor    & 50.0 \pm 0.2    & 370 \pm 37    &    39.46    &     21.7 \pm 2.2    \\
NGC 346   & 62 \pm 1        & 220 \pm 22      &    38.77    &    9.6 \pm 1.0     \\
Hubble X  & 490 \pm 40      & 160 \pm 16      &    38.21    &   10.0 \pm 0.02\\
Hubble V  & 490 \pm 40      & 130 \pm 13      &    38.3     &    9.8 \pm 0.03     \\
NGC 595   & 820 \pm 30      & 400 \pm 40      &    38.95    &   16.5 \pm 0.1     \\
NGC 604   & 820 \pm 30      & 400 \pm 40      &    39.42    &   17.5 \pm 0.30     \\
\bottomrule
\multicolumn{5}{l}{\FNa{}Value obtained from our observations.}
\end{tabular}\label{tab:regions-properties}
\end{center}
\end{table} 

Table~\ref{tab:regions-properties} summarizes the properties of the regions in our sample
with data taken from the literature. Further details on each individual source and their references are given in Appendix~\ref{sec:notes-individual-hii}.


\section{Plane-of-sky velocity fluctuations}\label{sec:met}

\begin{figure*}
 \centering
 \includegraphics[width=5in]{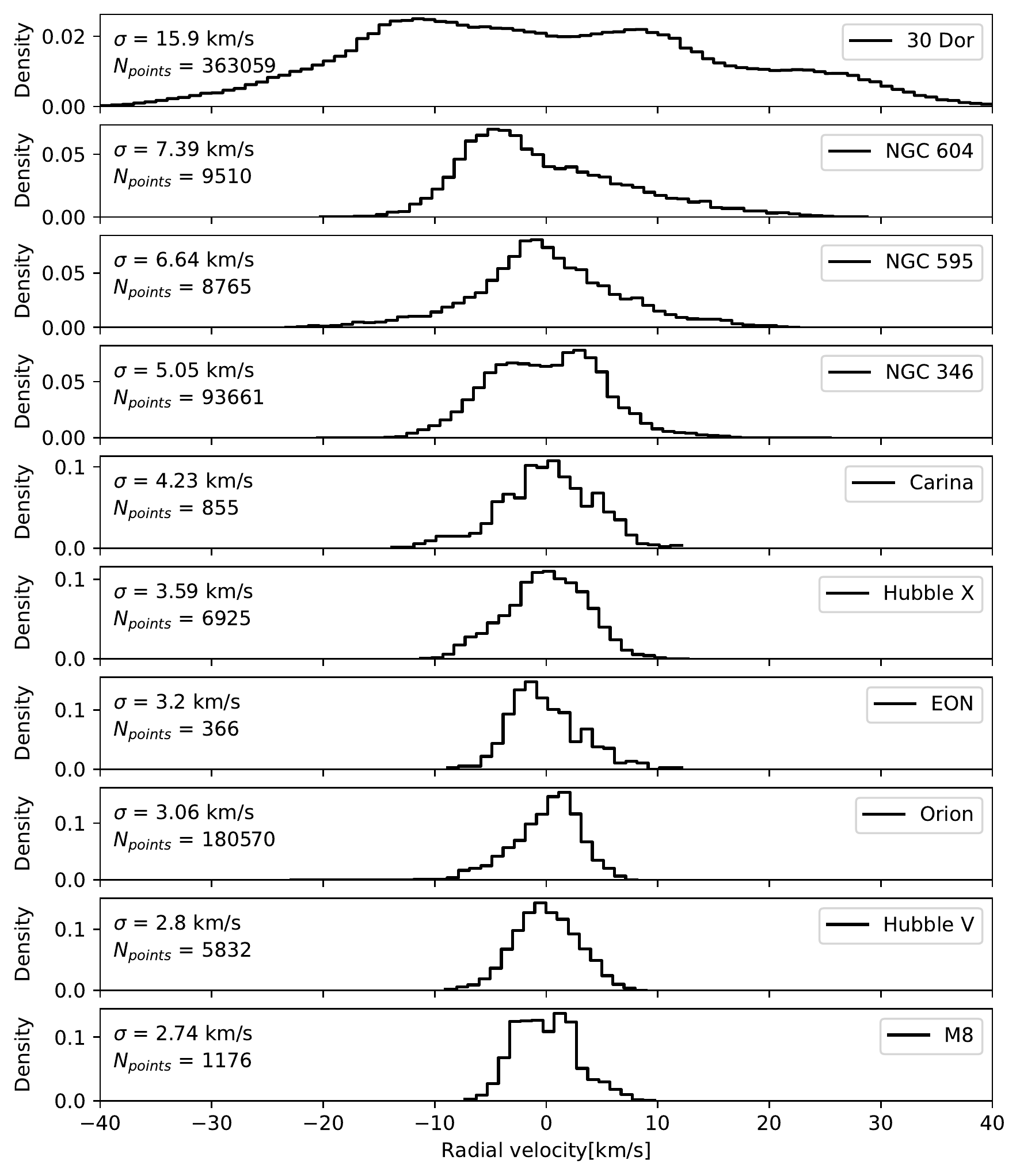}\par
 \caption{
   Histograms of the plane-of-sky variation in the \ha{} centroid velocities.
   Each panel shows the fraction of the observed spatial points
   (pixels or fiber positions)
   with a given velocity offset from the mean systemic velocity of each region.
   The RMS width \(\sigma\pos\) of each distribution is marked,
   as is the number of spatial points.
   The bin width is \SI{1}{km.s^{-1}} and the regions are arranged
   in order of decreasing \(\sigma\pos\).
 }
 \label{fig:pdfs}
\end{figure*}

Figure~\ref{fig:velocity-maps} shows maps of the \ha{} centroid velocity
\(V_c\) for each \hii{} region in our sample,
calculated as described above in section~\ref{sec:spectr-datas}
and after subtracting the mean systemic velocity,
\(\left\langle V_c\right\rangle\) in each case.
The overall amplitude of the plane-of-sky
velocity fluctuations can be characterized by a dispersion,
\(\sigma\pos\), defined as:
\begin{equation}
  \label{eq:sig-pos}
  \sigma^2\pos =
  \left\langle 
  \bigl[ V_c (\xx_i) -\langle V_ c\rangle  \bigr]^2
  \right \rangle ,
\end{equation}
where the average is performed over all observed points \(i\)
in a given map.
Note that \(\sigma\pos\) is also the RMS width of
the probability density function (PDF) of \(V_c\).
This PDF for each region is shown in Figure~\ref{fig:pdfs}
on a common velocity scale and arranged in order of decreasing \(\sigma\pos\).
It can be seen that the amplitude of the
velocity fluctuations varies considerably between regions,
with \(\sigma\pos\) in 30~Dor being more than 5 times higher than in 
nearby regions such as Orion.

All of the PDFs are significantly non-Gaussian
and many show evidence of a multi-modal structure.
Some, such as Orion core and Hubble~V, show a single dominant velocity component,
but most show several distinct peaks,
as many as four in the case of Carina.
There is no clear tendency for the number of identifiable velocity components
to increase with increasing velocity dispersion.
Instead, both the widths of the individual peaks and the separation between them
seem to increase in tandem.
For most regions the PDFs are approximately symmetrical,
but two regions show significant skewness:
Orion~core has a smooth asymmetrical tail towards the blue,
whereas NGC~346 has one towards the red.

\subsection{The second-order structure function}
\label{sec:second-order-struct}

In order to probe the dependence of the velocity fluctuations
on spatial scale,
the primary tool that we employ is
the second order structure function of differences in velocity centroids,
$B(r)$, which is a function of the scalar separation or lag, \(r\),
between two points on the plane of the sky:
\newcommand\Abs[1]{\vert #1\vert}
\begin{equation}\label{eq:Br}
  B(r) = \left\langle 
  \bigl[
  V_{c}(\xx_j) - V_{c}(\xx_i)
  \bigr]^{2} \right \rangle_{\Abs{\xx_j - \xx_i\!} \ \approx \ r} \ .
\end{equation}
The averaging is performed over all pairs of points
\((i, j)\)
whose scalar separation \(\Abs{\xx_j - \xx_i}\) is close to \(r\),
irrespective of the orientation of the separation vector.
In practice, we achieve this by binning the separations with a constant
logarithmic width of \SI{0.05}{dex}.

We will also employ the related quantity of the
normalized spatial autocovariance or autocorrelation function:
\begin{equation}
  \label{eq:autocovar}
  C(r) = \frac{1}{\sigma^2\pos}\left\langle 
  \bigl[
  V_{c}(\xx_j) \  V_{c}(\xx_i)
  \bigr] \right \rangle_{\Abs{\xx_j - \xx_i\!} \ \approx \ r} \ .
\end{equation}
If the fluctuations are perfectly spatially homogeneous, then the
two quantities are related \citep{1984ApJ...277..556S} as:
\begin{equation}\label{eq:functional}
  B(r) = 2\sigma^2\pos \bigl[   1 - C(r)\bigr] .
\end{equation}
In less ideal situations, then \(B(r)\) is to be preferred since it is
relatively unaffected by non-stationary effects
such as large-scale linear gradients 
\citep{1984ApJ...277..556S}.
Nonetheless, \(C(r)\) is more amenable to heuristic reasoning in some cases,
which we will take advantage of below.

\subsection{A heuristic model for the structure function}
\label{sec:methods-apply}

\begin{figure*}
 \centering
 \includegraphics[width=\linewidth]{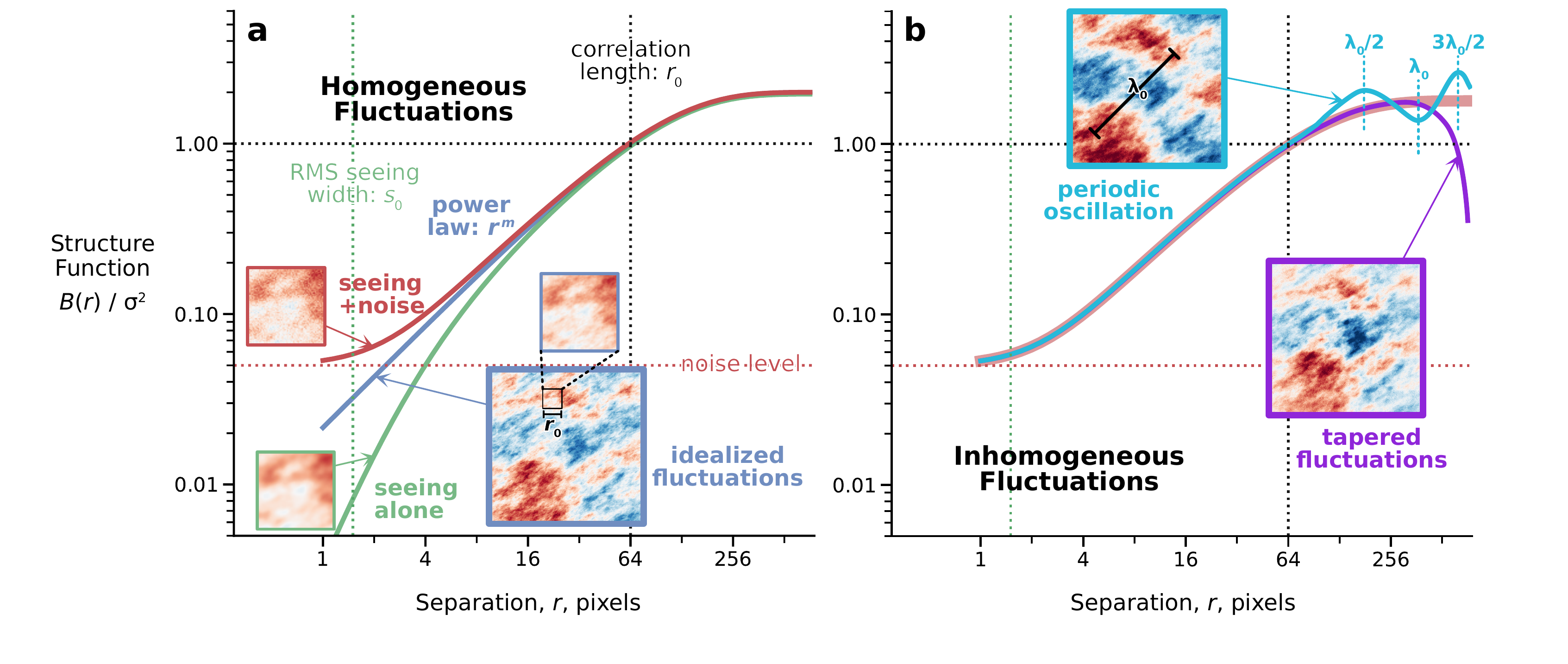}\par
 \caption{
   Example of the model structure function for homogeneous fluctuations,
   together with a realization of the corresponding velocity field
   on a \(512 \times 512\) pixel grid.  
   (a)~Idealized structure function (blue line)
   for the parameters \(m = 1\) and \(r_0 = \SI{64}{pixels}\).
   Red and green lines show the effects of observational limitations
   at small scales due to seeing and noise.
   (b)~Examples of inhomogeneous effects at large scales,
   which are not included in the model structure function.
   Note how hard it is to detect by eye the difference between these
   fields and the one shown in panel~a, despite the important changes
   in the large-scale end of the structure function.
 }
 \label{fig:model-strucfunc}
\end{figure*}

A common property of homogeneous fluctuating velocity fields is that neighboring points tend to have similar velocities
(\(C(r) \approx 1\) for small \(r\)),
whereas points that are far apart may have very different velocities
(\(C(r) \ll 1\) for large \(r\)).
The value of the separation that corresponds to
the transition between these two regimes
is called the correlation length, \(r_0\).
In the simplest case,
two points separated by \(r \gg r_0\) have totally uncorrelated velocities
in the sense that knowledge of the velocity at the first point is of
no help in predicting the velocity at the second point.
At scales smaller than \(r_0\), the fluctuations often show a power-law behavior
as a function of \(r\).

In order to capture these two behaviors,
we therefore propose the following idealized 2-parameter model
for the autocorrelation function:
\begin{equation}\label{eq:new-correlation-form}
  C\model(r;\ r_0, m) = 2^{- \left( r/r_0 \right)^m} 
\end{equation}
in which \(r_{0}\)\ is the correlation length (see above)
and \(m\) is the power-law slope at small scales.
This is constructed so that \(C\model(r) = 1/2\) at \(r = r_0\),
while the exponential form ensures that \(C(r)\) rapidly approaches zero
for larger separations.
We assume the validity of equation~\eqref{eq:functional} to determine the structure function
from this model autocorrelation function:
\begin{equation}
  \label{eq:model-strucfunc-ideal}
  B\model(r) = 2\sigma^2\pos \left[
    1 - 2^{- \left( r/r_0 \right)^m} 
  \right]
\end{equation}
This has the following properties:
\begin{enumerate}[1.]
 \item Small scales: \(B\model(r) \propto r^m\) for \(r \ll r_0\);
 \item Correlation scale: \(B\model(r_0) = \sigma\pos^2\);
 \item Large scales: \(B\model(r) \to 2 \sigma\pos^2\) for \(r \gg r_0\).
 \end{enumerate}
An example is shown by the blue line in Figure~\ref{fig:model-strucfunc}a.
 
In a previous paper, we used a different functional form
\citetext{See Fig.~13 of \citealp{arthur2016turbulence}}:
\(C(r) = 1/[1+(r/r_{0})^{m}]\), as proposed in \citet{1984ApJ...277..556S},
which behaves identically to equation~\eqref{eq:new-correlation-form}
in the first two regimes, but is more gradual in its approach
to the large-scale asymptote of \(2 \sigma\pos^2\).
However, we find that equation~\eqref{eq:new-correlation-form}
provides a much improved fit to the observed structure functions
of our sample regions (see following section). 

\newcommand\LL{\ensuremath{\mathcal{L}}}
Slightly different definitions of the correlation length
are found in the literature.
For instance the definition of \citet{Jaupart:2022i}
in two dimensions is
\begin{equation}
  \label{eq:2}
  \ell_c^2 = \frac{1}{4} \int C(r) \, d^2 r
  = \frac{\pi}{2} \int_0^\infty r C(r) \, d r
\end{equation}
while the ``integral length scale'' \citep{Pope:2000p} is defined
as
\begin{equation}
  \label{eq:3}
  \LL \equiv \int_0^\infty C(r)\, dr .
\end{equation}
For our model structure function, these evaluate to
\begin{equation}
  \label{eq:4}
  \LL = \frac{\Gamma(1/m)}{m\, (\ln 2)^{1/m}} \, r_0
\end{equation}
and
\begin{equation}
  \label{eq:4}
   \ell_c^2 = \frac{\pi\,  \Gamma(2/m)}{2m\, (\ln 2)^{2/m}} \, r_0^2
\end{equation}
where \(\Gamma\) is the usual Gamma function.
For example, if \(m = 1\) these become \(\LL \approx 1.443 r_0\)
and \(\ell_c \approx 1.808 r_0\). 

\subsubsection{Effects of observational limitations at small spatial scales}
\label{sec:effect-observ-limit}

Two observational effects can modify the observed structure function at the smallest scales.
The first is the blurring of the observed image by atmospheric seeing,
which tends to reduce \(B(r)\) for small \(r\).
In Appendix~\ref{sec:effects-seeing-struc}, we perform numerical experiments
with synthetic velocity fields and find that
a good approximation to the effects of seeing is to multiply
the model structure function by a factor:
\begin{equation}\label{eq:ffs}
  \startNEW
  S(r) =
  \left[ \frac{1}{1 + 1.25 s_0 / r_0} \right]
  \left[ \frac{1}{1 + (2.6 s_0 / r)^{1.5}} \right]
  ,
  \stopNEW
\end{equation}
where \(s_0\) is the RMS width of the seeing profile\footnote{%
  Note that the full-width half maximum (FWHM) seeing width is
  \(2 (2 \ln 2)^{1/2} s_0 \approx 2.35 s_0\).
},
and \(r_0\) is the correlation length.
An example is shown by the green line in Figure~\ref{fig:model-strucfunc}a.
The effect of seeing is to make the structure function 
bend down away from the idealized power law, 
becoming increasingly steep at scales smaller than \(s_0\),
but having a noticeable effect at scales up to \(10 \times s_0\).

The second effect is the presence of instrumental noise
in the observational measurements,
such as Poisson noise from photon-counting statistics. 
Although this affects all scales,
it is most noticeable for small separations,
where the intrinsic structure function is smallest.
If the noise is spatially uncorrelated
(equal and independent uncertainties in the velocity
measurement of each pixel),
then its contribution to the structure function is independent of separation,
which means it can be represented as a constant term \(B\noise\).

Combining both the effects of seeing and noise yields the corrected
model structure function:
\begin{equation}
  \tilde{B}\model(r) = B\model(r) \,  S(r) + B\noise
  \label{eq:sf-functional}
\end{equation}
An example of this equation is shown by the red line in Figure~\ref{fig:model-strucfunc}.
Note that seeing and noise have opposite effects on the slope
of the structure function at intermediate scales:
seeing tends to steepen the slope, while noise tends to flatten it.
This can lead to a degeneracy between these two parameters
when fitting observations over a limited range of separations.

\subsubsection{Effects of observational limitations at large spatial scales}
\label{sec:effects-observ-limit-large}

In Appendix~\ref{sec:finite-box-effects} we consider the effects
on the structure function of the finite size \(L\) of the observational map.
If the true correlation length \(r_0\) is more than a small fraction of the map size
(approximately, \(r_0 > 0.1 L\)),
then the apparent velocity variance measured directly from the data
is less than the true \({\sigma\pos}^2\) of the underlying fluctuations
(see Figure~\ref{fig:finite-box-effect}).
This is because the observed map is not large enough to sample
the full range of velocity variations.
At the same time, the apparent correlation length is smaller
than the true value.
However, this does not mean that the model structure function needs to be modified.
In the appendix we show that the model fitting is capable of determining the true fluctuation variance
and true correlation length (both to within 10\%),
so long as \(L > 3\, r_0\).
This is a significant improvement over direct empirical measurement,
which requires \(L > 10\, r_0\) in order to be accurate
(compare panels a and b of Figure~\ref{fig:finite-box-effect}).

\subsubsection{Additional effects omitted from the model structure function}
\label{sec:limit-model-struct}

For separations that are comparable to the diameter \(D\) of the \hii{} region,
then the supposition of homogeneity may break down.
For instance, if the amplitude of the velocity fluctuations is higher
in the core of the region than in the outskirts,
then the autocorrelation function will be U-shaped
since all separations with \(r > D/2\) must be between two points in the outskirts,
which will tend to be relatively well correlated simply because both velocities
will be close to the mean.
This yields a corresponding downturn in the structure function at the largest scales.

Another example is if there were a periodic velocity fluctuation,
with spatial wavelength \(\lambda\),
then the autocorrelation function would become negative for \(r \approx \lambda/2, 3 \lambda / 2, \dots\),
which yields periodic peaks in the structure function.
Both effects are illustrated in Figure~\ref{fig:model-strucfunc}b.
Neither of them can be captured by our model structure function,
which assumes that the autocorrelation is strictly positive
and monotonically decreasing with \(r\).
We deal with this limitation by not attempting to model scales larger than \(L / 2\),
where the influence of these effects will be greatest.

\section{Structure function fits}
\label{sec:results}

\begin{figure*}
  \centering
  \sffigggg{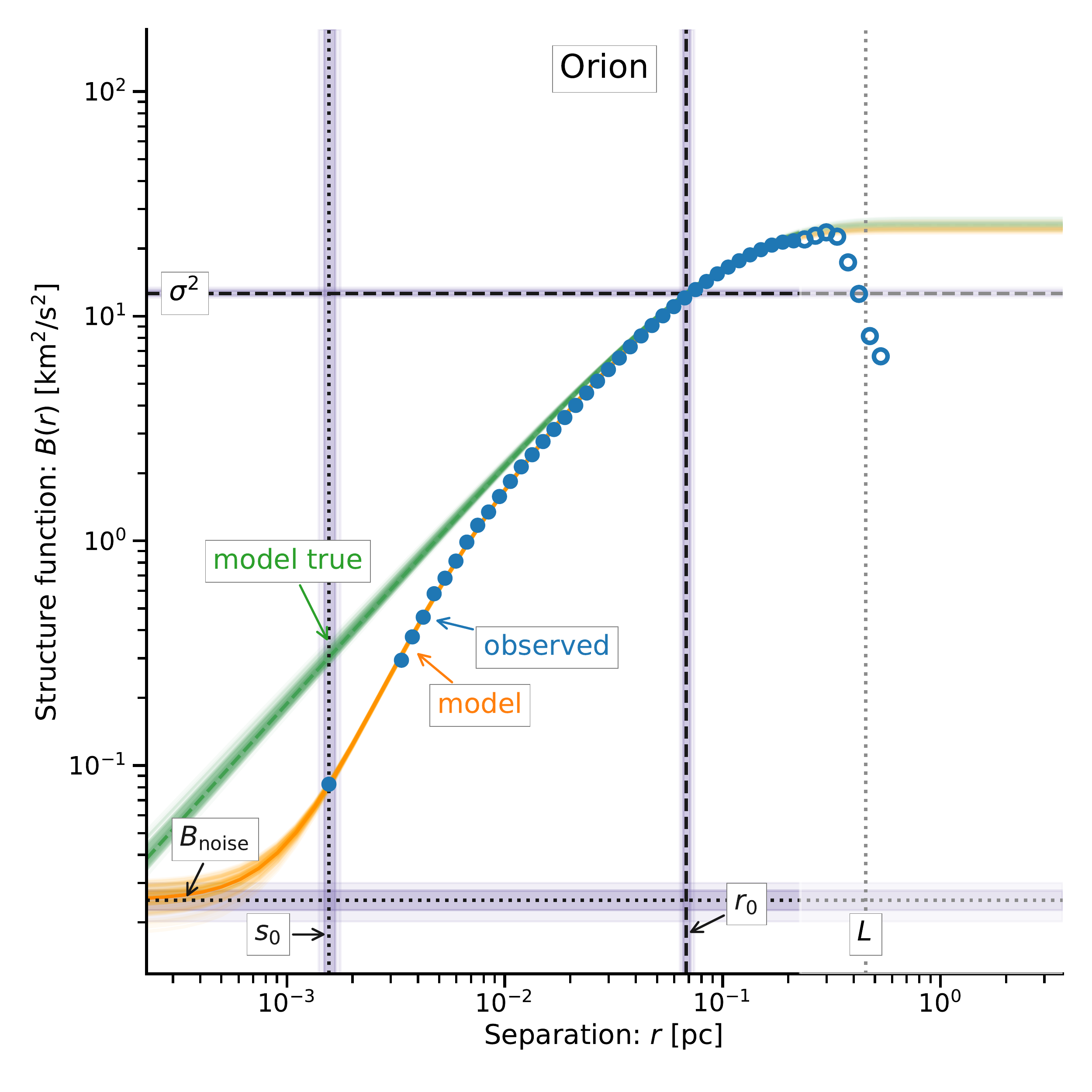}{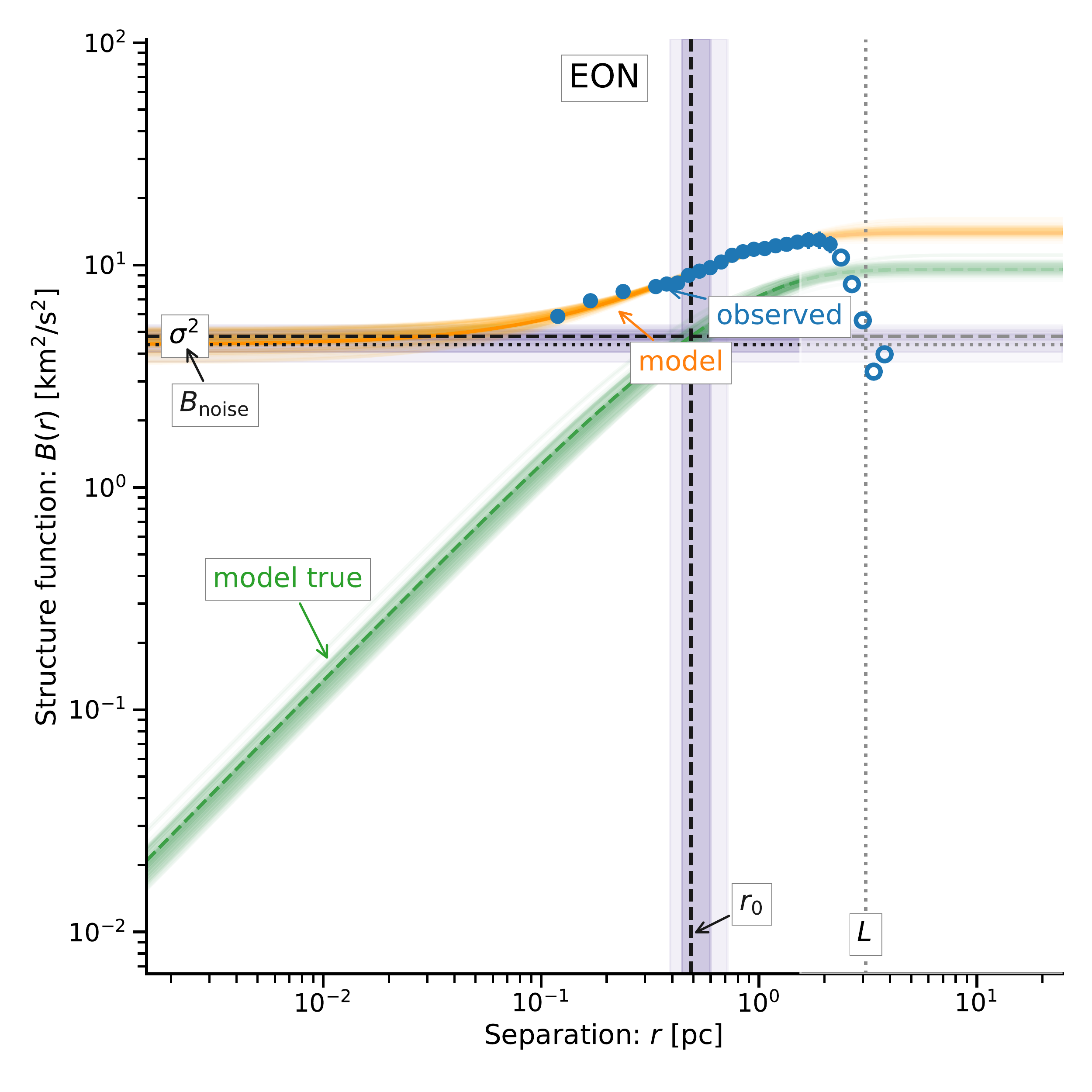}{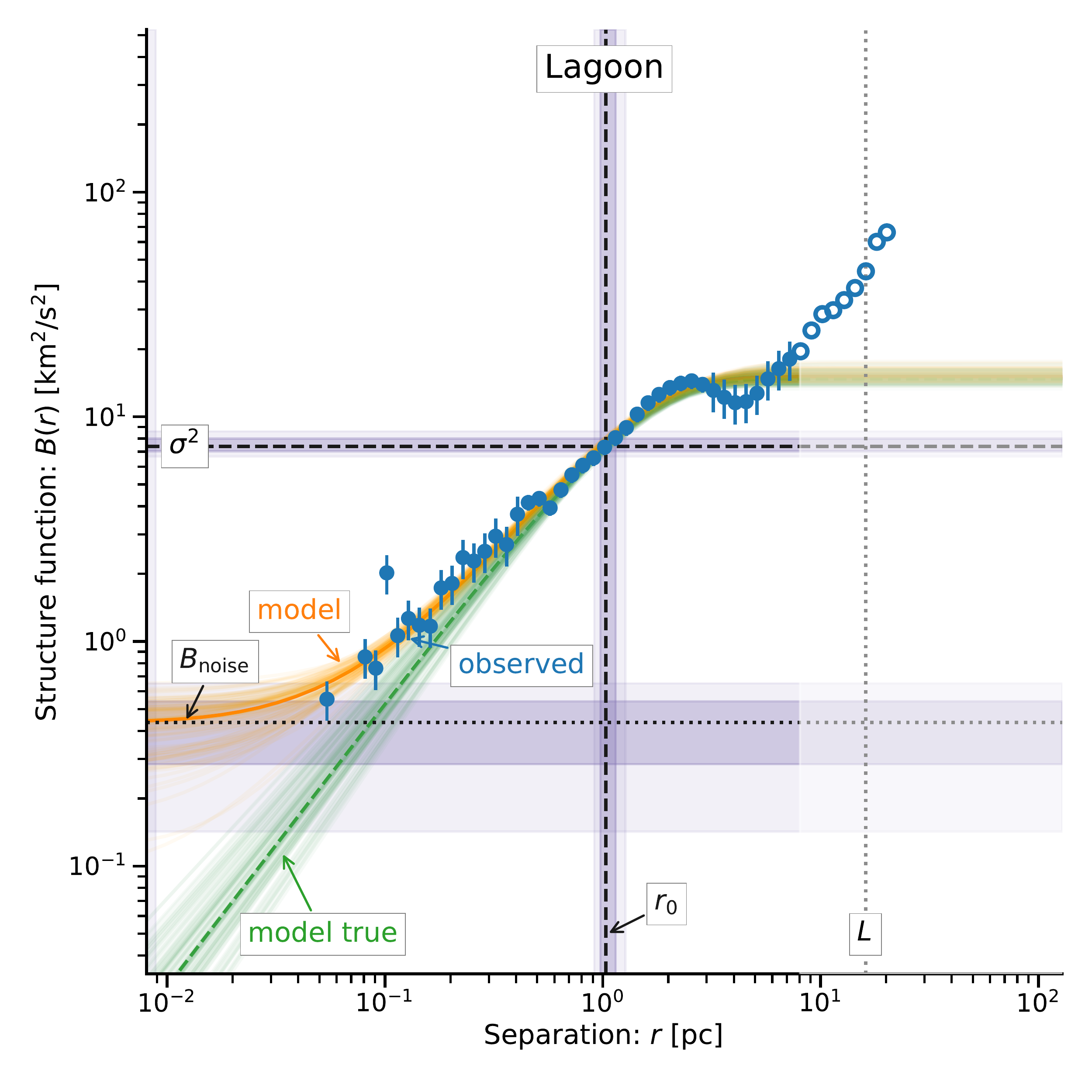}{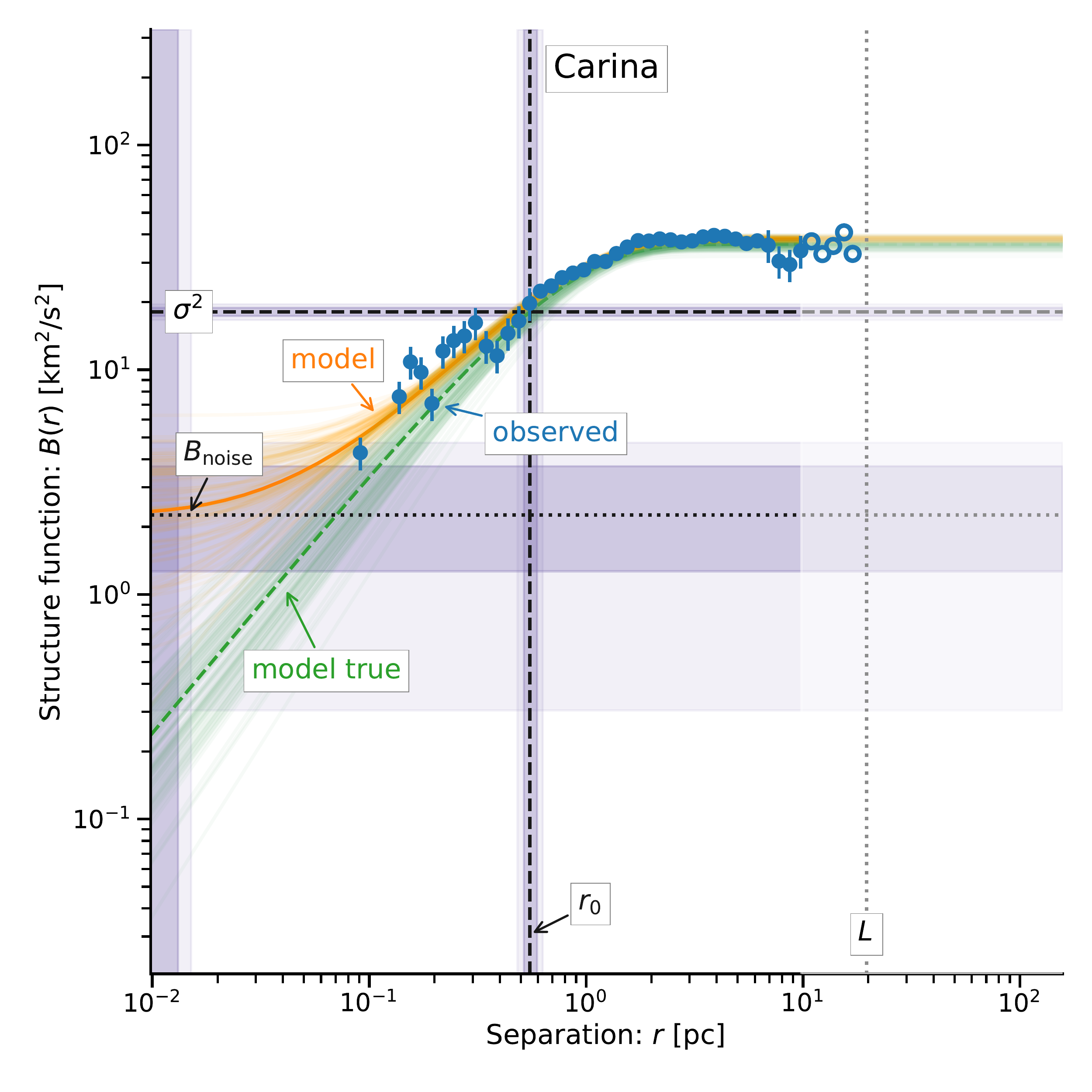}
  \caption{
    Second-order structure functions
    for velocity centroid images
    of the \ha{} emission line
    from Galactic \hii{} regions.
    The observed structure function \(B\obs(r)\) is shown by 
    blue symbols, with filled symbols indicating those points
    that are used to constrain the model fits.
    The full model is shown in orange,
    while the underlying ideal model
    (without the effects of seeing or noise)
    is shown in green.
    See text for details of the fitting process.
    (a)~Inner Orion Nebula.
    (b)~Extended Orion Nebula.
    (c)~M8, the Lagoon Nebula.
    (d)~Carina Nebula.
  }
  \label{fig:strucfunc-fit-Galactic}
\end{figure*}

\begin{table}
  \centering
  \caption{Bounds of allowed values for parameters in model fits}
  \label{tab:parameter-ranges}
  \newlength\partabwidth
  \setlength\partabwidth{0.8\linewidth}
  \begin{tabular*}{\partabwidth}{
    l @{\extracolsep{\fill}}
    r 
    r
    }
    \toprule
    Parameter & Lower & Upper\\
    \midrule
    \(\sigma^2\) & \(0.25\, \max [B\obs]\)& \(2\, \max [B\obs]\)\\
    \(r_0\) & \(0.01\, L\) & \(2\, L\)\\
    \(m\) & \(0.5\) & \(2.0\) \\
    \(s_0\) & \SI{0.1}{arcsec}& \SI{1.5}{arcsec}\\
    \(B\noise\) & \(0\) & \(3\, \min [B\obs]\) \\
    \bottomrule
    \multicolumn{3}{@{}p{\partabwidth}@{}}{
    Note: \(\max[B\obs]\) and \(\min[B\obs]\) are over all bins in the observed structure function with \(r < L/2\).
    }
  \end{tabular*}
\end{table}

\begin{figure}
  \centering
  \includegraphics[width=\SFwidth]{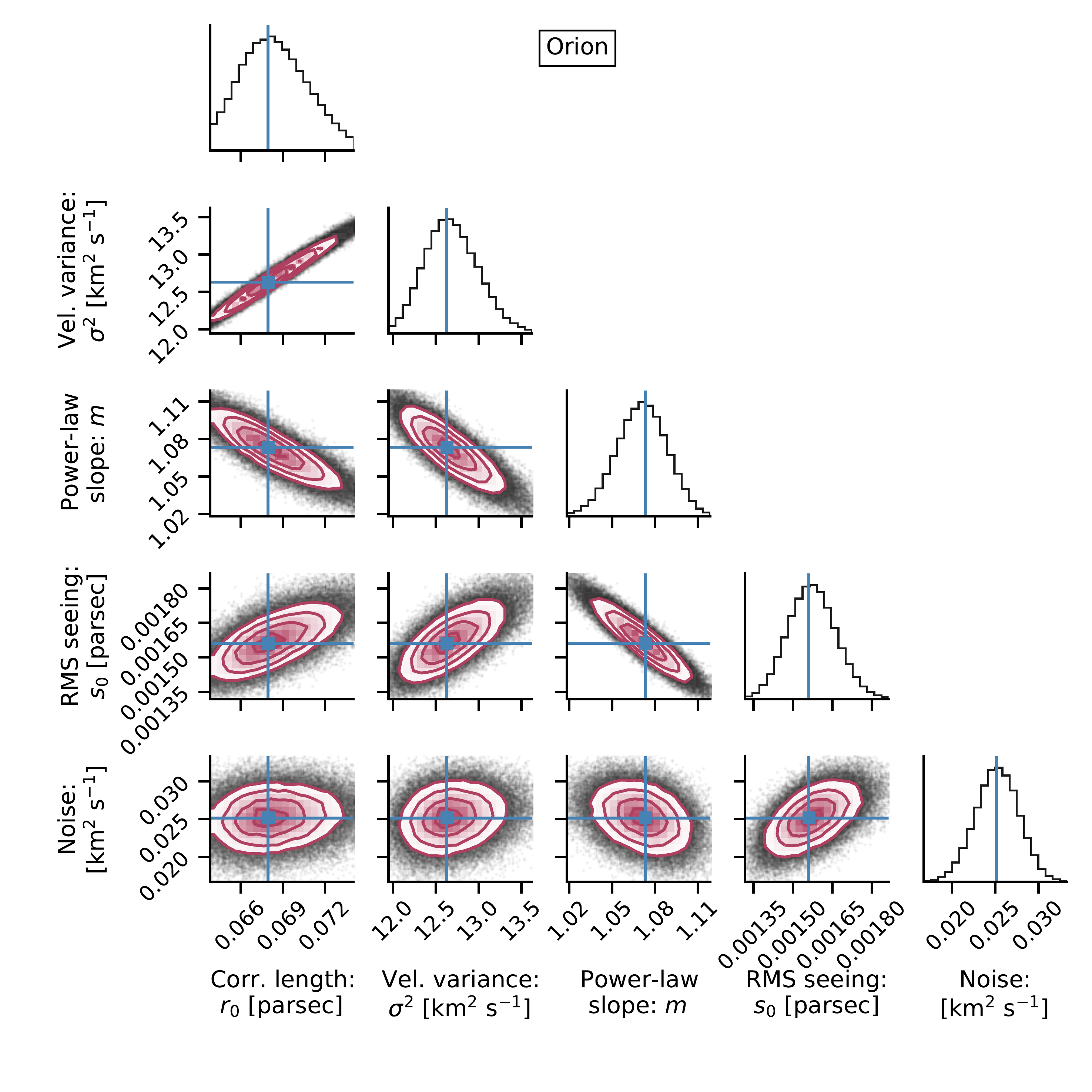}
  \caption{
    Example corner plot of covariances between
    model parameters of fits to the \ha{} structure function.
    The example shown is for Orion,
    while figures for the remaining regions are given in
    Appendix~\ref{sec:addit-covar-corn}.
    Plots on the diagonal show the 1-dimensional histogram
    of the posterior distribution of each parameter
    (labeled at bottom),
    as calculated by the MCMC method,
    assuming a uniform prior distribution
    within the limits given in Table~\ref{tab:parameter-ranges}.
    Off-diagonal plots show the 2-dimensional histogram of the
    joint posterior distribution each pair of parameters.
  }
  \label{fig:corner-example-Orion}
\end{figure}

\begin{figure*}
  \centering
  \sffigg{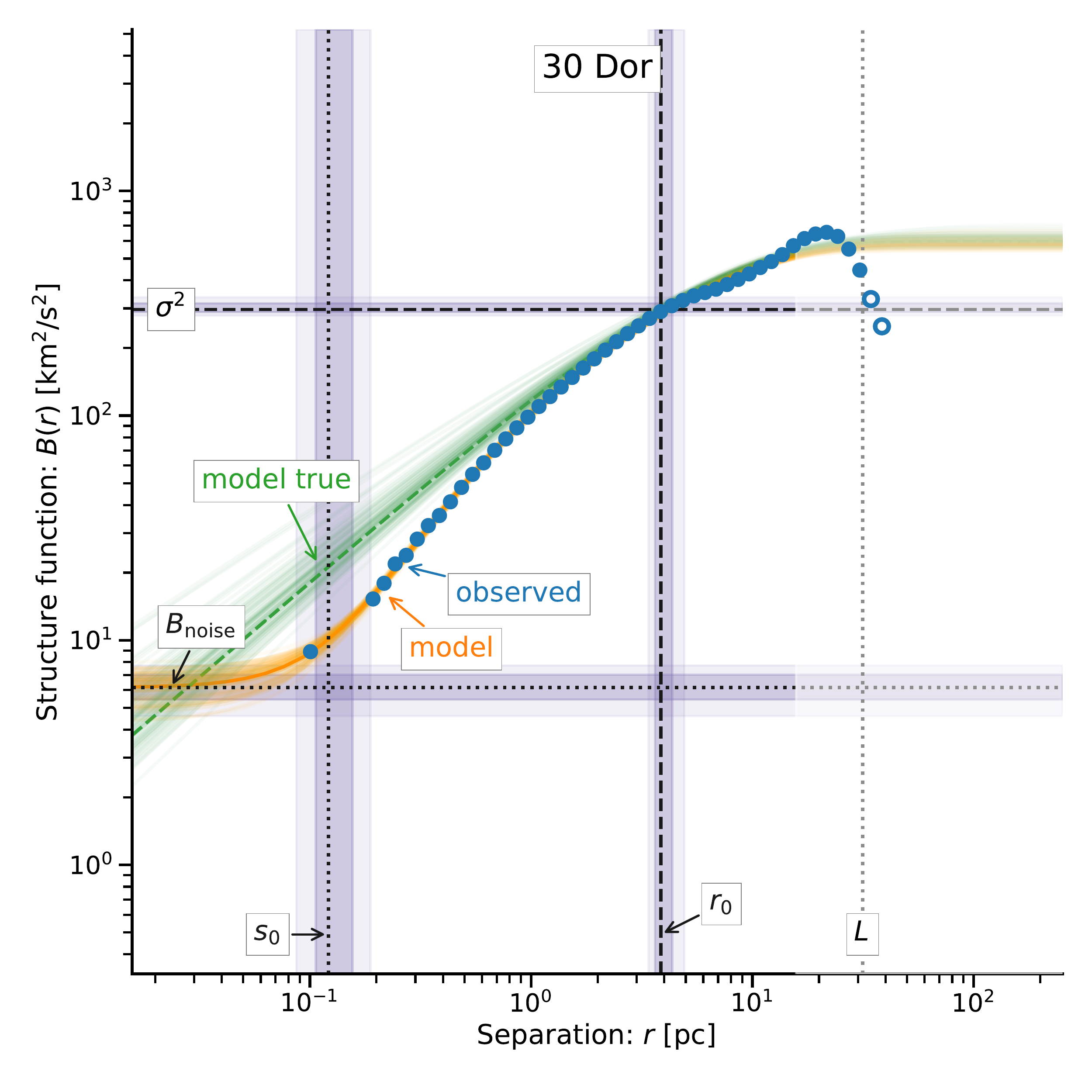}{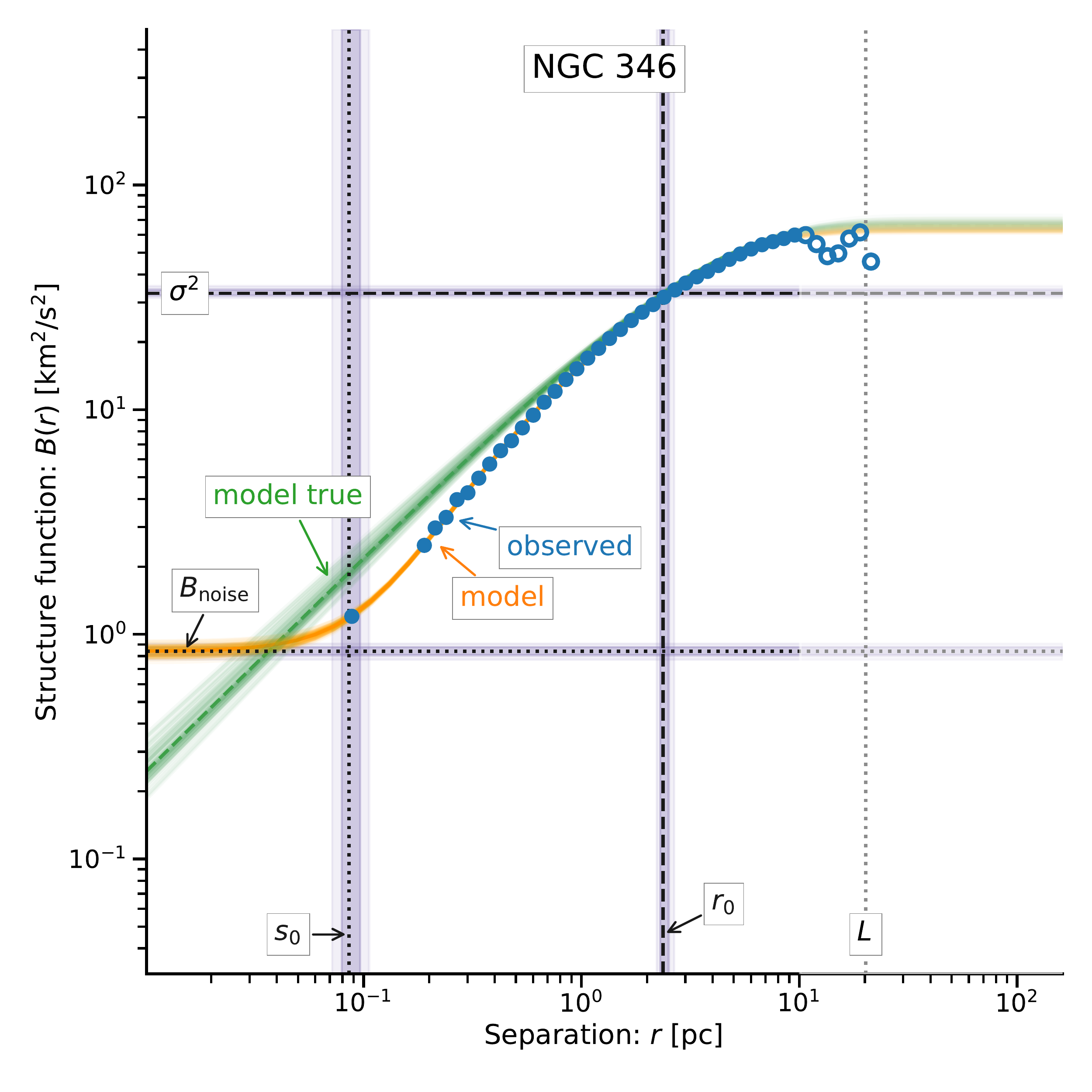}
  \caption{
    Same as figure~\ref{fig:strucfunc-fit-Galactic}
    except for Magellanic Cloud \hii{} regions.
    (a)~30 Doradus in the LMC.
    (b)~NGC~346 in the SMC.    
  }
  \label{fig:strucfunc-fit-MC}
\end{figure*}

\begin{figure*}
  \centering
  \sffigggg{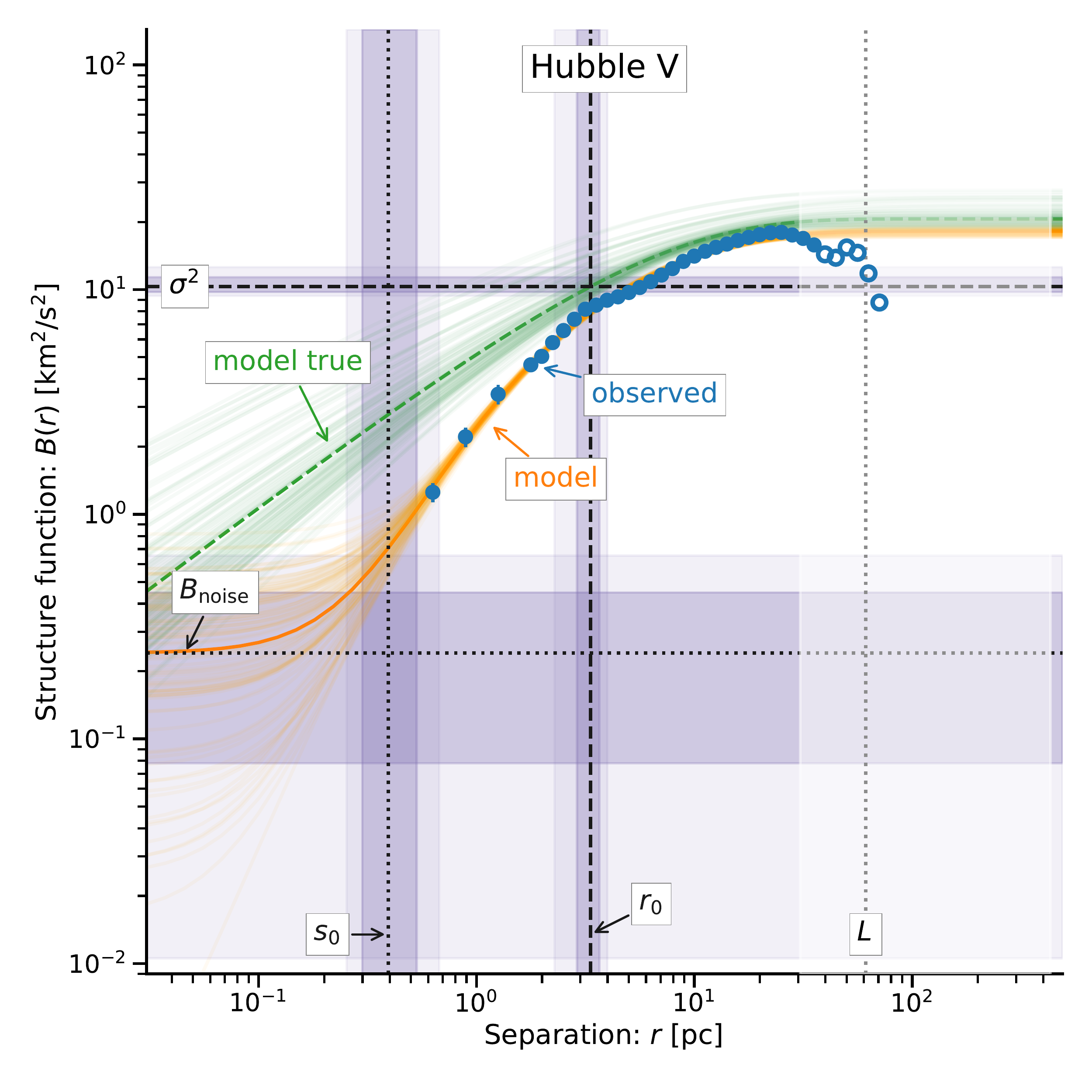}{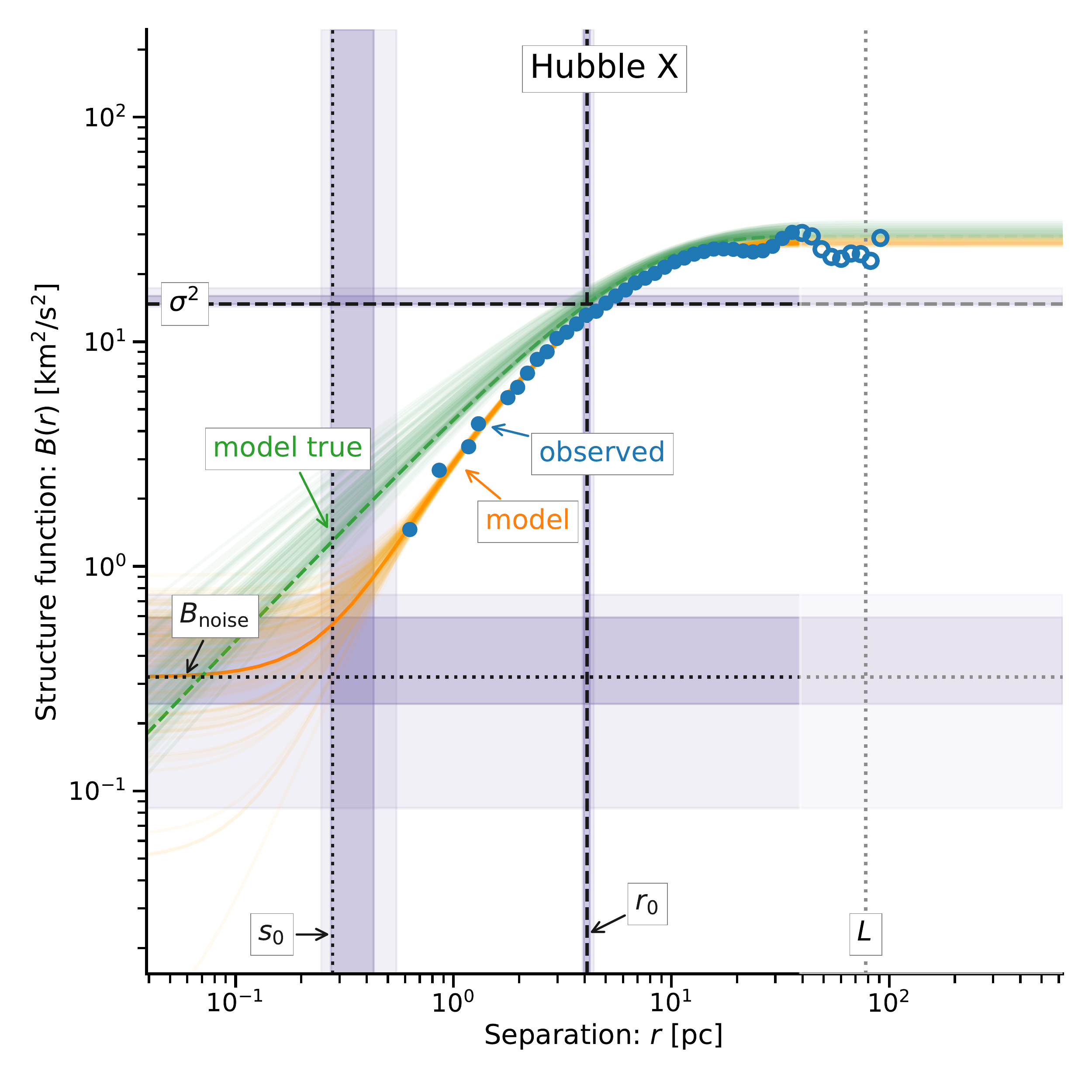}{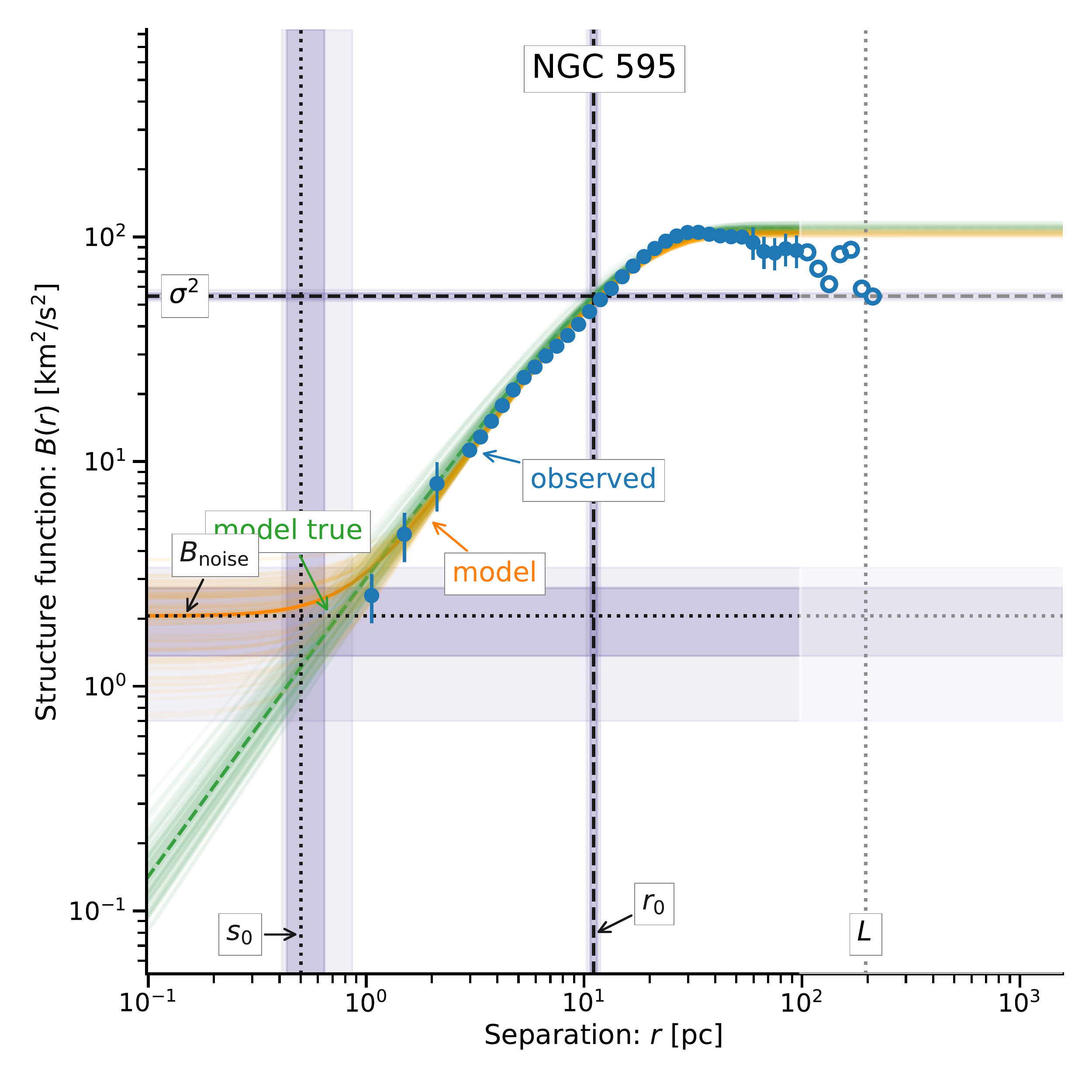}{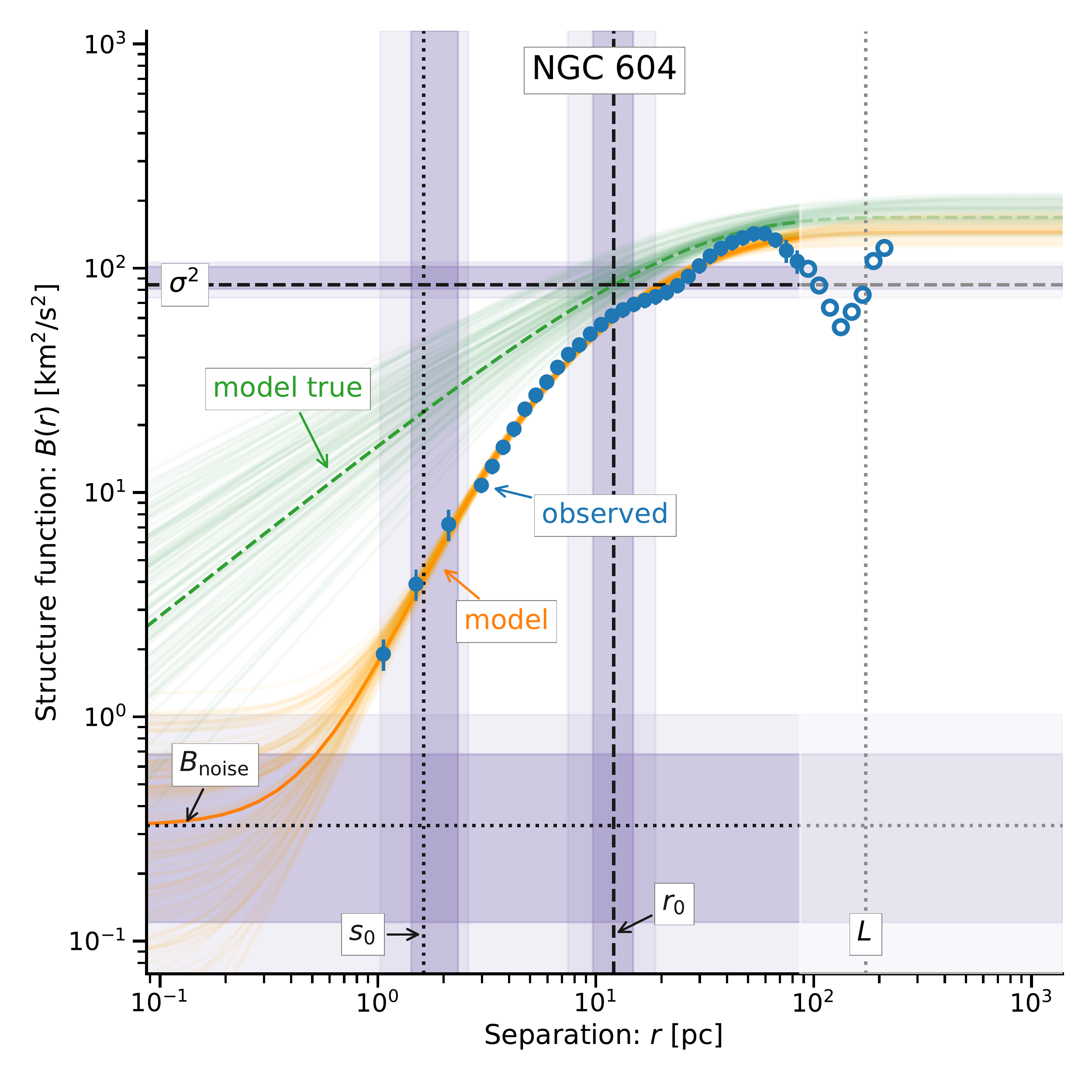}
  \caption{
    Same as figure~\ref{fig:strucfunc-fit-Galactic}
    except for \hii{} regions in more distant
    Local Group Galaxies.
    (a)~Hubble~V in NGC~6822.
    (b)~Hubble~X in NGC~6822.
    (c)~NGC~595 in M33.
    (d)~NGC~604 in M33.
  }
  \label{fig:strucfunc-fit-ExtraGal}
\end{figure*}

\newcommand\PM[2]{\ensuremath{\substack{+#1\\-#2}}}
\begingroup
\setlength{\tabcolsep}{6pt} 
\renewcommand{\arraystretch}{1.5} 
\begin{table*}
\begin{center}
  \caption{
    Best-fit model parameters and 95\% credibility intervals
    for fits to observed structure functions
  }
  \begin{tabular}{l RRRRRR  @{\hspace{6\tabcolsep}} RRR}
    \toprule
Region   & \sigma^2\pos            & B_{\text{noise}}       & s_0 (\text{RMS})          & r_0                    & L         & m                   & L/r_0 & s_0 / r_0 & s_0 (\text{FWHM}) \\
         & [\si{km^2.s^{-2}}] & [\si{km^2.s^{-2}}]     & [\si{pc}]                 & [\si{pc}]              & [\si{pc}] & [-]                 & [-]   & [-]       & [\text{arcsec}]   \\
\midrule
30 Dor   & 297\PM{40}{20}     & 6.2\PM{1.6}{1.6}       & 0.12 \PM{0.06}{0.03}       & 4.0\PM{1.0}{0.5}       & 31        & 0.85\PM{0.08}{0.13} & 8     & 0.03      & 1.2\PM{0.7}{0.4}  \\
NGC 604  & 84\PM{22}{10}      & <1.0                  & 1.6 \PM{1.0}{0.6}         & 12\PM{6}{5}            & 173       & 0.77\PM{0.20}{0.22} & 14    & 0.13      & 1.0\PM{0.5}{0.4}  \\
NGC 595  & 53\PM{5}{2}        & <3.3                   & <1.0         & 11\PM{1}{1}            & 196       & 1.36\PM{0.10}{0.15} & 18    & <0.04      & <0.5 \\
NGC 346  & 33\PM{3}{2}        & 0.8\PM{0.07}{0.07}       & 0.08 \PM{0.02}{0.02}      & 2.4\PM{0.3}{0.2}       & 20        & 0.95\PM{0.05}{0.07} & 8    & 0.04      & 0.7\PM{0.2}{0.1}  \\
Carina   & 18\PM{2}{2}        & <4.7                   & <0.010                    & 0.6\PM{0.1}{0.1}       & 20        & 1.16\PM{0.28}{0.18} & 33    & <0.01     & <3.4              \\
Hubble X & 15\PM{3}{1}        & <0.8                   & <0.55        & 4.0\PM{0.5}{0.2}       & 78        & 1.02\PM{0.06}{0.23} & 20    & <0.07      & <0.5 \\
Orion    & 13\PM{1}{1}        & 0.030\PM{0.005}{0.005} & 0.0020\PM{0.0002}{0.0002} & 0.068\PM{0.006}{0.004} & 0.5       & 1.07\PM{0.03}{0.04} & 7     & 0.02      & 1.8\PM{0.2}{0.2}  \\
Hubble V & 10\PM{3}{1}        & <0.65                  & <0.67        & 3.6\PM{0.5}{1.0}       & 61        & 0.81\PM{0.07}{0.28} & 17    & <0.14      & <0.65  \\
Lagoon   & 7\PM{1}{1}         & 0.45\PM{0.2}{0.3}      & <0.010                     & 1.0 \PM{0.2}{0.1}      & 16        & 1.26\PM{0.20}{0.20} & 16    & <0.01     & <3.5              \\
EON      & 5\PM{0.6}{0.4}         & 4.4\PM{0.9}{0.7}      & 0\FNa                     & 0.5\PM{0.2}{0.1}       & 3         & 1.0\FNa             & 6     & 0\FNa     & 0\FNa             \\
  \bottomrule
  \multicolumn{9}{l}{\FNa{}Assumed value.}
\end{tabular}\label{tab:Res}
\end{center}
\end{table*}
\endgroup

In order to provide a uniform description of the velocity fluctuations,
we fit the model structure function
\(\tilde{B}\model(r)\) of equation~\eqref{eq:sf-functional}
to each \hii{} region in our sample.
Results are shown in Figures~\ref{fig:strucfunc-fit-Galactic}, \ref{fig:strucfunc-fit-MC},
and~\ref{fig:strucfunc-fit-ExtraGal}, arranged in order of increasing distance.
Best-fit values for the model parameters,
together with 95\% credibility range,
are given in Table~\ref{tab:Res}.
After describing the methods used to
measure the structure function and fit the model (section~\ref{sec:techn-deta-model}),
we give details of the results in the example case of
the inner Orion Nebula (\ref{sec:example-results-orion})
and a broad overview of the results for other regions (\ref{sec:results-model-fitt}).
Further details of the results for individual regions are
given in Appendix~\ref{sec:notes-individual-hii}.

\subsection{Technical details of structure function measurement and fitting}
\label{sec:techn-deta-model}

For the observational values (\(B\obs(r)\); blue symbols),
the structure function is the average value of equation~\eqref{eq:Br}
over all pairs of points that contribute to each radial bin.
\startNEW
We judge the dominant source of error in \(B\obs(r)\) to be systematic,
rather than random, so we set the uncertainty of each bin (blue error bars)
as a constant fraction (2 to 8\%) of the observed value,
as listed in column 3 of Table~D1 (see supplementary material).
In some sources
(EON, and the regions observed using the FLAMES and the TAURUS instruments)
the uncertainty is increased by a factor of order two for a small number of bins
at the smallest and largest scales (columns 4 and 5 of Table~D1).
\stopNEW
For most sources a fixed bin size of \SI{0.05}{dex} is used,
but for those observed with multi-fiber spectroscopy
(Carina and M8, see section~\ref{sec:flames-multi-fiber}),
where coverage is sparse at small separations,
adjacent bins were merged to ensure at least 100 pairs contribute to each bin.
Bins with separation less than \(L/2\) are used to constrain the model fitting (filled symbols),
while larger separations (open symbols) are not used
since they are more likely to be affected by a breakdown
of the homogeneity assumption (see section~\ref{sec:limit-model-struct}).

Non-linear weighted least square fitting of the model is performed
using the Levenberg-Marquardt algorithm \citep{More:1978a} as implemented in the
\texttt{lmfit} Python library \citep{newville_matthew_2014_11813}.
The best-fit model \(\tilde{B}\model(r)\) is shown (heavy orange solid line),
together with the corresponding underlying ``true model'' \(B\model(r)\) (heavy dashed green line),
which does not include the effects of seeing or noise.
Note that the true model \(B\model(r)\) depends on only 3 parameters:
\(\sigma^2\), \(r_0\), and \(m\),
whereas \(\tilde{B}\model(r)\) additionally depends on
the observational nuisance parameters: \(s_0\) and \(B\noise\).
The best-fit values of each of these parameters are shown by
horizontal and vertical lines in the figure and are summarised in Table~\ref{tab:Res}.

The posterior distributions of model parameters
that are consistent with the observations for each region are estimated
using Markov Chain Monte Carlo (MCMC) ensemble sampling \citep{2010CAMCS...5...65G}
as implemented in the \texttt{emcee} Python library \citep{2013PASP..125..306F}.
A uniform prior distribution is assumed between upper and lower bounds
for each parameter, as given in Table~\ref{tab:parameter-ranges}.
To help ensure convergence of the MCMC algorithm,
we used chain lengths that exceeded \num{50} times the
estimated autocorrelation length for each parameter,
which typically required of order \num{50000} samples.
Thin translucent lines in the figures show structure functions using the parameters of
a random selection of \num{100} posterior samples from the MCMC chain,
both for \(\tilde{B}\model(r)\) (orange) and \(B\model(r)\) (green).
This gives an estimate in the uncertainty about the best-fit model.
Credibility intervals for the parameters are estimated from percentiles
of the posterior distribution and are indicated in the figures by shaded gray boxes
around the best-fit parameter values
(heavy shading for 68\% interval; light shading for 95\% interval).
The 95\% credibility interval for each parameter and for each source
is also given in Table~\ref{tab:Res}.
Figure~\ref{fig:corner-example-Orion} gives an example for Orion of
the pairwise correlations in the posterior distributions of the model parameters,
plotted using the \texttt{corner} Python library \citep{2017ascl.soft02002F}.
Corresponding plots for the remaining sources and a table for the fitting parameters are given in Appendix~\ref{sec:addit-covar-corn}.

\subsection{Example results of model fitting to the inner Orion Nebula}
\label{sec:example-results-orion}

This dataset is among the highest quality of those in our sample,
with more than \num{e5} spatial points and a factor of more than \num{500}
between the smallest and largest separations.
As a result, the observationally derived structure function (Figure~\ref{fig:strucfunc-fit-Galactic}a)
is very smooth and the model fit is very well constrained,
as evidenced by the tight credibility limits on the model parameters.
The derived correlation length \(r_0 = \SI{0.068}{pc}\) is \num{7} times smaller
than the box size, which indicates that there should be a moderate finite map effect
(section \ref{sec:effects-observ-limit-large} and Appendix~\ref{sec:finite-box-effects}),
but it is well within the range where the model fit can give reliable measurements
(see Figure~\ref{fig:finite-box-effect}).
The derived seeing parameter \(s_0 = \SI{0.002}{pc}\) is more than \num{30} times smaller
than \(r_0\), meaning that the seeing has a negligible effect at the correlation scale and above.
Figure~\ref{fig:corner-example-Orion}  shows  the covariances  between
parameters  in the  model fit,
which in some cases show significant correlations.
For example, \(\sigma^2\) and \(r_0\) are positively correlated,
whereas \(m\) and \(s_0\) are negatively correlated. 

\subsection{Results of model fitting to other sources}
\label{sec:results-model-fitt}

The structure functions for the remaining Galactic sources
are shown in Figure~\ref{fig:strucfunc-fit-Galactic}b--d.
The data quality for these sources is not so high as for the inner Orion Nebula,
with the result that the model parameters are not so well constrained.
In particular, the relatively coarse spacing between nearest-neighbor spatial points
(see sections~\ref{sec:fabry-perot-etalaon} and \ref{sec:flames-multi-fiber})
means that the seeing has no effect on the observed structure function,
so that \(s_0\) is indeterminate in the model fits
In addition, the effects of noise are also much greater,
as is apparent from visual inspection of the velocity fields
(top row of Figure~\ref{fig:velocity-maps}).
The principal effect of this on the model fits is to increase the uncertainty
in the power law slope \(m\).
The most extreme case is the Extended Orion Nebula (Fig.~\ref{fig:strucfunc-fit-Galactic}b),
where noise makes a significant contribution to the structure function at all scales.
The resultant degeneracy between parameters means that
is not possible to obtain a satisfactory fit
if all parameters are allowed to vary,
so we instead chose to fix the power slope at \(m = 1.0\),
which is close to the median value obtained for the other sources.

The structure functions for sources in the Magellanic Clouds are shown in Figure~\ref{fig:strucfunc-fit-MC}
and those in more distant galaxies in Figure~\ref{fig:strucfunc-fit-ExtraGal}.
The Magellanic cloud sources have generally high data quality due to the
large number of independent spatial points in the MUSE observations
(section~\ref{sec:muse-integral-field}).
The relatively poor spectral resolution means that the \(B\obs(r)\) cannot
be measured down to such low values as in the inner Orion Nebula,
but the larger amplitude of the fluctuations in these sources means
that the fitted model parameters are nevertheless tightly constrained.

For the more distant sources, the principal observational limitation is the seeing.
In three of the four sources (NGC 604, Hubble V and Hubble X),
the derived values of \(s_0\) exceed 10\% of the correlation length.
This implies that the true power law slope \(m\) is less steep than
would be naively inferred from the observations
(compare the green and orange curves in Figure~\ref{fig:strucfunc-fit-ExtraGal}),
but at the same time the degeneracies between parameters
lead to large uncertainties in the determination \(m\).
On the other hand, the values of \(r_0\) and \(\sigma^2\) are still well-constrained.

\subsection{On the reasonableness of the derived seeing widths and noise}
\label{sec:sanity-check-derived}
In the model fitting we allow the nuisance observational parameters
\(s_0\) and \(B\noise\) to vary freely within a wide range
(Table~\ref{tab:parameter-ranges}).
This is necessary because of the heterogeneous nature of our source datasets
and the fact that the details of the observational conditions
are not available to us in all cases.
However, it is worthwhile to perform a sanity check on the values that are implied by our model fits.
To that end, the last column in Table~\ref{tab:Res} lists the FWHM seeing width for each fit in units of arcsec.
Disregarding the 5 sources where only upper limits can be determined,
the mean and standard deviation are \SI{1.17(20)}{arcsec},
which is perfectly consistent with expectations for seeing from ground-based observations.

\startNEW
However, there are a few outlier values that deserve closer attention.
NGC~595, Hubble~V, and Hubble~X all have model-derived upper limits to the
seeing FWHM of about \SI{0.6}{arcsec}, which seems
unrealistically small, especially given the higher value
derived for NGC~604,
which was observed with the same instrument.
Unfortunately, we do not have access to the observing logs to check this,
so we suggest extra caution in interpreting the fit results for these objects. 
\stopNEW

The inner Orion Nebula has the largest inferred seeing FWHM of \NEW{\SI{1.8(2)}{arcsec}},
which is \NEW{nearly twice as high as} the typical seeing measured during the observations
\citep{Doi:2004a}.
However, this can be at least partially explained by the fact that the velocity map is constructed
by interpolating individual longslit observations onto a regular grid
(section~\ref{sec:longsl-echelle-spect}).
The seeing-limited resolution will only be achieved along the slits,
whereas the resolution in the perpendicular direction is determined by the slit spacing of \SI{2}{arcsec}.
Given this, the model-derived value of \(s_0\) for this source is not unreasonable.

On the other hand, it is perhaps significant that both
the inner Orion Nebula and 30~Doradus,
the two highest quality datasets in our sample,
should both have a fitting-derived seeing width that is
slightly larger than expected.
Since our intrinsic model \(B\model(r)\) assumes a single power law
for scales \(<r_0\),
the only way of accommodating a steepening of \(B\obs(r)\) at small separations
is via the seeing term \(S(r)\).
But if the true \(B(r)\) really does steepen at small scales,
then the seeing will be overestimated in the models.
Unfortunately, the currently available data are insufficient to
definitively decide this question.

For most of our sources \(B\noise / \sigma^2\pos < 0.03 \) and the noise
has almost no effect on the structure function,
apart from at the very smallest separations,
resulting in a negligible influence on the other model parameters.
The exceptions are the Galactic regions Lagoon, Carina,
and especially the Extended Orion Nebula,
for which the noise is sufficiently large that the slope
of \(B\obs(r)\) is significantly shallower than
the inferred slope \(m\) of the true model \(B\model(r)\).
These sources also show the largest fluctuations
\(B\obs(r)\) around the smooth fit
at intermediate scales
(see in particular Figure~\ref{fig:strucfunc-fit-Galactic}d),
which is probably due to the relatively small number of spectra.

\subsection{Evidence for inhomogeneity at the largest scales}
\label{sec:evid-inhom-at}

At scales larger than \(L/2\)
we see a variety of different shapes for the structure functions
of our sample regions.
These points are excluded from our model fits but we give
a qualitative description in this section.
In some cases,
such as Carina and Hubble~X,
\(B\obs(r)\) remains flat at a value of \(\approx 2\sigma^2\pos\),
which is consistent with uncorrelated homogeneous fluctuations
at the largest scales.

For other regions,
such as the Orion Nebula (both inner and outer) and 30~Doradus,
there is a clear downturn in \(B\obs(r)\) at the largest separations.
As discussed above in section~\ref{sec:limit-model-struct},
this is what would be expected if the velocity fluctuations
were inhomogeneous, with a larger amplitude in the center of the map
and a smaller amplitude in the outskirts.
A similar behavior, although not so marked, is shown by
Hubble~V and NGC~595.
The Orion Nebula is a particularly interesting case
since the \(\sigma^2\pos\) derived from the high-resolution dataset of the inner nebula on scales \(\approx \SI{0.1}{pc}\)
is roughly twice as large as that derived from the lower resolution
dataset of the Extended Orion Nebula on scales \(\approx \SI{1}{pc}\)
(compare panels a and b of Figure~\ref{fig:strucfunc-fit-Galactic}).
This is further evidence that the amplitude of velocity fluctuations
increases towards the center of the nebula in this source.

In contrast, other regions,
such as the Lagoon and NGC~346,
show \(B\obs(r)\) increasing at the largest separations,
reaching values significantly larger than \(2\sigma^2\pos\).
This can be evidence for a large-scale linear velocity gradient
across the region (see Figure~13 of \citealp{arthur2016turbulence})
or a periodic fluctuation with \(\lambda > r_0\)
(see section~\ref{sec:limit-model-struct} above).
This latter effect may also explain the behavior of NGC~604,
which shows oscillatory behavior of \(B\obs(r)\),
similar to the light blue line in Figure~\ref{fig:model-strucfunc}b.



\section{Discussion}\label{sec:discussion}

\subsection{Comparison with previous structure functions}
\label{sec:comp-with-prev}
Comparison with prior studies is complicated by the
diversity of methodologies that  have been employed.
There is almost universal agreement on how to measure
the velocity variance \(\sigma^2\pos\),
although even here there can be small differences according
to whether the centroid velocities come from gaussian fitting
or velocity moments,
and whether or not smooth large-scale trends are first removed.
Also, the field of view of the region would have an influence on \(\sigma^2\pos\).
\citet{arthur2016turbulence} compared multiple studies of the
inner Orion Nebula, finding agreement to within 20\%
in \(\sigma^2\pos\) determinations
for emission lines that trace the bulk of the fully ionized gas.

There is less agreement in the literature on the best way
define the correlation length, \(r_0\).
Our own definition corresponds to the lag where the autocorrelation function
has fallen to a value of one-half,
\(C(r_0) = 1/2\) (see section~\ref{sec:methods-apply}),
whereas other common choices are \(C(r) = e^{-1}\)
\citep{Mivi1995}
or the \textit{total decorrelation lag} \(\tau_0\) \citep{lagrois2011},
which corresponds to \(C(\tau_0) = 0\).
One disadvantage of this latter definition is that it
can be sensitive to non-homogeneous behavior of the
velocity field at large scales. 
The conversion between these different conventions will also depend
on the value of the power-law slope,
but we typically find that \(\tau_0 / r_0 = \num{2}\) to \num{3}.

The power law slope, \(m\), itself is probably the parameter that
is most sensitive to methodological assumptions.
The fundamental issue is that the observed structure function tends to
show a negative curvature in log-log space at intermediate scales:
\(d^2 \log B/ d^2 \log r < 0\).
This is caused primarily by the steepening due to seeing at small scales
together with the natural asymptotic flattening, 
\(B(r) \to 2 \sigma^2\pos\), at large scales.
As a result, the measured slope depends on the exact range of
separations over which the power law is fitted.
So long as the spatial dynamic range of the observations is high enough,
then a reliable slope can nevertheless be obtained,
as shown by \citet{arthur2016turbulence} for the case of the Orion Nebula.
However, this requires that \(r_0\) should be at least an order of
magnitude higher than the larger of the seeing width
or the minimum separation between spatial points.
This condition is not satisfied for roughly half of our sample regions,
and the same is true for many previous studies.
In such cases, our model-based approach is a more reliable way of
determining the slope. 

Specific comparisons for individual sources are given
in Appendix~\ref{sec:notes-individual-hii}.
In summary, we agree with previous studies in cases where there
is a good match in the exact area covered and in the angular resolution.
In other cases, large discrepancies are found,
which are probably caused by a combination of methodological differences
and real variations due to the inhomogeneity of the
velocity fluctuations on the largest scales.

\subsection{Scaling relations for turbulence parameters}\label{sec:scaling-relations}

We look for correlations and scaling relations 
between the physical parameters of the \hii{} regions
and the parameters of the fluctuating velocity fields
that we derived in section~\ref{sec:results}.
To accomplish this we implement a hierarchical Bayesian model for linear regression using the \texttt{linmix} Python library, which is based on the procedure of \cite{2007ApJ...665.1489K}.
We fit relations of the form
\(Y = a X + b\)
where \(X = \log_{10} x\) and \(Y = \log_{10} y\)
with \(x, y \in \{d, D_{\hii}, L(\ha), \sigma\pos, r_0, m\}\)
(taken from the physical properties in Table~\ref{tab:regions-properties}
and the derived turbulent parameters of Table~\ref{tab:Res}).
Thus the fitted relation between the original variables
is of the power-law form: \(y = 10^b\, x^a\).
Note that, unlike with an Ordinary Least Squares method,
the \citeauthor{2007ApJ...665.1489K} method
uses an errors-in-variables model that allows for
measurement uncertainty in both \(X\) and \(Y\).
Since the main focus of this paper is the turbulent velocity field,
we restrict consideration to correlations that include
one of the three structure function parameters,
\(\sigma\pos\), \(r_0\), or \(m\), as the  \(y\) variable. 
In addition, we fit a strictly linear relationship between
the plane-of-sky and line-of-sight velocity dispersions:
\(X = \sigma\pos\), \(Y = \sigma\los\).
The reason for not taking the logarithm in this instance is
both to facilitate comparison with previous studies
and to allow for a zero-point offset in the relation.

In the following analysis we omit the EON region since the
model fits were relatively poorly constrained by the observations.
For physical properties where uncertainties are not reported (size and luminosity), we assume an error of \num{10}\% of the value.
The same criteria is used for the uncertainty values of the mean non-thermal linewidths for 30 Doradus and NGC 346.
For the turbulent parameters we use the relative error
from our model fits, considering the 2-\(\sigma\) interval. 

\begin{figure}
\centering 
\includegraphics[width=3in]{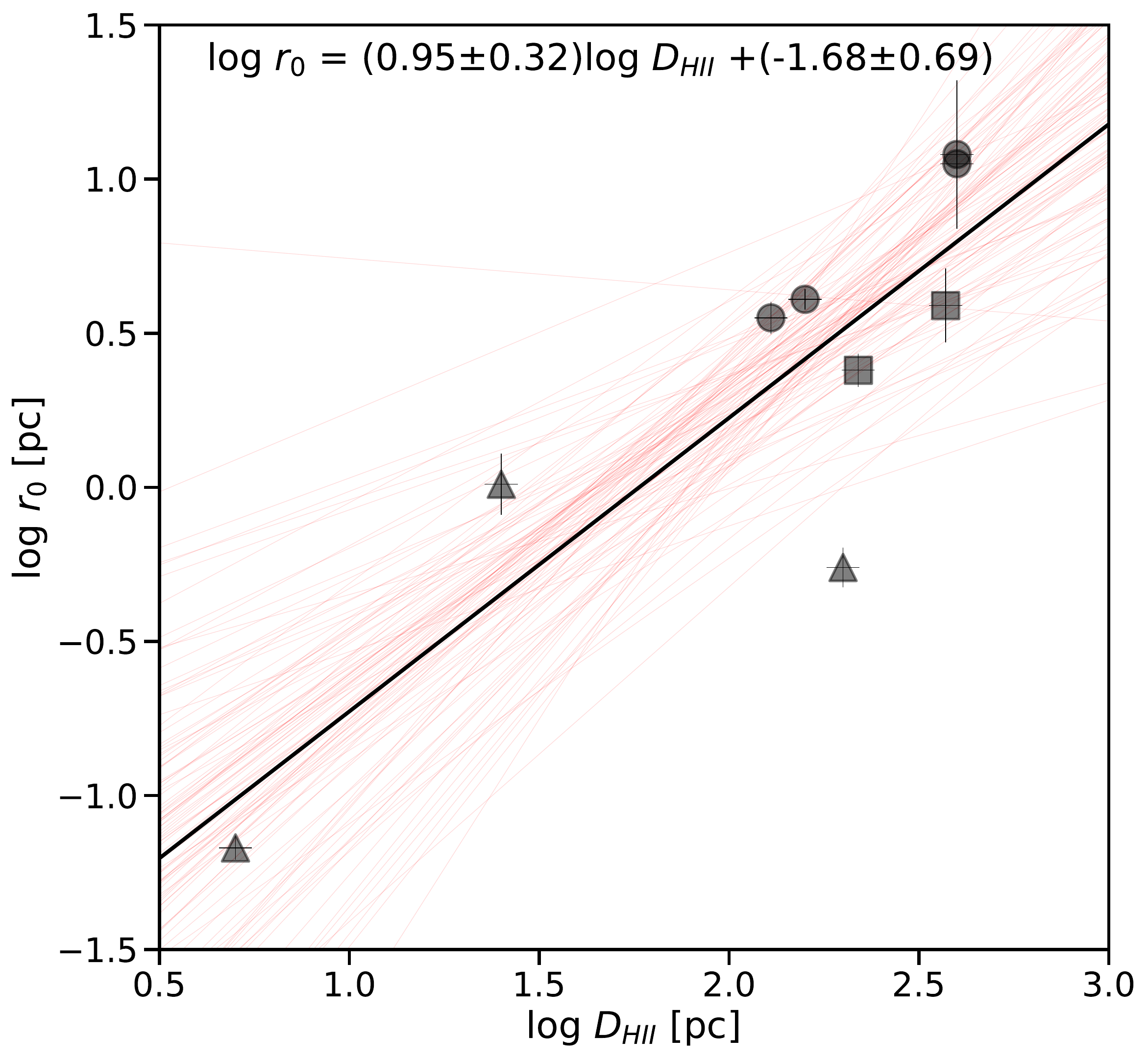}
\caption{
  The relationship between
  correlation length of velocity fluctuations,
  \(\log_{10} r_0\),
  and  \hii{} region diameter,
  \(\log_{10} D_{\hii}\)
  derived from our results.
  Symbols indicate Galactic (triangles),
  Magellanic (squares) and extragalactic (circles) regions.
}
\label{fig:rvsR}
\label{fig:first-correlation}
\end{figure}

\begin{figure}
\centering 
\includegraphics[width=3in]{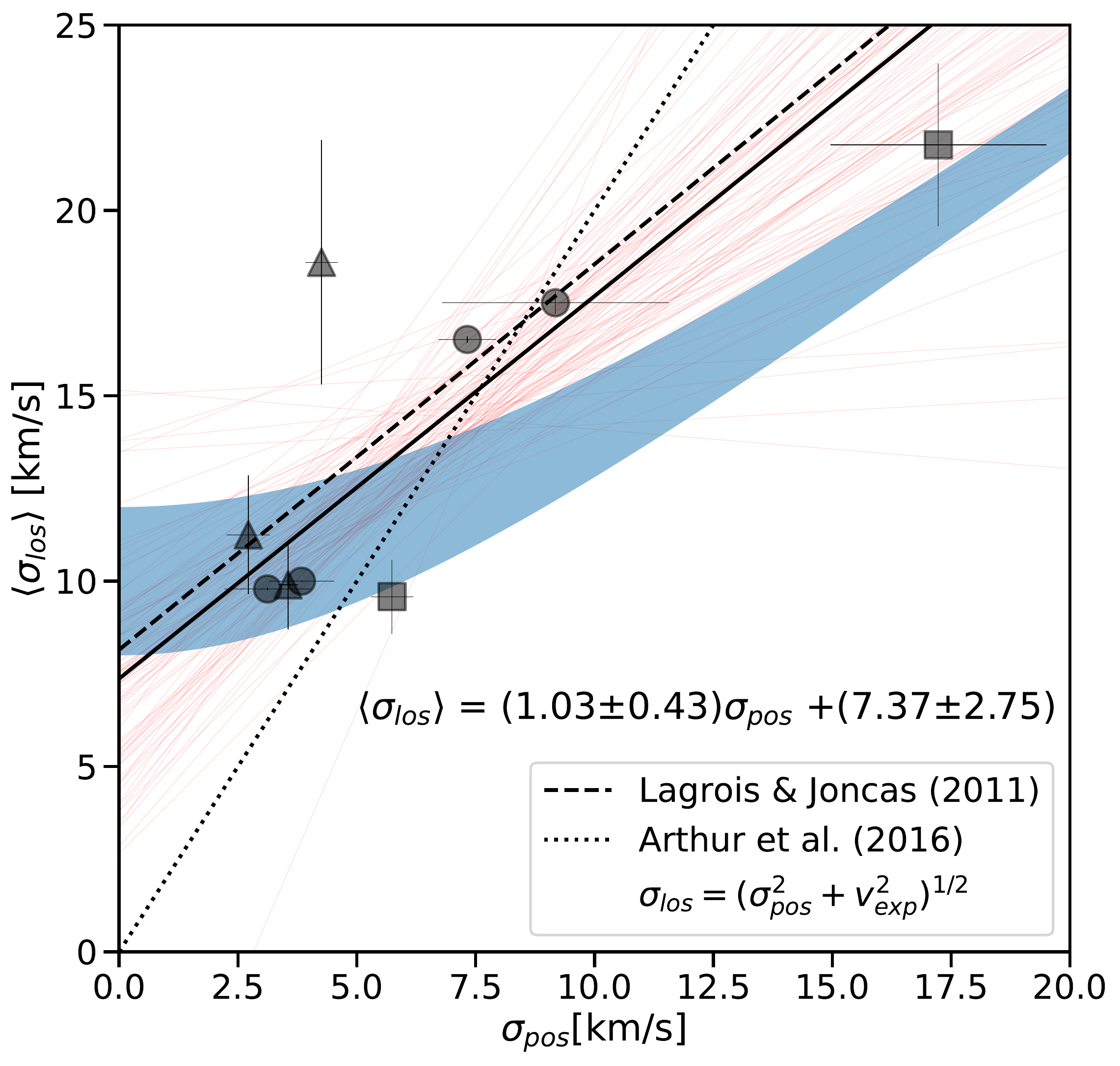}
\caption{
  The relationship between 
  line-of-sight non-thermal velocity dispersion,
  \(\sigma\los\),
  and plane-of-sky dispersion of velocity centroids,
  \(\sigma\pos\)
  derived from our results.
  Symbols indicate Galactic (triangles),
  Magellanic (squares) and extragalactic (circles) regions.
  The black solid line is the median linear regression,
  with thin red lines showing 100 random samples from the posterior
  distribution to indicate the uncertainty in the regression.  
  The dashed line shows the the linear fit obtained by
  \citet{2011MNRAS.413..705L} for a sample of
  small-to-intermediate size \hii{} regions in the M33 galaxy. 
}
\label{fig:los-vs-pos}
\end{figure}

\begin{figure}
\centering 
\includegraphics[width=3in]{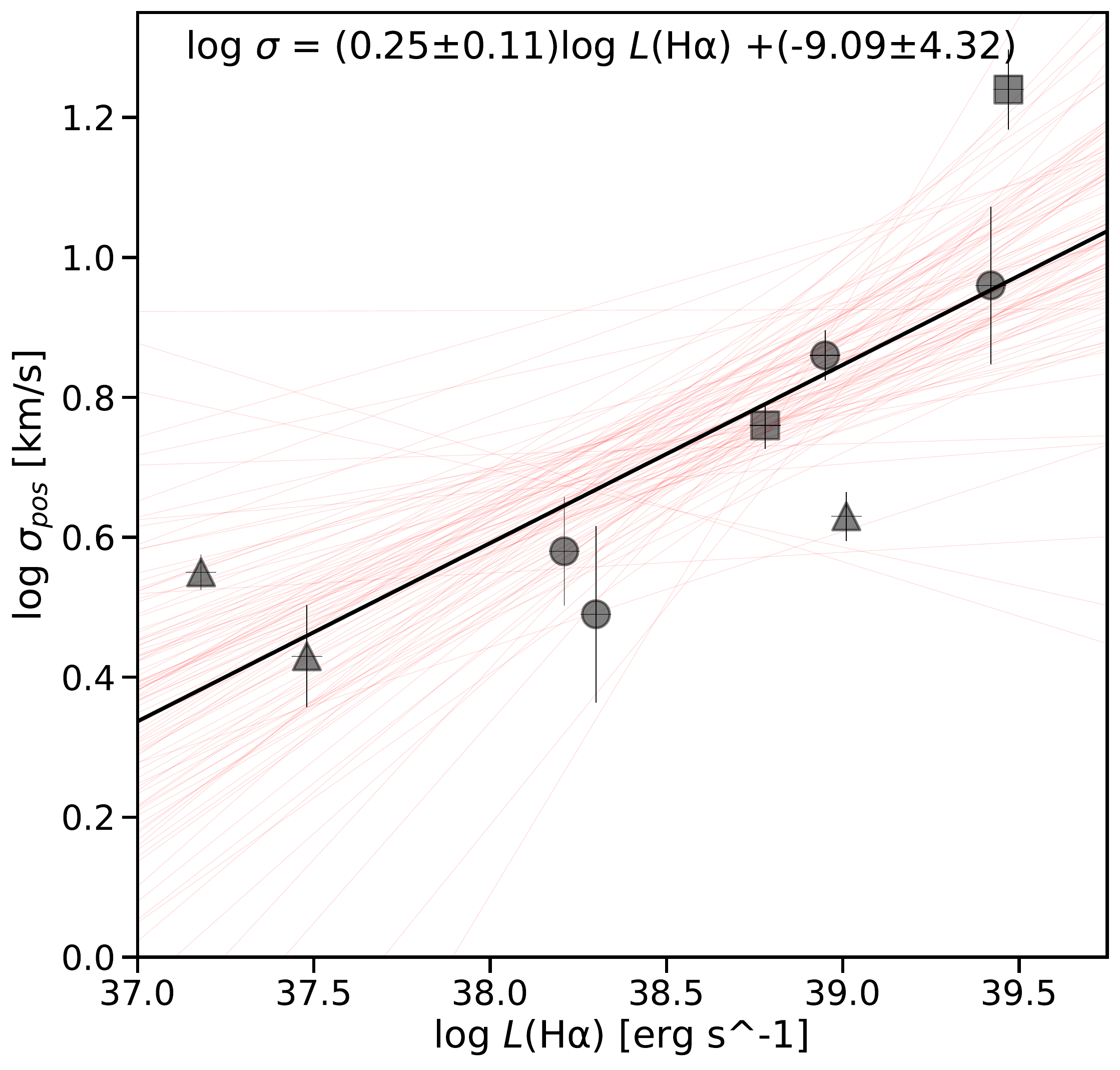}
\caption{
  The relationship between
  plane-of-sky dispersion of velocity centroids,
  \(\log_{10} \sigma\pos\),
  and \hii{} region luminosity,
  \(\log_{10} L(\ha)\),
  derived from our results.
  Symbols indicate Galactic (triangles),
  Magellanic (squares) and extragalactic (circles) regions.
}
\label{fig:sigvsl}
\label{fig:last-correlation}
\end{figure}

Results for the most significant correlations,
as measured by the Pearson correlation coefficient \(r\),
are given in Table~\ref{tab:RestStats} in descending order of \(r^2\).
We find three cases of highly significant correlations
with significance level \(p < 0.01\),
which are illustrated in
Figures~\ref{fig:first-correlation}--\ref{fig:last-correlation}:
\(r_0\)  versus \(D_{\hii}\), \(\sigma\los\) versus \(\sigma\pos\),
and \(\sigma\pos\) versus \(L(\ha)\).
Each of these is discussed in turn in the following sections.
The remaining correlations
are at best marginally significant (\(p > 0.05\)),
including all correlations of the power law slope \(m\)
with any other parameter,
but this is itself interesting and is
also given its own section below.


\begin{table*}
\begin{center}
\caption{Linear regressions values in the form Y = aX + b between our turbulent parameters obtained using the chi-square statistic and properties of each region (Table \ref{tab:regions-properties}). The fifth column, $r$, is the Pearson correlation coefficient and the last column is the $p$-value. This results were obtained using the procedure in \citet{2007ApJ...665.1489K}.}
\begin{tabular}{RRRRRR}
  \toprule
  Y &                   X &                 a &                 b &       r &      p \\
  \midrule
  \log r_0 &         \log D_{\hii} &   0.95 \pm 0.32 &  -1.68 \pm 0.69 &   0.86 &  \mathbf{0.003} \\
  \log \sigma\pos &        \log L(\ha) &    0.25 \pm 0.11 &  -9.09 \pm 4.32 &   0.81 &  \mathbf{0.008} \\
  \sigma\los &  \sigma\pos &   1.03 \pm 0.43 &   7.37 \pm 2.75 &   0.78 &   \mathbf{0.010} \\[\smallskipamount]
  \log \sigma\pos &         \log D_{\hii} &   0.26 \pm 0.18 &   0.19 \pm 0.39 &   0.64 &   0.06 \\
  \log \sigma\pos &   \log r_{0} &    0.18 \pm 0.19 &   0.67 \pm 0.14 &   0.50 &  0.17 \\
  \log m &  \log d &  -0.02 \pm 0.04 &   0.04 \pm 0.08 &   -0.41 &   0.27 \\
  \log m &  \log  \sigma\pos &  -0.10 \pm 0.20 &   0.08 \pm 0.15 &  -0.37 &  0.33 \\
  \log m &  \log  r_{0} &   -0.02 \pm 0.07 &   0.01 \pm  0.05 &  -0.29 &  0.45 \\
  \bottomrule
\end{tabular}\label{tab:RestStats}
\end{center}
\end{table*}



\subsubsection{Diameter vs correlation length}\label{sec:D-vs-r0}

Figure~\ref{fig:rvsR} shows the relation between the
correlation length \(r_0\) of the velocity fluctuations
and the diameter \(D_{\hii}\) of the \hii{} region. 
The relationship we find is
\(\log r_0 = (0.95 \pm 0.32) \log D_{\hii} - (1.6 \pm 0.7)\),
which is very close to linear,
implying that the correlation length is a constant fraction
of the diameter:
\( r_0 /  D_{\hii} = 0.02 \pm 0.01\).

However, \(r_0\) is also well correlated with
the observed map size, \(L\),
so it is important to check that the correlation lengths that we
measure are true properties of the regions,
and are not just artifacts of the finite map sizes.
In Appendix~\ref{sec:finite-box-effects} we show that \(L\)
has a significant effect on the naive apparent \(r_0\) when \(r_0 > 0.1 L\),
but that an accurate \(r_0\) can still be recovered from model fitting
so long as \(r_0 < 0.20 L\).
From our results (Table~\ref{tab:Res}) we find that \(r_0 \approx (0.08 \pm 0.01) L\),
which is comfortably small enough that the measured correlation lengths should
be reliable.
A clear indication that the derived \(r_0\) is not real would
be a structure function that never flattens off, but rather keeps
rising at all scales.
We do not see such a behavior in any of our regions.
The most similar is the Lagoon Nebula
(Figure~\ref{fig:strucfunc-fit-Galactic}c),
where the structure function does flatten at intermediate scales,
but then rises again for separations of order \(L\).
It may be that there is a second, outer correlation scale in this region,
which we fail to measure because the observed map is to small.

The small value of \( r_0 /   D_{\hii}\) that we obtain means that it
would be feasible to study the spatial variation of the turbulence parameters
within a given region. This is because parts of the nebula that are separated
by a distance \(\gg r_0\) are essentially independent.
In principle,  the variation of \(\sigma\pos\) could be
measured in ``macro-pixels'' of size \(2 r_0\),
but a minimum size of  \( 3 r_0\) is necessary to reliably determine \(r_0\)
itself (Appendix~\ref{sec:finite-box-effects}),
and it is quite possible that \(r_0\) might vary
across the face of the nebulae.
The best that could be done with the current \(L \times L\) velocity maps
would therefore be about \(3\times3\) pixels, but this could be increased to
about \(15\times15\) pixels if the entire \(D_{\hii} \times D_{\hii}\) region were mapped.



The energy injection or driving scale of turbulence,
\(\ell_E\), is commonly identified
with the ``outer scale'' of the fluctuations 
\citep{Haverkorn:2004r, 2010ApJ...710..853C}. 
The outer scale has been given diverse definitions \citep{Klipp:2014a}
but most commonly is either the integral scale \LL{},
or the scale at which the structure function becomes flat.
These are both of the same order as the correlation length \(r_0\)
as we define it in section~\ref{sec:methods-apply},
but are typically larger by a factor of 1.5 to 2
(for example, equation~\eqref{eq:4} for the integral scale).
However, high resolution numerical simulations
of driven isothermal turbulence \citep{Federrath:2021r}
show a somewhat larger gap of a factor of 4 between the correlation length
and the driving scale.  
In either case, our results imply that the turbulence
in \hii{} regions is driven on relatively small scales,
between 2\% and 5\%  of the diameter of the \hii{} region.

\subsubsection{The constancy of the power law slope}
\label{sec:constancy-power-law}
The slope \(m\) of the power-law portion of the structure function
shows no significant correlation with any other parameter
(see Table~\ref{tab:RestStats}),
with a weighted mean value of \(m = 1.03\).
The dispersion in fitted \(m\) values is 1.6 times higher than expected
on the basis of the confidence limits for the individual regions,
implying that there may be a real intrinsic dispersion of \(\pm 0.17\) in the slopes.
However, a \(\chi^2\) test shows that this is only marginally significant
(\(p = 0.11\)), so we cannot rule out that all the slopes are identical.

\newcommand\DDD{\ensuremath{_{\mathrm{3D}}}}
For incompressible Kolmogorov turbulence,
the predicted slope of the three-dimensional structure function is \(m\DDD = 2/3\),
but intermittency and compressibility \citep{Schmidt:2008a}
can increase this to \(m\DDD \approx 0.8\) in the subsonic regime,
and up to \(m\DDD = 1\) in the supersonic regime.
Comparison with our measurements is complicated by projection smoothing
\citep{1984ApJ...277..556S},
which implies that \(m\DDD\) is only
observed directly for separations larger than the line-of-sight depth
of the emitting regions.
For smaller separations, one potentially observes a steeper slope:
\(m = m\DDD + 1 + \delta\kappa\),
where \(\delta\kappa\) represents the effects of emissivity fluctuations\footnote{%
  Note that literature on turbulence in molecular clouds discuss fluctuations of \emph{density}
  since they are mainly concerned with emission lines
  whose emissivity is approximately linearly proportional to density.
  In the case of photoionized regions, the emissivity of important lines such as \ha{}
  is proportional to \emph{density-squared},
  so in adopting the results of \citet{Brunt:2004a} we substitute emission measure
  (or its proxy, \ha{} surface brightness)
  in place of column density. 
}
along the line of sight \citep{Brunt:2004a},
and varies from \(\delta\kappa = 0\) in the incompressible limit
to \(\delta\kappa = -1\) in the case of strongly driven supersonic turbulence.
In this latter case, the effects of projection smoothing and emissivity fluctuations
cancel out, leading once more to \(m \approx m\DDD\).

The \hii{} regions of our sample span the transonic regime,
from subsonic velocity fluctuations in the Galactic sources
to supersonic fluctuations in the more luminous extragalactic sources. 
It is therefore not surprising that we see no significant correlation
between \(m\) and \(\sigma^2\pos\) 
(\(p = 0.33\), Table~\ref{tab:RestStats}).
Although theoretically one might expect a positive correlation of \(m\DDD\)
with the amplitude of the velocity fluctuations
due to the transition from subsonic
to supersonic turbulence \citep{Galtier:2011a},
this will be offset by a decrease in \(\delta\kappa\) 
if the amplitude of the density fluctuations
(and hence emissivity fluctuations) increases along
with the velocity fluctuations.

\begin{figure}
\centering 
\includegraphics[width=\linewidth]{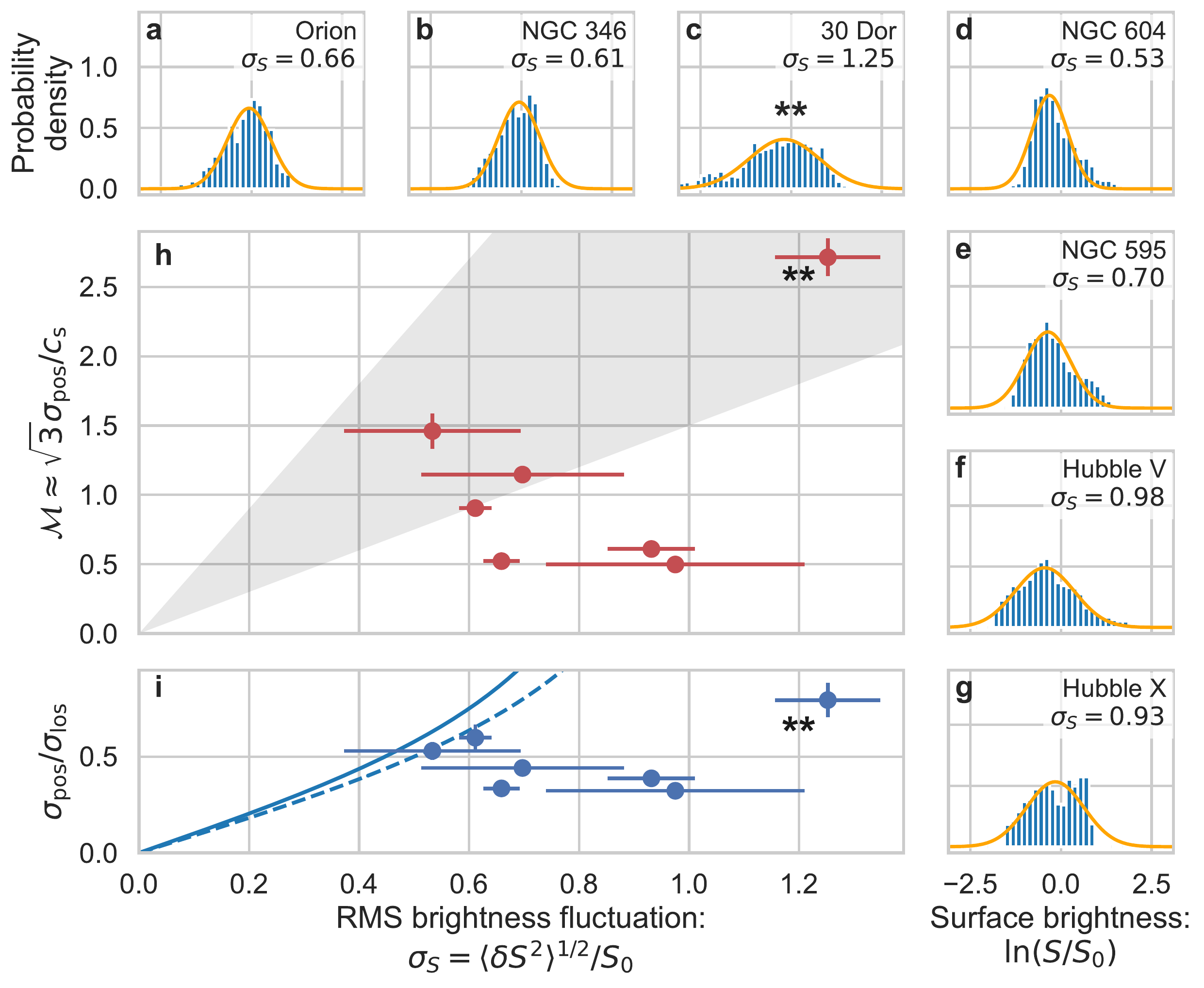}
\caption{
  Panels a--g: Histograms (blue) of the probability density function (PDF)
  of the \ha{} surface brightness of \hii{} regions in our sample.
  Smooth lines (orange) show fitted log-normal distributions.
  In order to suppress fluctuations on scales
  smaller than the correlation length,
  surface brightness maps were rebinned by averaging
  into pixels of size \(0.5 r_0\) before calculating the PDF.
  \startNEW
  Panel~h: RMS Mach number of three-dimensional
  velocity fluctuations
  as a function of the RMS fractional brightness fluctuation calculated from the PDFs.
  The gray shaded area shows the range of predicted values
  for the brightness fluctuations if they are entirely due to density
  fluctuations formed by driven turbulence with the observed Mach number.
  Panel~i: Ratio of plane-of-sky to line-of-sight velocity dispersions
  as a function of the
  RMS fractional brightness fluctuations calculated from the PDFs.
  The solid and dashed lines show the expected relationship for
  the toy models of Appendix~\ref{sec:appar-veloc-fluct}. 
  \stopNEW
}
\label{fig:brightness-pdfs}
\end{figure}

Evidence for just such an increase is shown in Figure~\ref{fig:brightness-pdfs},
which shows how the fractional width of the probability distribution of
large-scale \ha{} surface brightness fluctuations
varies with the RMS turbulent Mach number, \Mach{},
of the three-dimensional velocity fluctuations.
The Mach number is estimated by assuming that the plane-of-sky velocity dispersion is
a good estimate of the turbulent fluctuations in one dimension
and that the turbulence is isotropic, yielding:
\(\Mach = \sqrt{3} \sigma\pos / \csound\),
where an ionized sound speed of \(\csound = \SI{11}{km.s^{-1}}\)
is assumed for all sources.
The RMS fractional width of density fluctuations in a turbulent medium,
\(\shortsig{\rho} \equiv \longsig{\rho}\),
is predicted to depend linearly on the Mach number as
\(\shortsig{\rho} = b \Mach\),
where \(b\) depends on the nature of the turbulent driving \citep{Federrath:2010z},
varying between \(b = 1/3\) for solenoidal driving and \(b = 1\) for compressive driving.
This is shown by gray shading\footnote{\label{fn:brightness-fluctuations}%
  In the figure, it is assumed that the relative fluctuations in surface brightness
  and density are equal, which is a result of the
  cancellation between two effects. First, that the volumetric \ha{} emissivity \(E\)
  is proportional to density squared, so that \(\shortsig{E} = 2 \shortsig{\rho}\). 
  Second, that fluctuations in surface brightness \(S\) are related to those in emissivity
  by \(\shortsig{S} = R^{1/2} \shortsig{E}\), where \(R\)
  is the 2D-to-3D variance ratio \citep{Brunt:2010b}.
  For an emissivity power spectrum \(P(k) \sim k^{-3}\)
  and assuming a ratio of map size to correlation length
  of \(L / r_0 = 10\) (typical value from Table~\ref{tab:Res}),
  we find \(R \approx 0.2\) and hence \(\shortsig{S} \approx \shortsig{\rho}\).
  We calculate \shortsig{S} by fitting a log-normal
  function to the probability distribution function
  of the surface brightness map after filtering out fluctuations
  on scales smaller than the velocity correlation length.
}
in Figure~\ref{fig:brightness-pdfs}h,
where it can be seen that many of the sources with subsonic velocity dispersions
(\(\Mach < 1\)) fall to the right of the theoretical prediction,
implying that other factors in addition to turbulence are responsible for
producing the fluctuations in density and surface brightness in these regions
(see also section 4.5 of \citealp{arthur2016turbulence}).
For our highest luminosity source of 30~Dor on the other hand,
the velocity dispersion is clearly supersonic (\(\Mach \approx 2.7\))
and, although the surface brightness fluctuations are also larger than in the other sources,
they are consistent with being caused by turbulence-induced density fluctuations.

We find no clear evidence for a broken power-law in any of our
structure functions, as has been claimed in previous studies.
Although in some regions we do observe a steepening at the smallest scales
(for example Figures~\ref{fig:strucfunc-fit-Galactic}a, \ref{fig:strucfunc-fit-MC}ab),
this can be accounted for in our model fits by
the spatial smoothing caused by atmospheric seeing.
This has a noticeable effect on the structure function at scales of
up to 5 times the FWHM of the seeing (Appendix~\ref{sec:effects-seeing-struc}).
There is a hint in the highest quality datasets that there may also be
a real steepening of the structure function at small scales
(see section~\ref{sec:sanity-check-derived}),
but higher spatial resolution observations are required in order to study this. 


\subsubsection{Plane-of-sky vs line-of-sight
  velocity dispersions}
\label{sec:sigmapos-vs-sigmalos}
\newcommand\siglosm{\ensuremath{\langle\sigma\los\rangle}}

In Figure~\ref{fig:los-vs-pos} we show the
mean non-thermal line-of-sight velocity dispersion
\(\langle\sigma\los\rangle\)
(obtained from the spectral line profiles)
versus the plane-of-sky dispersion of mean velocities
\(\sigma\pos\)
(obtained from our structure function fits).
The relationship we obtain is
\( \langle \sigma\los \rangle = (1.03 \pm 0.43) \sigma\pos + (7.4 \pm 2.8)\),
which can be compared with previous results
using different types of samples.
\citet{2011MNRAS.413..705L} surveyed
a sample of small-to-intermediate size \hii{} regions in M33,
finding
\(\langle \sigma\los \rangle = (1.04 \pm 0.14) \sigma\pos + (8.15 \pm 0.93) \)
for their sub-sample with higher resolution observations
(21 sources).
This is shown as the dashed line in Figure~\ref{fig:los-vs-pos},
which can be seen to be almost identical to our own result. 
Both fits are consistent with a slope of unity and an offset of
\(\SI{8}{km.s^{-1}}\).
\citet{arthur2016turbulence} calculate the relationship
between \(\langle\sigma\los\rangle\) and \(\sigma\pos\) for different optical emission lines
([\ion{S}{2}], [\ion{N}{2}], \ha, [\ion{O}{3}])
in the Orion Nebula, finding \(\langle\sigma\los\rangle \approx 2 \sigma\pos\),
which is shown by the dotted line in Figure~\ref{fig:los-vs-pos}.
Although this line is a fair representation of the low-\(\sigma\)
points in our sample, it fails badly for highest dispersion source,
30~Doradus.

At least three different explanations for the relation
between \(\sigma\pos\) and \(\siglosm\) have been proposed previously.
The classic interpretation 
in terms of homogeneous incompressible turbulence \citep{von1951methode}
is that \(\sigma\pos\) is reduced with respect to the ``true''
internal velocity dispersion
due to averaging over multiple independent turbulent cells
along the line of sight,
which is another manifestation of projection smearing, 
as discussed above in section \ref{sec:constancy-power-law}.
Assuming that \(\siglosm\) measures the true dispersion,
then one would expect to find \(\sigma\pos / \siglosm = (\LL /H)^{1/2}\),
where \(H\) is the line-of-sight depth of the emitting gas
and \(\LL\) is the integral length scale of
the fluctuations defined in section~\ref{sec:methods-apply},
which yields \(H / r_0 \approx 1.44 (\siglosm / \sigma\pos)^2\).
From the results of Figure~\ref{fig:los-vs-pos} we would therefore find
\(H / r_0 \approx 6\) for the lower dispersion sources,
falling to \(H / r_0 \approx 2\) for the higher dispersion sources.
However, such an analysis ignores the effect of density fluctuations,
which would tend to reduce the effectiveness of projection smoothing
compared with the incompressible case (see discussion of \(\delta\kappa\) above),
so it is unclear if the results remain valid
for the sources with supersonic velocity dispersion.

\newcommand\vexp{\ensuremath{v_{\text{exp}}}}
An alternative interpretation,
as discussed in \citet{2011MNRAS.413..705L},
is that \(\siglosm\) includes a contribution due to
ordered large-scale velocity gradients
(for instance, radial expansion)
in addition to the turbulence,
whereas \(\sigma\pos\) includes only the turbulent motions. 
Assuming a constant value \vexp{} for the ordered contribution
and that projection smoothing is negligible,
one would predict \(\siglosm = (\sigma\pos^2 + \vexp^2)^{1/2}\),
which is shown by the blue shaded-area in Figure~\ref{fig:los-vs-pos}
with an upper limit of \SI{12}{km.s^{-1}} and lower limit of \SI{8}{km.s^{-1}} for \(\vexp\).
It can be seen that such a relationship reproduces the 
observationally derived values almost as well as the linear fit.
On the other hand, we can rule out the possibility that
\vexp{} might increase in proportion to the turbulent velocity dispersion
since that would predict that \(\siglosm \propto \sigma\pos\),
which is inconsistent with the observations as discussed above.

However, as pointed out by \citet{arthur2016turbulence},
the previous analysis is oversimplified since the presence
of density fluctuations (or, more generally, emissivity fluctuations)
means that the plane-of-sky dispersion of centroids \(\sigma\pos\)
is not independent of the 
ordered large-scale velocity gradients along the line of sight.
Indeed, the neglect of any projection smoothing for the turbulent contribution
in the \citeauthor{2011MNRAS.413..705L} analysis is equivalent to
assuming \(\delta\kappa = -1\), which \emph{requires} substantial density fluctuations. 
This leads to a third potential explanation of the
\(\sigma\pos\)--\(\siglosm\) relationship.
In the limit that the direct contribution of
turbulent velocity fluctuations is negligible
and that \(\siglosm\) is entirely due to ordered expansion along the line of sight,
then a non-zero \(\sigma\pos\) can arise only from emissivity fluctuations,
which cause different parts of the ordered velocity field to be picked out
for different lines of sight.
A simple model of this process is explored in Appendix~\ref{sec:appar-veloc-fluct},
where a relation between the surface brightness dispersion and the apparent
velocity dispersions is derived.
This is compared with the behavior of the ratio
\(\sigma\pos / \siglosm\) for
our observed sample in panel~i of Figure~\ref{fig:brightness-pdfs},
in which the solid and dashed lines show results for different assumptions
about the effects of velocity gradients on \(\siglosm\).
Although the model does qualitatively reproduce the increase
in \(\sigma\pos / \siglosm\) with increasing \(\sigma_S\), the quantitative agreement
is poor, with the model substantially overestimating the magnitude of
plane-of-sky fluctuations as compared with the line-of-sight widths.
This is evidence that the combination of a velocity gradient with
emissivity fluctuations is insufficient on its own,
and that we require an additional disordered velocity component
in order to explain our observations.

In summary, our results are most consistent with a combination
of expansion and disordered motions, in which the disordered component
becomes increasingly dominant for the sources with higher velocity dispersion.
We cannot rule out a role for either projection smoothing or emissivity fluctuations,
but the contribution is probably minor in both cases.

\subsubsection{Luminosity vs centroid velocity dispersion}\label{sec:L-vs-sigmapos}

Figure~\ref{fig:sigvsl} shows the observed relation between \ha{} luminosity
and plane-of-sky dispersion of velocity centroids for our sample,
for which we obtain the power-law fit
\(\log \sigma\pos = (0.25 \pm 0.11) \log L(\ha) - (9.1 \pm 4.3)\). 
It is common to study the \(L\)--\(\sigma\) relationship,
not only in individual \hii{} regions, but also in \hii{} galaxies,
both in the local universe and at high redshift
\citep{terlevich1981, Chavez:2014a}.
In such studies, the velocity dispersion is usually derived from the
line width  of the entire region (\(\sigma\los\)) since the spatially resolved observations
necessary for measuring \(\sigma\pos\) are not available.%
\NEW{\footnote{
  \startNEW
  The linewidth \(\sigma\los\) for the integrated emission of an \hii{} region
  will be larger than the mean of the spatially resolved linewidths \siglosm{}
  since it also includes the plane-of-sky fluctuations \(\sigma\pos\).
  Given the results of section~\ref{sec:sigmapos-vs-sigmalos},
  and assuming that the linear relationship of Figure~\ref{fig:los-vs-pos}
  can be extrapolated to higher values, then the approximate
  relation is \(\sigma\los = \sqrt{2}\,\siglosm\) for \(\sigma\los \gg \SI{10}{km.s^{-1}}\).
  \stopNEW
}}
A reliable empirical \(L\)--\(\sigma\) relationship allows the use of bright photoionized
regions as standard candles in order to constrain cosmological parameters
\citep{Chavez:2012a, 2020ApJ...888..113W, Gonzalez-Moran:2021d}.
The luminosities of the regions used in these cosmological studies range from
that of our own brightest source up to 100 times more luminous
(\(L(\ha) = \num{e40}\) to \SI{e42}{erg.s^{-1}}),
although other studies cover intermediate luminosity regions
that bridge the two ranges \citep{moiseev2012, Yu:2019a}.

Expressed in the customary form \(L \sim  \sigma^\alpha\), our result corresponds to
\(\alpha = 4.0 \pm 1.6\), which is smaller than the value of \(\alpha \approx 5\) that is typically found
for more luminous regions
\citep{Moiseev:2015a, 2020ApJ...888..113W}
but not significantly so, given the uncertainty.
Studies that rely on \(\siglosm\) generally find that
\(\sigma\) becomes insensitive to luminosity below about \(L(\ha) = \SI{e40}{erg.s^{-1}}\),
reaching a constant value of \(\sigma = \SI{15}{km.s^{-1}}\) \citep{Moiseev:2015a, Yu:2019a}.
On the other hand,
by using \(\sigma\pos\) instead of \(\siglosm\),
we succeed in extending the same power-law relation
down to \(L(\ha) = \SI{2e37}{erg.s^{-1}}\) and \(\sigma = \SI{3}{km.s^{-1}}\).
Given our analysis of the previous section, one possible explanation for this
is that \(\sigma\pos\) is a better measure of the disordered turbulent motions within
the \hii{} regions than \(\siglosm\) since the latter
is contaminated by large-scale ordered motions,
which are always at least as large as the ionized sound speed.

\section{Conclusions}\label{sec:conclusions}

We have used the second-order structure function of velocity centroids
on the plane of the sky to carry out a systematic study of the turbulent kinematics
in a sample of 9 \hii{} regions spanning a wide range of luminosities and sizes.
Our principal findings are as follows:
\begin{enumerate}[1.]
\item The velocity fluctuations range from subsonic (Mach number \(\Mach \approx 0.5\))
  in the case of low-luminosity Galactic regions, such as the Orion Nebula,
  up to supersonic (\(\Mach \approx 2.5\)) in the case of giant extragalactic regions
  such as 30~Doradus in the Large Magellanic Cloud. 
\item The correlation length of the velocity fluctuations is always a small fraction
  (roughly 2\%) of the \hii{} region diameter (section~\ref{sec:D-vs-r0}),
  implying that turbulent energy
  is injected on scales smaller than about 10\% of the size of each region.
\item The power law slope of the structure function at scales smaller than
  the correlation length is \(m \approx 1.0 \pm 0.2\) for all regions,
  with no significant correlation with size, luminosity, or velocity amplitude
  (section~\ref{sec:constancy-power-law}).
  This can be explained as due to a fortuitous cancellation
  between, on the one hand, a steepening underlying velocity spectrum and,
  on the other hand, a reduced role for projection smoothing as the turbulent fluctuations
  pass from the subsonic to the supersonic regime. 
\item The non-thermal component of the emission line widths (\(\siglosm\)) is found
  to be proportional to the amplitude of the plane-of-sky velocity centroid fluctuations
  (\(\sigma\pos\)) with a slope of unity, but an offset such that
  \(\siglosm \approx \SI{10}{km.s^{-1}}\) when \(\sigma\pos = 0\)
  (section~\ref{sec:sigmapos-vs-sigmalos}).
  This can be explained if \(\siglosm\) is affected by ordered velocity fields,
  such as radial expansion, which leave no imprint on \(\sigma\pos\),
  implying that \(\sigma\pos\) is a better diagnostic of turbulent fluctuations than
  \(\siglosm\), even though it is more challenging to measure.
\item The amplitudes of \ha{} surface brightness fluctuations
  and turbulent velocity fluctuations 
  are only weakly correlated (Figure~\ref{fig:brightness-pdfs}h).
  For regions with supersonic turbulence
  these are consistent with the
  predictions for turbulence-induced density fluctuations,
  but for regions with subsonic turbulence the inferred density fluctuations
  are too large to be caused by turbulence alone. 
\item On scales much larger than the correlation length,
  a variety of behaviors are seen (section~\ref{sec:evid-inhom-at}).
  Some regions, such as Carina and Hubble~X show uncorrelated homogeneous fluctuations
  at the largest scales, whereas others, such as Orion and 30~Doradus show evidence
  for radial gradients in the amplitude of the velocity fluctuations
  with more vigorous turbulence in the core of the nebula than in the outskirts.
  In other regions, such as the Lagoon Nebula and NGC~346, there is evidence
  for quasi-periodic oscillations on scales similar to the size
  of the mapped region. 
\end{enumerate}

\section*{Acknowledgements}

Based in part on observations obtained at the Kitt Peak National Observatory,
which is operated by the Association of Universities for Research in Astronomy, Inc.,
under cooperative agreement with the National Science Foundation.
Based in part on observations made with the MUSE and FLAMES spectrographs
on VLT telescopes at the La Silla Paranal Observatory, ESO, Chile
under programme IDs 188.B-3002, 076.C-0888 and 098.D-0211.
Based in part on observations made with the William Herschel Telescope
operated by the Isaac Newton Group of Telescopes,
located at the Spanish
\foreignlanguage{spanish}{Roque de los Muchachos}
Observatory of the
\foreignlanguage{spanish}{Instituto de Astrofisica de Canarias}
on the island of La Palma. 
We gratefully acknowledge financial support provided by
\foreignlanguage{spanish}{%
  Dirección General de Asuntos del Personal Académico,
  Universidad Nacional Autónoma de México},
through grant
\foreignlanguage{spanish}{%
  Programa de Apoyo a Proyectos de Investigación
  e Inovación Tecnológica}
IN109823.  
We are grateful to Norberto Castro Rodríguez for providing maps of emission line velocity moments
for 30~Doradus derived from MUSE-VLT observations.
JGV acknowledges and thanks CONACyT-Mexico for a PhD research scholarship. 
A sincere thanks to the anonymous referee who help us improve the quality of our work.


\section*{Data availability statement}
\label{sec:data-avail-stat}
All data and accompanying analysis programs used in this paper are available
from the github repository \url{https://github.com/JavGVastro/PhD.Paper}.

\bibliographystyle{mnras}
\bibliography{bibphd}



\appendix

\renewcommand\textfraction{0.0}
\renewcommand\topfraction{1.0}
\renewcommand\bottomfraction{1.0}

\startNEW
\section{Degradation of the structure function due to observational limitations}
\label{sec:degr-struct-funct}
\stopNEW

To characterize the impact of the finite map size
and the atmospheric seeing on the structure function we perform a series of experiments using artificial turbulent velocity maps.
We quantify how these observational limitations affect the determination of
the structure function parameters, such as
the correlation length \(r_0\) and the velocity field variance \(\sigma\pos^2\). 
The synthetic velocity maps are created using a modified version of the \NEW{\texttt{make\textunderscore{}3dfield}} command from the \texttt{turbustat} Python library \citep{Koch2019AJ....158....1K},
which creates a \NEW{three-dimensional} fractional Brownian Motion field
\citep{Miville-Deschenes:2003a}
based on a power-law energy spectrum in wavenumber \(k\), \(E(k) \propto k^{-n\DDD}\),
and with random phases. 
Our modification multiplies the energy spectrum by a
low-wavenumber exponential taper:
\(e^{-1 / 2 \pi r_0 k}\).
As a result, the generated fields become uncorrelated at scales larger than \(r_0\)
and the resulting structure function closely approximates our model form
(section~\ref{sec:methods-apply}).

\startNEW
In order to create a 2D plane-of-sky velocity map, we first
create 3D cubes of fluctuating velocity and emissivity
with \texttt{make\textunderscore{}3dfield},
then transform to a PPV (position-position velocity) cube
using the \texttt{make\textunderscore{}ppv} command from \texttt{turbustat},
assuming thermal broadening with FWHM of \SI{20}{\kilo\meter\per\second},
as appropriate for a hydrogen line at \(T \approx \SI{e4}{K}\).
Finally, the first velocity moment map is obtained
by integrating along the velocity axis.

For the emissivity field, we assume a log-normal distribution
with RMS fractional width
\(\shortsig{E} = 1.0\), which is typical of our sources
(see footnote~\ref{fn:brightness-fluctuations} and Figure~\ref{fig:brightness-pdfs} above).
We further assume that there is no correlation between the emissivity
and velocity fluctuations and that their spatial correlation length
and power law indices are equal.\footnote{%
  \startNEW
  We have investigated departures from these assumptions and found that they
  make very little difference to the results.
  In particular, we have varied \(\shortsig{E}\) between
  0.0 (constant emissivity) and 2.0 (very strong fluctuations,
  as seen in 30~Doradus), and found that the fitting parameters
  for the effects of seeing (see below) change by less than 20\%
  \stopNEW
}

\newcommand\DD{\ensuremath{_{\mathrm{2D}}}}
The 3-dimensional 2nd-order structure function slope is related to the power law index as \(m\DDD = n\DDD - 3\), while the equivalent relation in two dimensions is \(m\DD = n\DD - 2\) for separations smaller than the line-of-sight
depth of the emitting region.
So long as the emissivity fluctuations are weak
and uncorrelated with the velocity fluctuations, then \(n\DD = n\DDD\)
\citep{Miville-Deschenes:2003a, Levrier:2004a},
but if these conditions are not satisfied, then the structure function
of the velocity centroids does not purely reflect the velocity power spectrum
\citep{Brunt:2004a, Esquivel:2005a, Ossenkopf:2006a, Esquivel:2007b}
and one has \(m\DD = n\DDD - 2 + \delta\kappa\) where \(\delta\kappa\) is a correction factor
that accounts for the contribution to the structure function from
column density fluctuations and cross terms (see section \ref{sec:constancy-power-law}).

\citet{Brunt:2004a} found that \(\delta\kappa\) is a function of the Mach number for
hydrodynamic turbulent box simulations.  In principle,
we could calculate \(\delta\kappa\) for
the fractional Brownian motion models that we use here
by using the analytic machinery of \citet{Esquivel:2007b},
but we prefer instead to use a more empirical approach.
We vary \(n\DDD\) until the
resultant structure function slope matches a particular value in the
range \(m\DD = 0.8 \to 1.2\)
that encompasses our observational results.\footnote{%
  Note that \(m\DD\) is written simply as \(m\) in the main text.
}
At each step, we estimate \(m\DD\) by fitting our idealized model structure function
(equation~\eqref{eq:model-strucfunc-ideal}, without the seeing and noise terms)
to the structure function of the simulated velocity map.
In the case of a uniform emissivity field (\(\shortsig{E} = 0\)),
we find \(\delta\kappa \approx 0\),
which is maximal projection smoothing, exactly as expected for a case that
mimics incompressible turbulence (see Appendix of \citealt{Miville-Deschenes:2003a}).
As the amplitude of the emissivity fluctuations increase,
we find that \(\delta\kappa\) becomes increasingly negative,
with \(\delta\kappa \approx -0.3\) for \(\shortsig{E} = 1\) (the case that we illustrate here)
and \(\delta\kappa \approx -0.55\) for \(\shortsig{E} = 2\).
This is qualitatively consistent with the results in
Figure~12a of \citet{Brunt:2004a}, although a quantitative
comparison is hard to make, since they are not holding \(m\) constant
in the way we do here. 
\stopNEW

The maps generated by \texttt{turbustat} are spatially periodic,
but we make them non-periodic by always dividing the full map into at least
four tiles, which are each analyzed separately. 

\startNEW
\subsection{Finite box effects}
\label{sec:finite-box-effects}
\stopNEW

\begin{figure}
  \begin{tabular}{@{} l @{}}
    (a)\\
    \includegraphics[width=\linewidth]{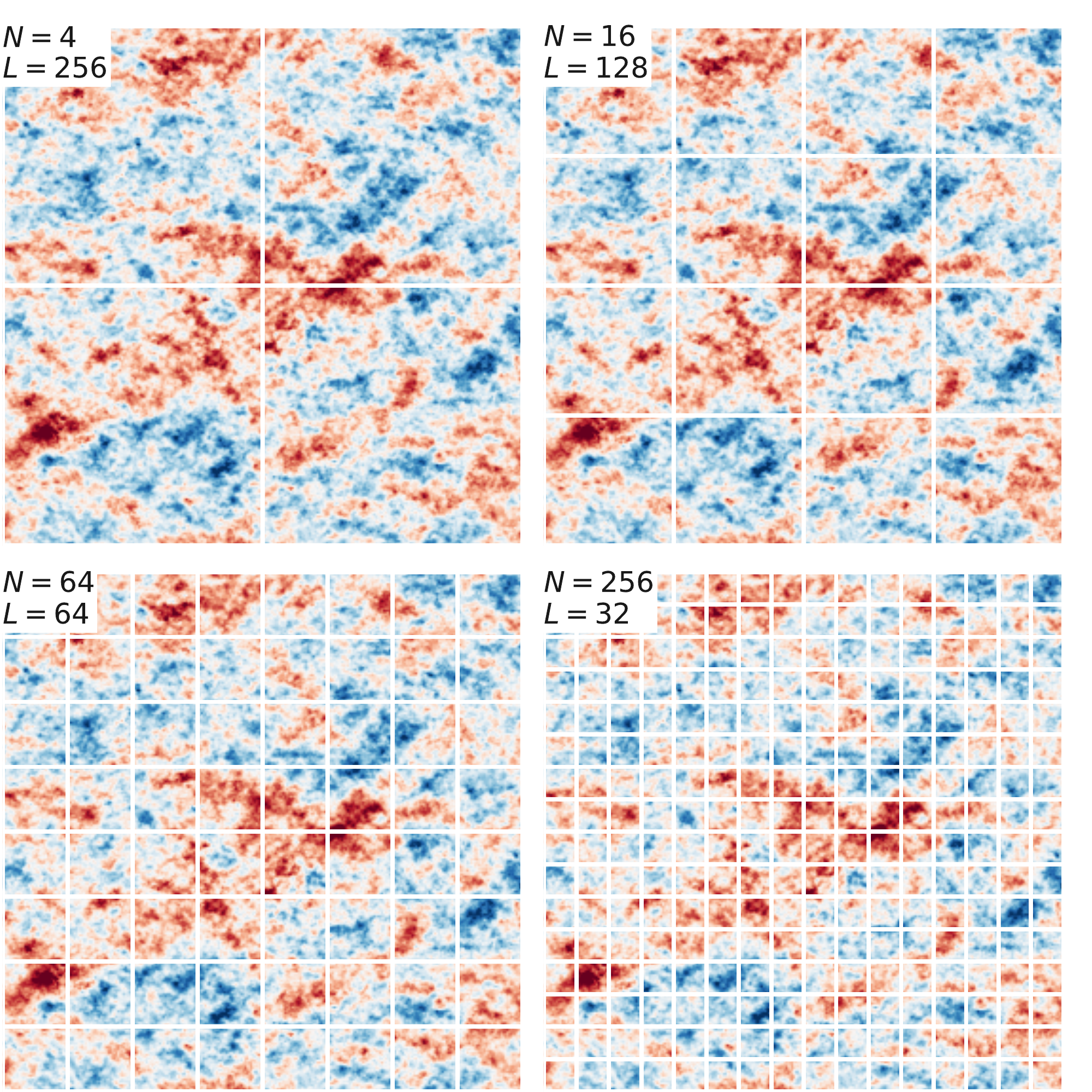}
    \\[\bigskipamount]
    (b)\\ \includegraphics[width=\linewidth]{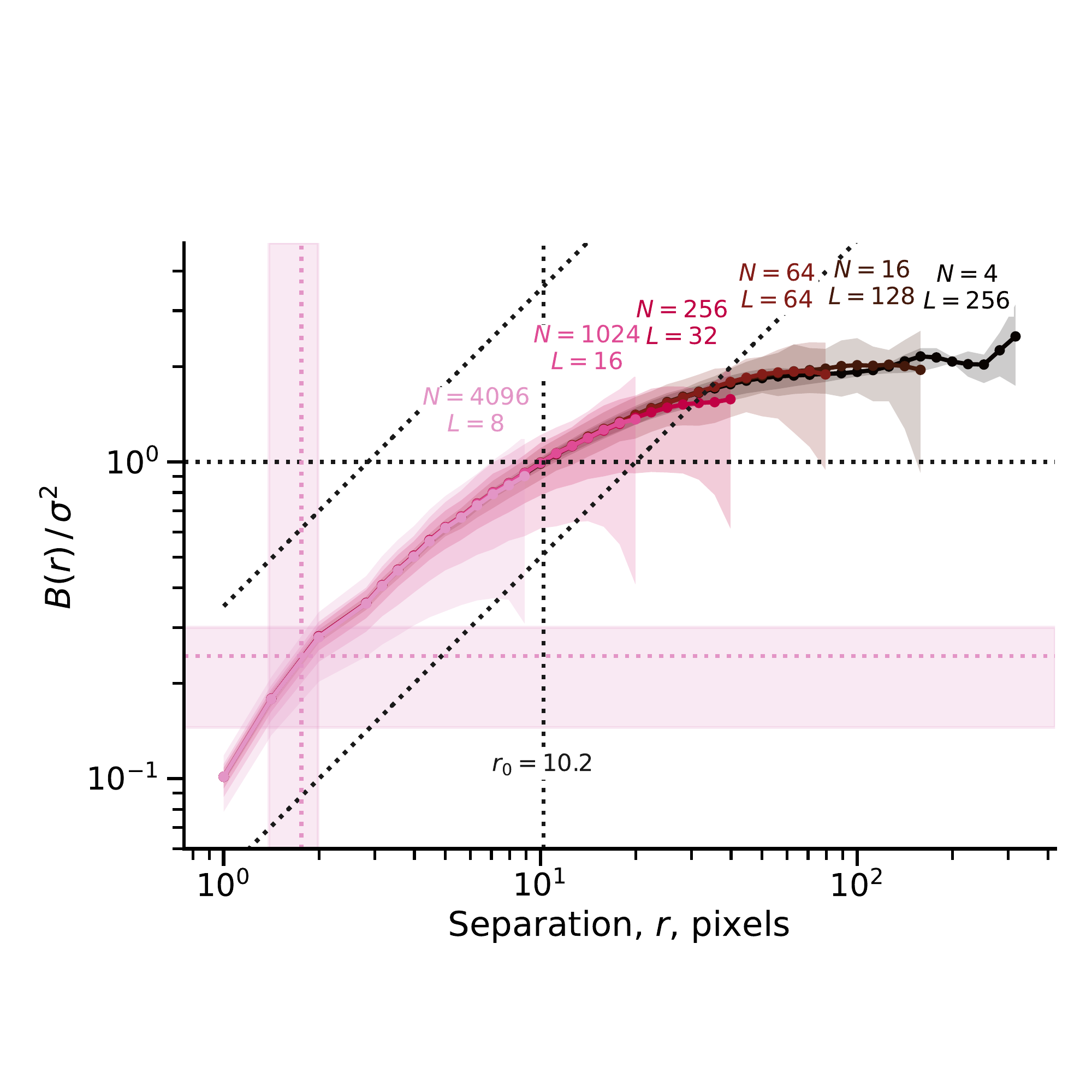}
  \end{tabular}
  \caption{Effects of finite box size on the structure function.
    (a)~Construction of simulated turbulent velocity fields of different sizes
    by repeated division of an initial field of size \(512 \times 512\) pixels.
    The \(j\)th level of division yields \(N = 4^j\) fields,
    each of linear size \(L \times L\) where \(L = 2^{9 - j}\).
    The color scale is the same for all maps and it considers \(- 3\) to \(+ 3\) times the standard deviation.
    (b)~Resultant structure functions.
    Colored lines show the average \(B(r)\) over all sub-images,
    while the shaded areas show the standard deviation of \(B(r)\).
  }
  \label{fig:finite-box}
\end{figure}

\begin{figure}
  \begin{tabular}{@{} l @{}}
    (a)\\ \includegraphics[width=0.95\linewidth]{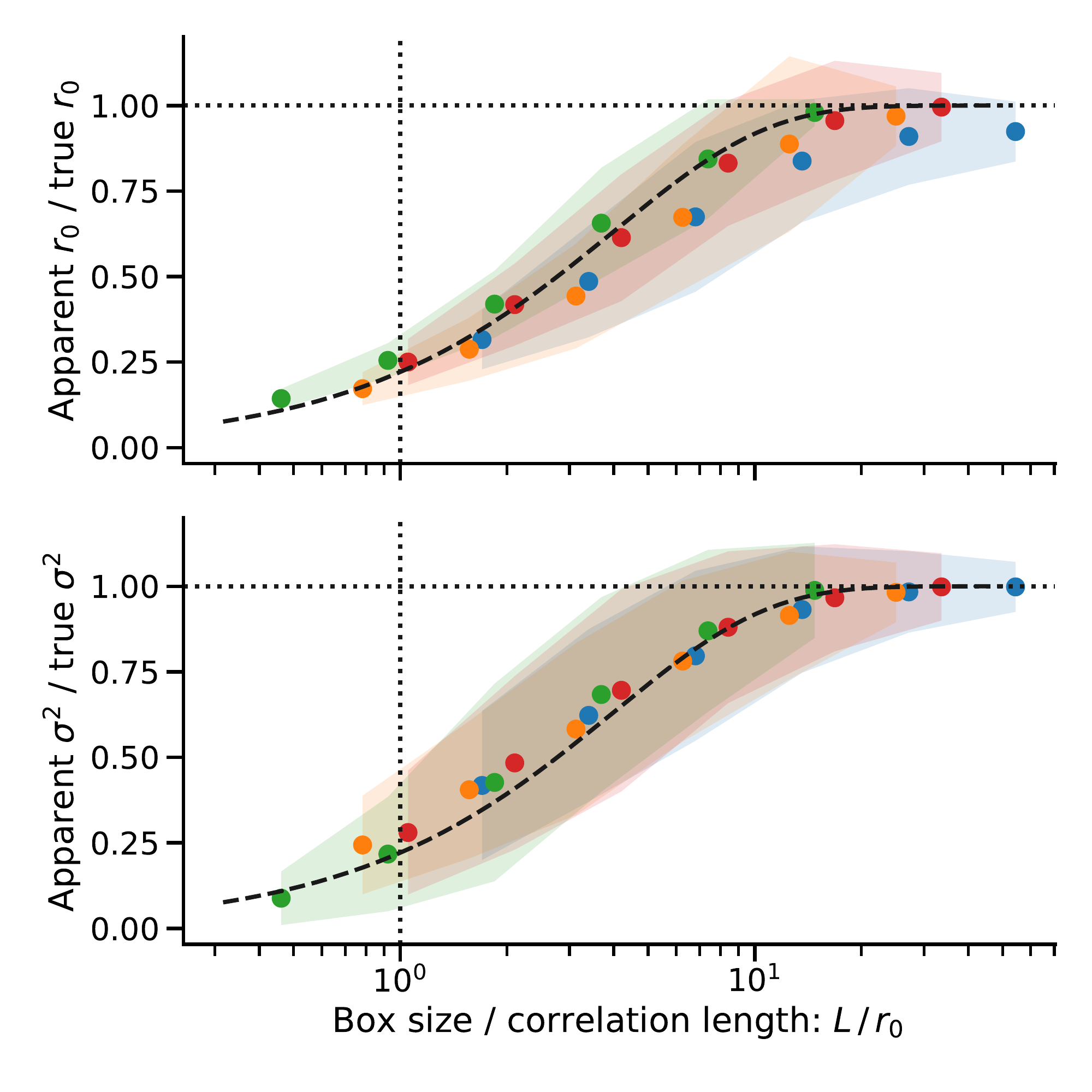}
    \\[\bigskipamount]
    (b)\\ \includegraphics[width=0.95\linewidth]{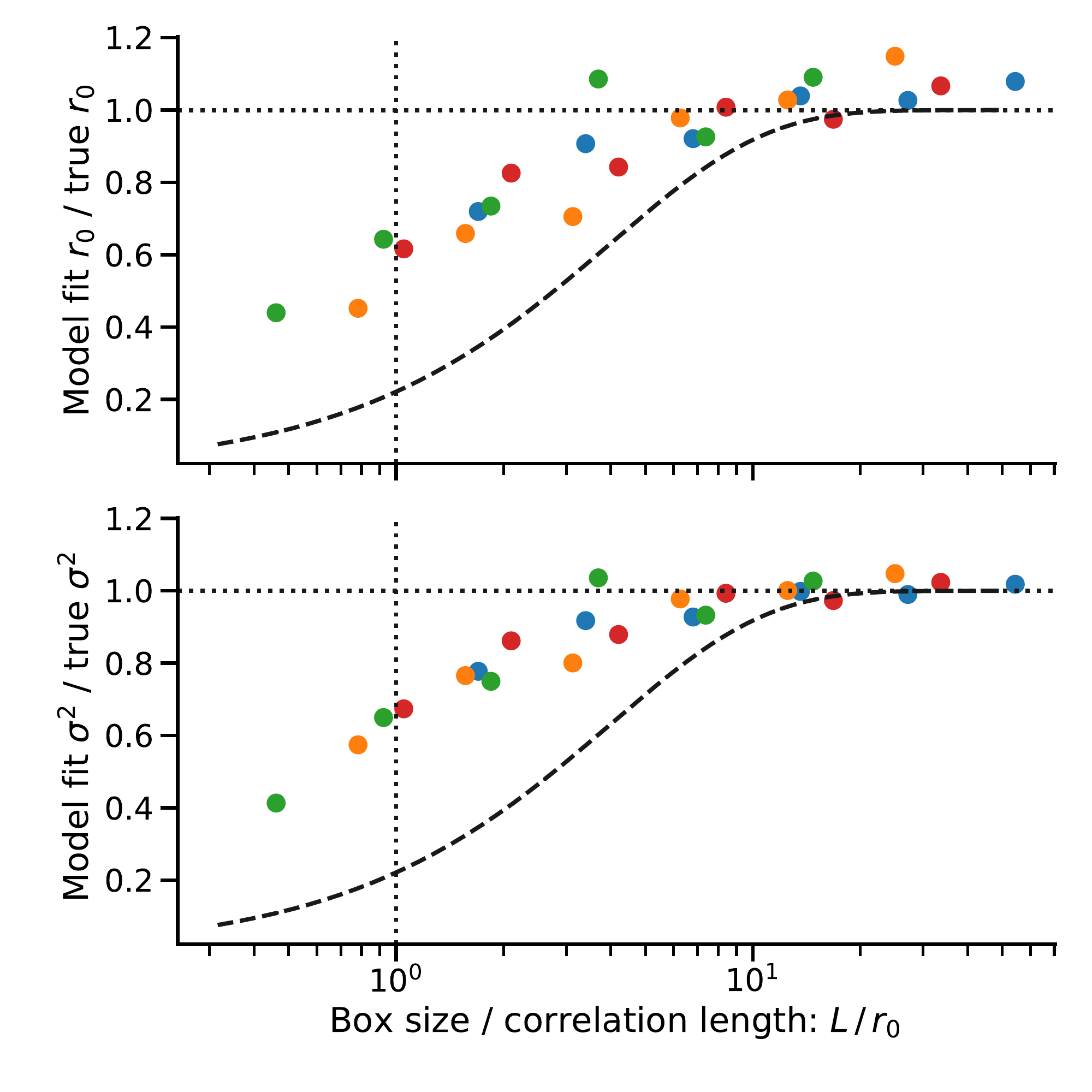}
  \end{tabular}
  \caption{
    Reduction of apparent velocity variance and correlation length
    due to finite box size.
    (a)~Empirical measurements of \(\sigma^2\) and \(r_0\).
    Colored dots show mean results from different simulated velocity fields,
    with different power law slopes.
    Colored shading shows the inter-quartile range over the different sub-images.
    The dashed line is an approximate analytic fit to these results
    (\NEW{\(1 - e^{L / 4.0r_0}\)}).
    (b)~Determination of \(\sigma^2\) and \(r_0\) from fitting the model
    of equation~\eqref{eq:new-correlation-form}.
    Colored dots are mean results from different simulated velocity fields,
    while the dashed line is the same as in (a).
  }
  \label{fig:finite-box-effect}
\end{figure}

\NEW{Figure~\ref{fig:finite-box}a} shows how we test for finite map effects by taking
a simulated \NEW{1st moment map of} size \(512 \times 512\) pixels and iteratively
dividing the full map into \(2 \times 2\) tiles.
On the \(j\)th iteration we have \(N = 4^j\) tiles,
each of size \(L \times L \) pixels with \(L = 2^{9 - j}\).
The figure shows the result for \(j = 1\), 2, 3, and 4, in which the full map
was generated with a correlation length \NEW{\(r_0 = 32\)~pixels}.
We then calculate the individual structure function for each tile and
find the mean and standard deviation for each level of subdivision,
showing the results in \NEW{Figure~\ref{fig:finite-box}b}.
It can be seen that the mean structure functions for each level
of subdivision (shown by colored dots and solid lines)
overlap very closely with one another, apart from deviations
due to edge effects when \(r > L/2\).
On the other hand, there is considerable dispersion between
individual tiles (shown by light colored shading) for the cases
where \(L < r_0\).

We next empirically measure the velocity variance and
apparent correlation length for each tile.
The velocity variance \(\sigma^2\) is found by
directly averaging over the pixels in the tile using equation~\eqref{eq:sig-pos},
while the apparent correlation length \(r_0\)
is found as the separation \(r\) for which \(B(r) = \sigma^2\),
using the tile's own empirically measured structure function and variance.
In \NEW{Figure~\ref{fig:finite-box-effect}a} we show how these
quantities deviate from the ``true'' values
(as measured on the full \(512 \times 512\) map)
as a function of the tile size \(L\).
It can be seen that the true values are successfully recovered
so long as the map size is more than ten times the correlation length,
but the apparent values of \(r_0\) and \(\sigma^2\) are smaller than the true values
when \(L < 10 r_0\), increasingly so as \(L\) becomes smaller.
This can be intuitively understood as due to the fact that a tile that
is small compared with the correlation length will
fail to sample the full range of velocity fluctuations,
leading to an underestimate of \(\sigma^2\).
Then the fact that \(B(r)\) increases with \(r\) in the power law regime
means there will be a concomitant underestimate of \(r_0\).

However, it is possible to do a considerably better job
of recovering the true \(\sigma^2\) and \(r_0\) by fitting our
heuristic model structure function of equation~\eqref{eq:model-strucfunc-ideal}.
This is shown in \NEW{Figure~\ref{fig:finite-box-effect}b}, where it can be seen
that the true parameters are reliably recovered so long as \NEW{\(L > 5 r_0\)},
which is a roughly \SI{0.5}{dex} improvement over the purely empirical method.
In order for the model fitting to work, the tile size must be large enough
that the structure function begins to show some flattening at the largest scales,
which is why the results are still inaccurate for the smallest values of \(L\)
that sample only the rising power-law portion of \(B(r)\). 

\startNEW
\subsection{Effects of seeing}
\label{sec:effects-seeing-struc}
\stopNEW

\begin{figure}
  \begin{tabular}{@{} l @{}}
    (a)\\
    \includegraphics[width=\linewidth]{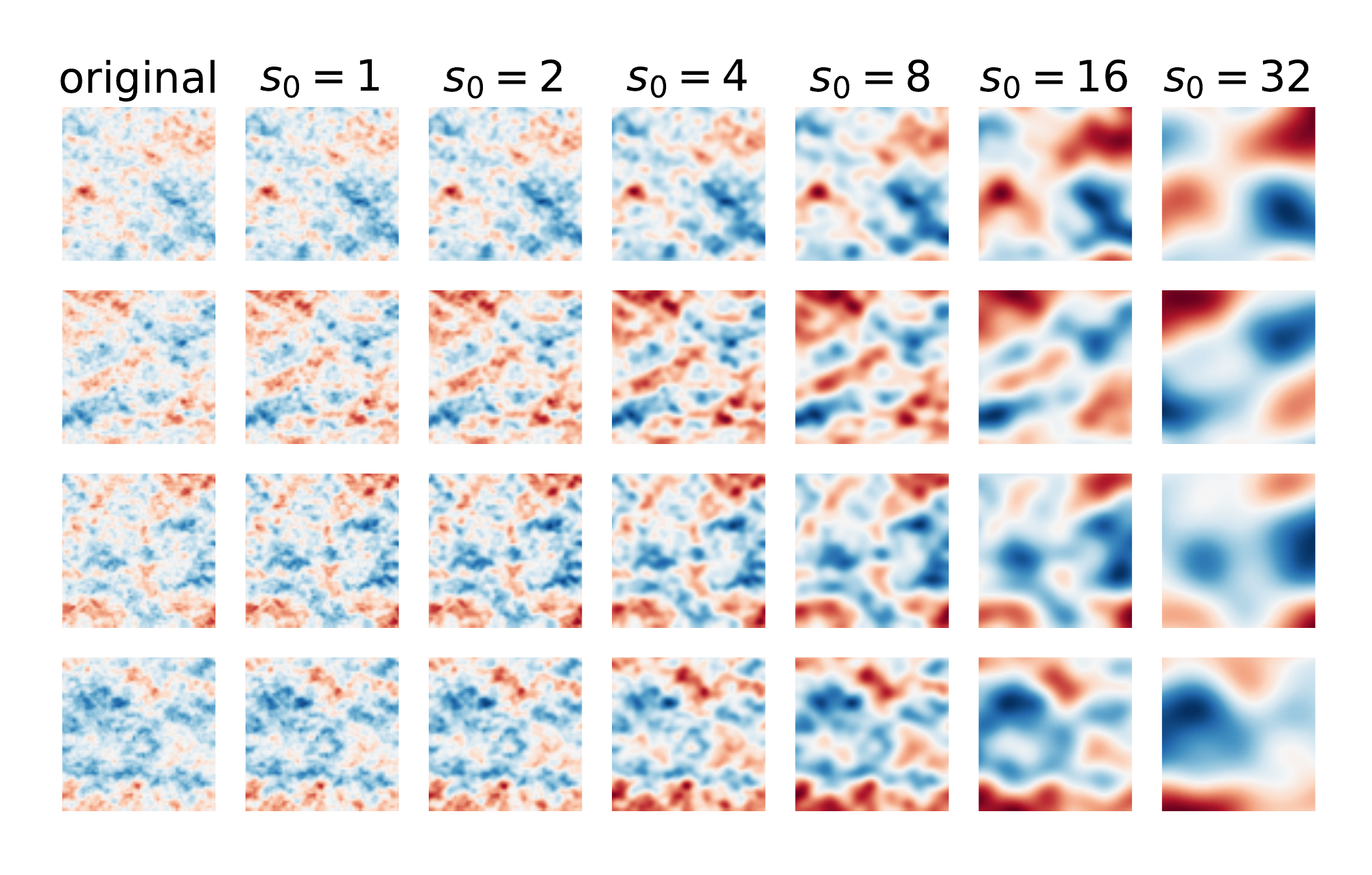}\\
    (b)\\
    \includegraphics[width=\linewidth]{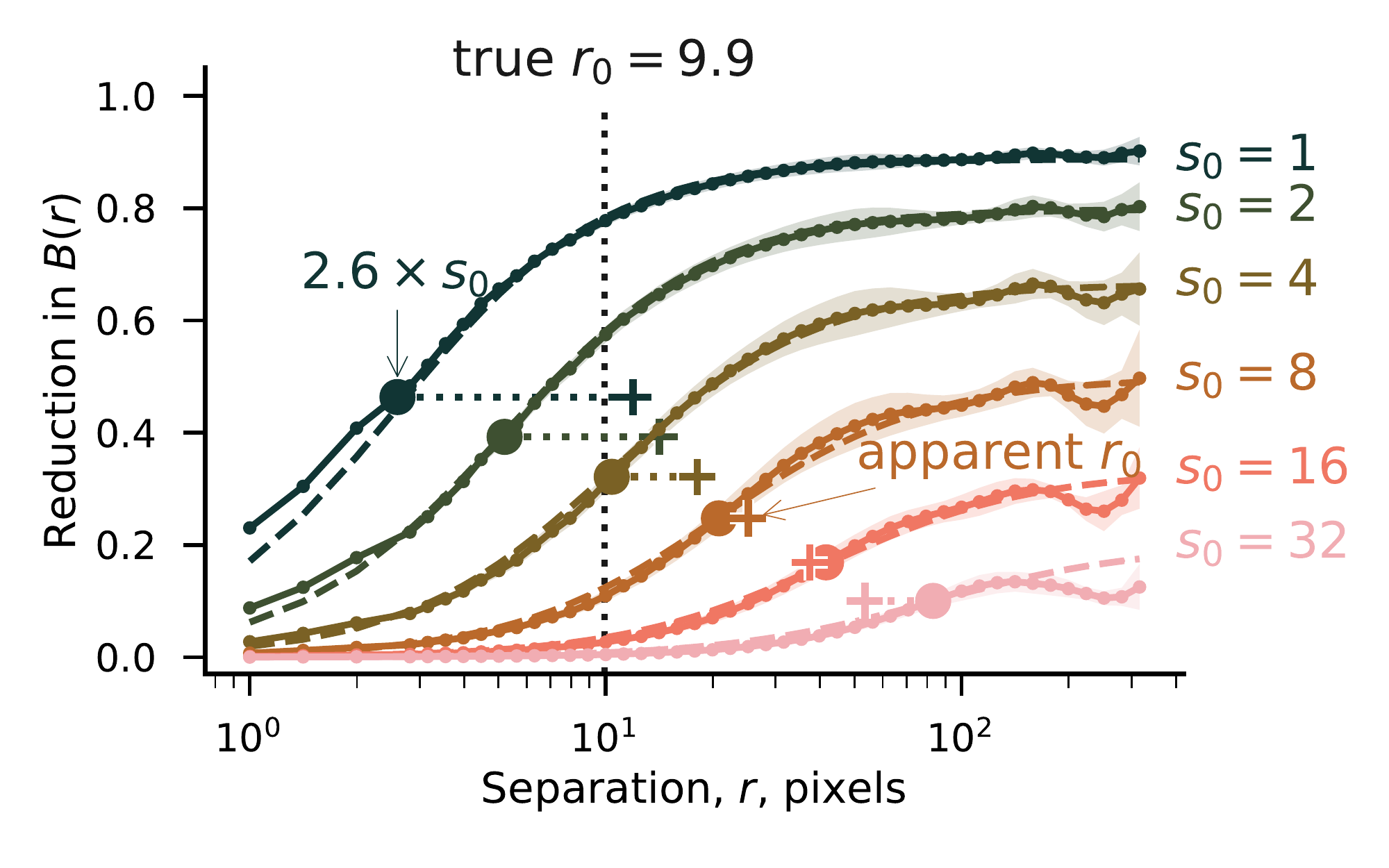}
  \end{tabular}
  \caption{Effects of seeing on the structure function.
    (a)~Each row shows a different simulated random velocity field
    on a \(256^2\) grid (see text for details).
    The left column shows the original field,
    while the remaining columns show the effects of smoothing by a gaussian kernel
    with RMS width \(s_0\) from 1 to 32 pixels.
    The same color scale is used for all columns, ranging from \(-3\) to \(+3\) times
    the standard deviation of the unsmoothed map.
    (b)~Relative change in the second-order structure function due to the gaussian smoothing.
    Solid lines with symbols show the average over the 4 maps of
    \(B(r, s_0) / B(r, s_0 = 0)\) as a function of separation \(r\),
    with a separation of \NEW{\(2.6 s_0\)} indicated by a large filled circle on each curve.
    Dashed lines shows the empirical fit discussed in the text.
    The dotted line shows the correlation length of the original fields,
    while colored plus symbols show the apparent correlation length of the smoothed fields.
  }
  \label{fig:seeing-reduction}
\end{figure}

Ground-based optical observations are affected by atmospheric seeing,
which limits the angular resolution of any observations.
This changes the observed structure function by suppressing fluctuations
between the most closely separated points.
\citet{Bensch:2001l} studied the  analogous problem
of beam smearing in the context of the \(\Delta\)-variance method \citep{Stutzki:1998a}, 
deriving a leading order approximation for the effect,
but an exact analytic treatment is difficult.
We therefore prefer to empirically study the effect of seeing using the
synthetic velocity maps described above.

\NEW{Figure~\ref{fig:seeing-reduction}a} illustrates how four different
synthetic \NEW{1st-moment maps of} size \(256\times256\) pixels
and correlation length \NEW{\(r_0 = \SI{32}{pixels}\)}
are affected by progressively broader gaussian seeing,
with RMS width \(s_0 = 1\) to \SI{32}{pixels}.
\NEW{%
  Note that the seeing is applied to each velocity channel of the
  PPV emissivity cube before integrating to find the 2D velocity map.
}
We calculate the structure function \(B(r)\) for each map and show
in Figure~\ref{fig:seeing-reduction}b the average ratio between
\(B(r)\) and the value in the absence of seeing,
with different colors representing different values of \(s_0\).
Results are shown for a power-law slope of \NEW{\(m = 1.0\)}
but we find very similar behaviour for other values of \(m\) within the range
seen in our observations.
For small values of \(s_0\), the principle effect is seen at
the smallest separations but the structure function is noticeably depressed
for separations as large as \(r = 10 s_0\).
As \(s_0\) increases to become comparable with \(r_0\),
even the largest separations are significantly affected.
At the same time the apparent correlation length begins to increase,
as indicated by the plus symbols in the figure.

We find that the following simple analytic function reproduces
these results to within better than 10\%:
\begin{equation}\label{eq:seeingemp2}
  \startNEW
  S(r; s_0, r_0) = 
  \left [
    \left (1+a_1\frac{s_0}{r_0} \right) 
    \left(1+ \left (a_2 \frac{ s_0}{r} \right) ^{a_3} \right)
  \right ]^{-1} ,
  \stopNEW
\end{equation}
which is shown by dashed lines in \NEW{Figure \ref{fig:seeing-reduction}b}.
\NEW{%
  The exact values of the constants \(a_1\), \(a_2\), \(a_3\)
  are found to depend slightly on the power law slope \(m\)
  and the magnitude of the emissivity fluctuations \(\shortsig{E}\).
  We adopt the compromise values \(a_1 = 1.25\),
  \(a_2 = 2.6\), and \(a_3 = 1.5\), which fit the numerical results to
  within 20\% for \(0.8 < m < 1.2\) and \(0.5 < \shortsig{E} < 2.0\).}
We therefore use this equation to represent the seeing
in our structure function fits (section~\ref{sec:results}),
with \(s_0\) as an additional free parameter for the models.

\FloatBarrier

\section{Apparent velocity fluctuations from emissivity fluctuations}
\label{sec:appar-veloc-fluct}

\begin{figure}
  \centering
  \includegraphics[width=\linewidth]{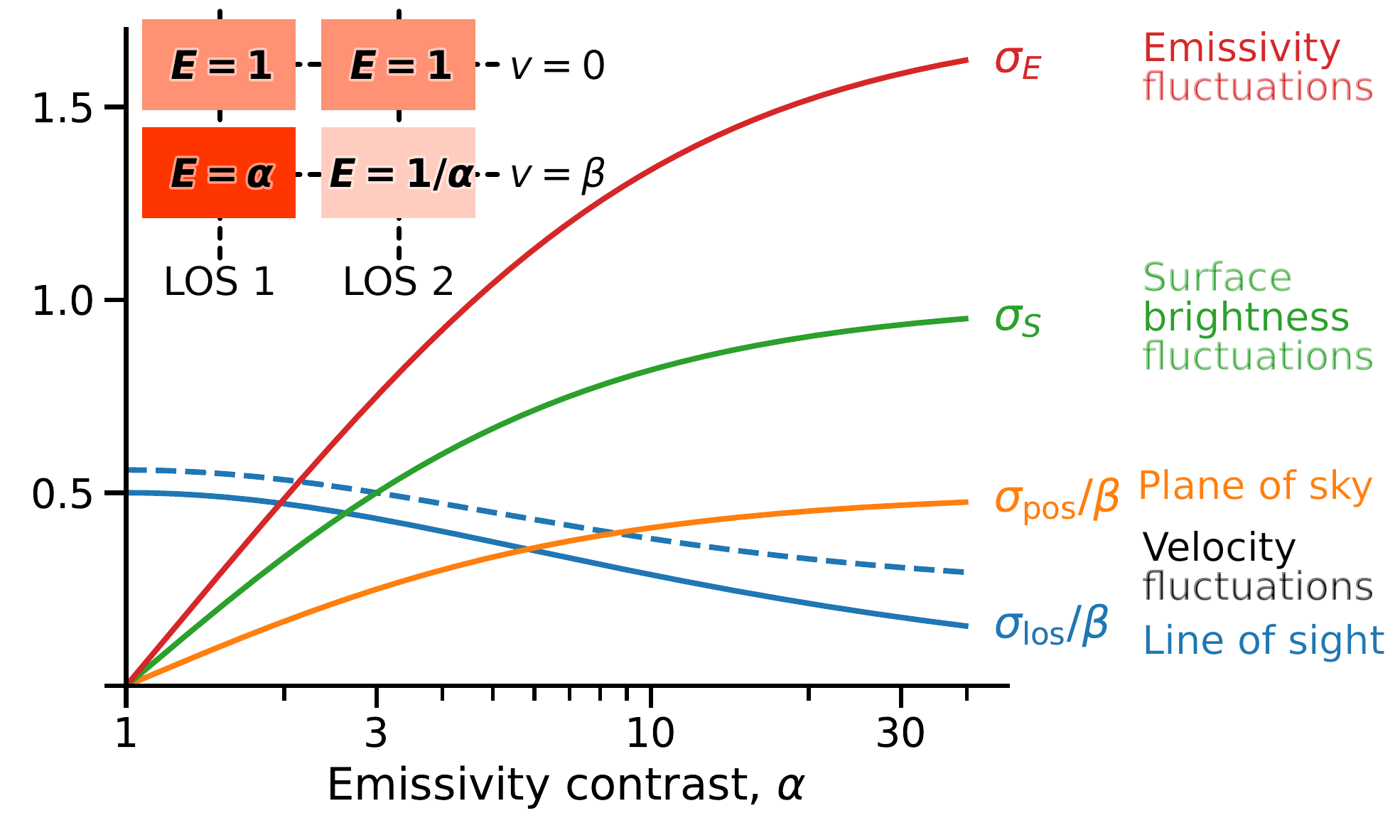}
  \caption{
    Toy model of emissivity variations in an ordered velocity field,
    showing co-variation of fractional dispersion of
    emissivity, \(\sigma_E\), and surface brightness, \(\sigma_S\),
    together with plane-of-sky dispersion in mean velocity,
    \(\sigma\pos\),
    and the mean RMS line width, \(\siglosm\),
    both in units of the inter-layer velocity difference \(\beta\).
  }
  \label{fig:bright-to-vel-fluct}
\end{figure}

In this appendix we describe a simple toy model to illustrate
how fluctuations of emissivity,
combined with an ordered velocity gradient along the line of sight,
can give rise to apparent fluctuations on the plane of the sky
of the mean velocity.
The model is illustrated in the inset of Figure~\ref{fig:bright-to-vel-fluct}.
It consists of two overlapping emission layers, A and B,
with velocities along the line of sight \(v_A = 0\) and \(v_B = \beta\), respectively,
which are observed along two lines of sight (LOS), 1 and 2.
The emissivity of layer A is \(E_{A1} = E_{A2} = 1\) for both LOS,
while the emissivity of layer B is \(E_{B1} = \alpha\) for LOS~1
and \(E_{B2} = \alpha^{-1}\) for LOS~2, where \(\alpha\) is a positive constant.
Both layers are assumed to have unit depth along line of sight,
such that the surface brightness for the two LOS are given by
\(S_1 = 1 + \alpha\) and \(S_2 = 1 + \alpha^{-1}\).

It then follows that the fractional standard deviation of the emissivity
and surface brightness are given by:
\begin{align}
  \label{eq:toy}
  \sigma_E &=
  \frac{
    \bigl[ 3\bigl(\alpha + \alpha^{-1}\bigr)^2 - 4\bigl(1 + \alpha + \alpha^{-1}\bigr)\bigr]^{1/2}
  }{
    2 + \alpha + \alpha^{-1}
  }
  \\
  \sigma_S &=   \frac{\vert\alpha - 1\vert}{1 + \alpha} ,
\end{align}
while the RMS line width and standard deviation of the mean velocities on the plane of the sky
are given by
\begin{align}
  \label{eq:toy}
  \sigma\pos &=\frac{\vert\alpha - 1\vert \beta}{2 (1 + \alpha)}
  \\
  \siglosm &=   \frac{\alpha^{1/2}\beta}{1 + \alpha}.
\end{align}
These are all shown in Figure~\ref{fig:bright-to-vel-fluct} as a function
of the emissivity contrast \(\alpha\).
It can be seen that \(\sigma_E\), \(\sigma_S\), and \(\sigma\pos/\beta\) all increase with \(\alpha\),
reaching asymptotic values of \(\sqrt{3}\), \(1\), and \(0.5\),
respectively as \(\alpha \to \infty\).\footnote{%
  To reach higher relative brightness dispersion \(\sigma_S > 1\) would require
  including more than two lines of sight in the model.
}
The line width \(\siglosm/\beta\) behaves in the opposite sense, falling from
a maximum value of \(0.5\) when \(\alpha = 0\) towards zero as \(\alpha \to \infty\),
a result of the emission becoming increasingly dominated by one
or the other of the two layers for high emissivity contrast.
The dashed line shows the effect of including a small velocity gradient
within each layer that contributes an additional width of \(\beta/4\),
which is assumed to add in quadrature with \(\siglosm\).

The predictions of this model are compared with observations in
panel~i of Figure~\ref{fig:brightness-pdfs}.

\section{\boldmath Notes on individual \hii{} regions}
\label{sec:notes-individual-hii}
For each region or pair of regions in our sample,
we give a more detailed description of the physical properties
(supplementing section~\ref{sec:regions-milky-way})
together with additional details of the structure function
(supplementing section~\ref{sec:results})
and comparison with previous results from the literature
(supplementing section~\ref{sec:comp-with-prev}).

\startNEW
\subsection{Orion Nebula}
\label{sec:orion-nebula}
\stopNEW

The Orion Nebula (M42) is the prototype of a galactic \hii{} region located at a distance of 440 pc \citetext{\SI{1}{\arcsecond} = \SI{0.002}{pc}; \citealp{2008AJ....136.1566O}}.
The O7 V star \(\uptheta^{1}\) Ori C is the most luminous and hottest star with an ionizing flux \(Q_0\) of \SI{1.3e49}{photons.s^{-1}} \citep{2006A&A...448..351S},
\startNEW
dominating the ionization of the nebula, which has a
\stopNEW
luminosity in \ha{} of \SI{1.5e37}{erg.s^{-1}}.
With 3 other B-type stars they form the Trapezium cluster.
The bright region that surrounds these massive ionizing stars is called the Huygens region.
The position of the Orion Nebula \NEW{on the near side of} the Orion Molecular Cloud (OMC-1)
\startNEW
makes it an ongoing site of star formation with a large population of young stars,
\stopNEW
some of which are sources of stellar jets and Herbig-Haro objects \citep{1993ApJ...410..696O}.
Its physical properties and kinematics, along with its stellar population is documented in \citet{2001ARA&A..39...99O}.

\startNEW
\subsubsection{Observed structure function and model fit}
\label{sec:observ-struct-funct-orion}
\stopNEW

\paragraph*{Orion core}
 
Figure~\ref{fig:strucfunc-fit-Galactic}a shows the structure function of the Orion core.
A detailed description of the model fit is given in section~\ref{sec:example-results-orion} and will not be repeated here.
\startNEW
At scales larger than the correlation length,
a maximum \(B(r)\) value of \SI{23}{km^{2}.s^{-2}} is seen at \(r = \SI{0.30}{pc}\),
beyond which the structure function declines to \SI{7}{km^{2}.s^{-2}} at \(r =  \SI{0.5}{pc}\). 
This ``hook'' behavior can be explained by a radial gradient in the
amplitude of the velocity fluctuations (see section~\ref{sec:limit-model-struct}).

An alternative approach to determine the velocity variance and correlation length
is to empirically measure \(\sigma\pos^2\) from the observations and then to use linear
interpolation on the measured structure function to find the point where \(B(r_0) = \sigma\pos^2\).
This yields \(\sigma\pos^2 = \SI{9}{km^{2}.s^{-2}}\) and \(r_0 = \SI{0.05}{pc} \),
as compared to the model-derived values of \(\sigma\pos^2 = \SI{13}{km^{2}.s^{-2}}\) and \(r_0 = \SI{0.07}{pc} \).
The main reason for the discrepancy is the finite box effect
discussed in section~\ref{sec:finite-box-effects} above, where it is shown
that for homogeneous fluctuations it is necessary to sample a map of size \(L > 10 r_0\)
in order to empirically determine \(\sigma\pos^2\).
In the case of the Orion core observations, \(L = 7 r_0\),
so the fluctuations are not fully sampled, leading
to the empirical \(\sigma\pos^2\) being an underestimate.
However, given the evidence for a radial gradient in the amplitude of the velocity fluctuations,
it could be argued that the finite box effect is not so important as it would be for homogeneous fluctuations
and it is instead the model-derived value of \(\sigma\pos^2\) that is an overestimate.
\stopNEW


\paragraph*{EON}

Figure~\ref{fig:strucfunc-fit-Galactic}b shows the structure function of the Extended Orion Nebula, which is ten times larger than the core in linear size,
but observed at a much lower angular resolution
(see Figures~\ref{fig:hii-regions} and~\ref{fig:velocity-maps}).
The data for this region are the lowest quality of all the regions in
our sample, and there are too many degeneracies between the parameters
to achieve a credible fit using the full model.
We therefore fix the power slope at \(m = 1\),
which implies that noise must make a significant contribution at all scales
in order to reproduce the observed flat structure function.
In turn, this implies that
\startNEW
the true velocity variance is
\stopNEW
\(\sigma^2\pos = \SI{5}{km^2.s^{-2}}\),
which is lower than the \SI{7}{km^2.s^{-2}}
that would be naively derived from the observations
(compare green and orange lines in Figure~\ref{fig:strucfunc-fit-Galactic}b).
\startNEW
Comparing the extended nebula with the core for the size scales where they overlap
(\num{0.1} to \SI{0.5}{pc}), one sees that \(B(r)\) is roughly four times smaller
in the EON than in the core for the same value of \(r\).
This is further evidence for a decline in the fluctuation amplitude with radius,
as mentioned above.
Also consistent with this is a hook-like downturn in \(B(r)\) for \(r > \SI{2}{pc}\)
for the EON region.
The fact that the derived \(r_0\) is larger for the EON region than the core
(\SI{0.5}{pc} versus \SI{0.07}{pc}) means that the correlation length is also inhomogeneous,
tending to increase from the center to the outskirts of the nebula. 
\stopNEW

\startNEW
\subsubsection{Comparison with previous studies}
\label{sec:comparison-orion}
\stopNEW

\citet{arthur2016turbulence}
\startNEW
calculated the structure function of different emission lines in the inner Orion Nebula with varying degrees of ionization.
They found that higher ionization lines show a larger velocity dispersion
and a steeper structure function slope.
\stopNEW
They also make a comparison with previous work (see Table~5 of their paper), finding a broad
\startNEW
agreement in derived structure function parameters
and their dependence on ionization energy.
\stopNEW
The power-law index obtained by \citet{arthur2016turbulence} is
\startNEW
\(m \sim 1.17 \pm 0.08\) for \halpha,
with a correlation length of \SI{\sim 0.05}{pc} and velocity variance
\(\sigma\pos^2= \SI{9.4}{km^{2 }.s^{-2}}\).
This compares with our own values of
\(m = 1.07 \pm 0.03\), \(r_0 = \SI{0.068 \pm 0.004}{pc}\)
and \(\sigma\pos^2 = \SI{13 \pm 1}{ km^{2}.s^{-2}}\). 
Since they are based on the same observational data,
the differences of order 30\% between the \citet{arthur2016turbulence}
results and our own are entirely due to the different methodologies employed.
In the case of \(\sigma\pos^2\), their value is lower than ours since
they masked out the locations of known high-velocity Herbig-Haro flows, whereas we do not.
\stopNEW

\startNEW
\subsection{Lagoon Nebula}
\label{sec:lagoon-nebula}
\stopNEW

The Lagoon Nebula (M8, NGC~6523) is located at a distance of \SI{1250}{pc}  \citetext{\SI{1}{\arcsecond} = \SI{0.006}{pc} ; \citealp{2005A&A...430..941P}}.
It contains the young stellar cluster NGC~6530 and the region is illuminated by different massive stars of spectral types O and B, the hottest being 9~Sgr, type O4V \citep{Damiani:2017b}.
\startNEW
The optically brightest part of the nebula is denoted the ``Hourglass nebula'',
which is illuminated by the O7 star Herschel~36 \citep{1986AJ.....91..870W}.
\stopNEW
These stars gives an ionizing flux \(Q_0 = \SI{2.6e49}{photons.s^{-1}}\) (2 times higher than in the Orion Nebula) and the luminosity in \ha{} is \SI{2.95e37}{erg.s^{-1}} \citep{1984ApJ...287..116K}.
The mean non-thermal linewidth in \ha{} is \SI{11.7 \pm 1.6}{km.s^{-1}} \citep{1973ApJ...183..851B}.
\startNEW
Molecular gas and ongoing star formation are present in the region around
Herschel~36 \citep{Arias:2006e, Tiwari:2018a} and the lower-luminosity region
M8~East \citep{1984ApJ...278..170S, Tiwari:2020a}.
\stopNEW
The properties of the nebula are reviewed by \citet{2008hsf2.book..533T}.

\startNEW
\subsubsection{Observed structure function and model fit}
\label{sec:observ-struct-funct-lagoon}

Figure~\ref{fig:strucfunc-fit-Galactic}c shows the structure function of the Lagoon nebula.
The model fit gives \(\sigma\pos^2 = \SI{7 \pm 1}{km^{2}.s^{-2}}\), \(r_0 = \SI{1 \pm 0.2}{pc} \), and \(m = 1.26 \pm 0.2\).
In this case, there is good agreement between the model value and empirical value of \(\sigma\pos^2\),
indicating that finite box effects are unimportant.
On the other hand, the model-derived \(m\) is steeper than would be naively derived
from the observations, which is a symptom of the effects  of noise.
At separations smaller than the correlation length, the structure function shows irregular oscillations about the mean ascending trend,
which are a symptom of the relatively small number of spatial points in this source.

At scales much larger than the correlation length
the structure function begins to increase again, reaching a value of \SI{66}{km^{2}.s^{-2}} at \SI{20}{pc},
which is greater than \(9 \sigma\pos^2\).
As discussed in section~\ref{sec:evid-inhom-at}, this may be a sign of periodicity
or a large-scale gradient in the velocity field.
\stopNEW

\startNEW
\subsubsection{Comparison with previous studies}
\label{sec:comparison-lagoon}
The first investigation of the Lagoon nebula structure function is from \citet{1970A&A.....8..486L},
who use Fabry-Perot observation of the \ha{} line to probe scales
between \SI{0.7}{pc} and \SI{7}{pc}.
They find a rising power-law behavior with slope of \(m = 0.64\),
which is slightly smaller than the value we derive.
However, given the very coarse binning in separation of their structure function,
and the fact that they do not account for noise or the flattening of the structure function at large scales,
their results are broadly consistent with ours,
with the value of \( \sigma\pos^2 = \SI{16}{km^{2}.s^{-2}}\) at \(r = \SI{3.5}{pc}\) being almost
identical to our own value at that scale.

More recent measurements of the structure function of the Lagoon Nebula
have used the [\ion{O}{3}] \Wav{5007} line and have probed smaller scales.
\citet{1987ApJ...317..676O} used longslit observations at a small number of position angles
to study the structure function over an area \(\num{2} \times \SI{2}{arcmin^2}\),
centered on the nebula's principal ionizing star, 9~Sgr.
However, it is hard to compare their results to ours, as they carry out
a Gaussian decomposition of the line profile into two components, A and B,
and analyse the structure function of the two components separately.
The total variance of component A (the stronger and narrower of the two)
is \SI{2}{km^{2}.s^{-2}} at scales \(< \SI{1}{pc}\),
which is significantly smaller than our value of \SI{7}{km^{2}.s^{-2}},
but this is not surprising since they are probing scales smaller than the
correlation length and therefore suffer from a severe finite-box effect.
Component~B on the other hand shows a much larger variance
of \SI{13}{km^{2}.s^{-2}} and an essentially flat structure function.

Another [\ion{O}{3}] \Wav{5007} study \citep{Chakraborty:1999a}
calculated the structure function in the Hourglass region of the nebula:
an area of \(\num{1} \times \SI{1}{arcmin^2}\) centered at Herschel~36.
As compared to the data used in this paper, which spans between \num{0.1} and \SI{10}{pc}),
their results cover smaller scales, between \num{0.012} and \SI{0.3}{pc}.
\citet{Chakraborty:1999a} report a rising power-law behavior with slope of \(m = 0.46\) over the entire range.
No flattening of the structure function at the largest scales was seen, but that is to be expected
if the correlation length is larger than the size of the region mapped.
The amplitude of the fluctuations is much higher (\(\sigma^2\pos = \SI{130}{km^2.s^{-2}}\))
than we find for the wider nebula (\(\sigma^2\pos = \SI{7}{km^2.s^{-2}}\)),
suggesting an even more extreme version of the inhomogeneity that we find in the Orion Nebula.
This is supported by the fact that the Hourglass region also shows
much larger line widths than the rest of the nebula \citep{Chakraborty:1999a},
which is probably a symptom of the ongoing active star formation in that region.
\stopNEW

\startNEW
\subsection{Carina Nebula}
\label{sec:carina-nebula}
\stopNEW

The Carina nebula is a large star-forming complex located at a distance of \SI{2 350}{pc} \citetext{\SI{1}{\arcsecond} = \SI{0.01}{pc}; \citealp{2006ApJ...644.1151S}}.
It contains young clusters like Trumpler~14 and~16, and Collinder~228 from the Car~OB1 association, one of the largest OB associations in the Galaxy.
This OB association has more than 65 stars earlier than B0, and several Wolf-Rayet (WR) stars and the most luminous star is the blue variable $\eta$~Car.
The ionizing flux \(Q_0\) is \SI{8.9e50}{photons.s^{-1}} \citep{2008hsf2.book..138S} (60 times higher compared to the Orion Nebula) and the luminosity in \ha{} of \SI{e39}{erg.s^{-1}}.
As a high-mass star formation region, it presents several phenomena related to young stars,
such as evaporating protoplanetary discs, erosion of large dust pillars, and the triggering of a second generation of embedded stars and Herbig-Haro objects \citetext{see \citealp{2008hsf2.book..138S} and reference therein}.
\startNEW
This region is at an intermediate scale between nearby regions like Orion
that are dominated by a single star, and giant star formation regions such as 30~Doradus.
\stopNEW

\startNEW
\subsubsection{Observed structure function and model fit}
\label{sec:observ-struct-funct-carina}
\stopNEW

Figure~\ref{fig:strucfunc-fit-Galactic}d shows the structure function of the Carina nebula.
\startNEW
The model fit gives \(\sigma\pos^2 = \SI{18 \pm 2}{km^{2}.s^{-2}}\), \(r_0 = \SI{0.6 \pm 0.1}{pc} \), and \(m = 1.2 \pm 0.3\).
Again there is good agreement between the model value and empirically measured
value of \(\sigma\pos^2\),
indicating that finite box effects are unimportant. In addition, both noise and seeing are
found to be unimportant. 
Similarly to the Lagoon nebula,
the structure function of Carina also shows a non-uniform and irregular behavior
below the correlation length, which can again be ascribed to the sparse spatial sampling.
As a result of this, the power-law slope determination is not very precise.

On scales larger than the correlation length,
the structure function of the Carina nebula is remarkably flat
for \(r = 2\) to \SI{6}{pc},
in good agreement with the prediction of homogeneous fluctuations
(Figure~\ref{fig:model-strucfunc}a).
But at the largest scales accesible to our observations, there is evidence for low-amplitude
periodic fluctuations with a wavelength of \(\lambda \approx \SI{8}{pc}\),
which corresponds to the typical separation
between the individual star clusters that make up the complex
(for instance, between Tr~14 and Tr~16).
\stopNEW

\subsubsection{Comparison with previous studies}
\label{sec:comparison-carina}

We have found no previous studies of Carina's structure function in the literature.

\startNEW
\subsection{30 Doradus}
\label{sec:30-doradus}
\stopNEW

The giant \hii{} region 30~Doradus (Tarantula Nebula) is located at a distance of \SI{49.9}{kpc} \citetext{\SI{1}{\arcsecond} = \SI{0.24}{pc}; \citealp{2013Natur.495...76P}}.
It is a luminous star-forming region in the LMC (Large Magellanic Cloud) and is the most luminous complex in the Local Group \citep{1984ApJ...287..116K}.
\startNEW
The star cluster that energizes the region is NGC~2070, with a central core (known as R136)
with ionizing luminosity \(Q_0 = \SI{2.5e51}{photons.s^{-1}}\)  \citep{2020MNRAS.499.1918B},
which produces an \ha{} luminosity of \SI{2.9e39}{erg.s^{-1}} (200 times the Orion Nebula)
for the central ionized region of size \SI{40}{pc}.
\stopNEW
The mean non-thermal linewidth in \ha{} is \SI{22}{km.s^{-1}} \citep{2013A&A...555A..60T}.
\startNEW
Star formation is still ongoing in NGC~2070 \citep{2013AJ....145...98W},
although the star-formation rate peaked about \SI{4}{Myr} ago
\citep{2015ApJ...811...76C} and it currently hosts some of the most massive stars known
\citep{Kalari:2022p}
and several Wolf-Rayet stars \citep{2011A&A...530A.108E}.
\stopNEW
This nebula is commonly used as a comparison for extragalactic star-forming regions.

\startNEW
\subsubsection{Observed structure function and model fit}
\label{sec:observ-struct-funct-30dor}

The 30~Doradus structure function is shown in Figure~\ref{fig:strucfunc-fit-MC}a.
The model-derived parameters are \(\sigma\pos^2 = \SI{297 \pm 40}{km^{2}.s^{-2}}\),
\(r_0 = \SI{4\pm 1}{pc} \)), and \(m = 0.85 \pm 0.1\).  The empirically-derived variance is about
15\% smaller, indicative of a moderate finite-box effect. On either measure,
the region is a clear outlier in our sample, with a value of \(\sigma\pos^2\) that is
more than three times larger than any other of our regions,
as can be graphically appreciated from the velocity histograms in Figure~\ref{fig:pdfs}. 

At scales larger than the correlation length, the structure function has a maximum
at \(r = \SI{20}{pc}\) and then falls at the largest scales.
This behavior is similar to that seen in the Orion Nebula and might
likewise be evidence for a decline of the fluctuation amplitude towards the outskirts.
On the other hand, the sharp upward turn towards the maximum in \(B(r)\) is more reminiscent
of the expectations for a periodic fluctuation (Figure~\ref{fig:model-strucfunc}b),
in which case the wavelength would be of order the size of the map.
Evidence for such a pattern can be seen directly in the velocity map
(Figure~\ref{fig:velocity-maps}) as a red-blue asymmetry between the south-east
and north-west, which is also seen in the kinematics of the O~stars
\citep{Henault-Brunet:2012h}
and has been interpreted as evidence for rotation of the cluster.
\stopNEW

\startNEW
\subsubsection{Comparison with previous studies}
\label{sec:comparison-30dor}

\citet{1961MNRAS.122....1F} investigated the structure function in 30~Doradus by using
spectra at \(N = 37\) different points, spread over a region of diameter \SI{150}{pc}
and with minimum separation of \SI{4}{pc}.
Velocities were measured by combining results from multiple spectral lines in
the blue spectral region, the strongest being of [\ion{O}{2}], [\ion{O}{3}], and \hb{}.
They found an essentially flat \(B(r)\) for \(r > \SI{10}{pc}\) with
\(\sigma\pos^2 = \SI{130}{km.s^{-1}}\), which is roughly half the value that we determine for the
central region of the nebula (of diameter \SI{30}{pc}).
This implies that the amplitude of the fluctuations is falling with radius,
similar to what we inferred above for the Orion Nebula.
For the smallest separations, of order \SI{4}{pc}, they found that \(B(r)\) declined
significantly, although an exact measurement was hampered by uncertainties in the
correction for the effects of noise.
This behavior is consistent with our own value of \(r_0 = \SI{4}{pc}\),
suggesting that the correlation length is not greatly different
in the wider area covered by their study.
This is a clear difference from the Orion Nebula,
where we find a significantly larger correlation length
in the extended nebula than in the core. 

\citet{Melnick:1987a} used Fabry-Perot observations in \ha{} and [\ion{O}{3}]
from \citet{Smith:1972a} to calculate
the structure function over a region of diameter \SI{50}{pc},
which is smaller than in \citet{1961MNRAS.122....1F}
but with a slightly larger number of spatial samples (\(N = 44\)).
They found an approximately flat \(B(r)\) for \(r > \SI{20}{pc}\)
and a shallow rising power law, \(\text{index} \approx 0.16\), for \(r = 5\) to \SI{20}{pc},
with \(\sigma\pos^2 \approx \SI{280}{km.s^{-1}}\).
This is similar to our own results over the same range of separations,
including the local peak in \(B(r)\) at \(r = \SI{20}{pc}\),
which is clearly visible in Figure~A2 of \citet{Melnick:1987a}.
The reason that \citeauthor{Melnick:1987a} measure such a shallow slope for their smallest
separations is that they are only sampling the range from \(1.25 r_0 \to 5 r_0\),
in which the structure function is tending toward its asymptotic value of \(2 \sigma\pos^2\).
In order to measure the power-law portion of the structure function,
it would be necessary to sample separations as small as \SI{1}{pc} or \SI{4}{arcsec}.

\citet{Melnick:2021x} used \ha{} spectra from VLT FLAMES/GIRAFFE (see section~\ref{sec:flames-multi-fiber}) for \(N = 1668\) distinct fiber positions
covering a region of diameter \SI{300}{pc}
\citep{Torres-Flores:2013t}.
They claim to find a completely flat structure function on scales from \num{3} to \SI{200}{pc},
which seriously conflicts with our own observations at the lower end of this range.
In particular, their two smallest separation bins are centered on \(r = 2.5\) and \SI{7.5}{pc},
for which our own results yield \(B(r) / \siglosm^2 \approx 0.7\) and \(1.3\),
respectively (our Figure~\ref{fig:strucfunc-fit-MC}a),
whereas they find \(B(r) / \siglosm^2 \approx 2.0\) at all separations (see their Figure 15).
We are at a loss to explain this discrepancy, which is inconsistent not only with our own results,
but also with those of \citet{1961MNRAS.122....1F} and \citet{Melnick:1987a},
who both found at least some evidence for a downturn in \(B(r)\) for \(r < \SI{10}{pc}\). 
The result also seems to contradict the evidence of their own Figures~7 and~13,
which show spatial coherence in the velocity field.
If the structure function were truly flat, then blue shifted and red shifted profiles
would be randomly mixed spatially, which is not observed. 
\stopNEW

\startNEW
\subsection{NGC 346}
\label{sec:ngc-346}
\stopNEW

NGC 346 is the most active star-formation region in the SMC (Small Magellanic Cloud) located at a distance of 62 kpc \citetext{\SI{1}{\arcsecond} = \SI{0.30}{pc} ; \citealp{2001ApJ...562..303D}}. 
It has more than 30 O stars that ionize N66, the largest \hii{} region \citep{2011ApJ...740...10D}.
\NEW{The cluster}
has an ionizing output \(Q_0\) of \SI{5.1e50}{photons.s^{-1}} (40 times higher than the Orion Nebula) and \NEW{the \hii{} region has} an \ha{} luminosity of \SI{5.9e38}{erg.s^{-1}} \citep{2010A&A...517A..39H,1984ApJ...287..116K}.
\startNEW
A supernova remnant is present in the region \citetext{J59.4\(-7210\), \citealp{Ye:1991d}}
but it does not overlap with the MUSE field that we observe here \citep{Maggi:2019q}.
High-resolution \ha{} spectroscopy of the \hii{} region
gives a non-thermal velocity width of \(\siglosm = \SI{10.1}{km.s^{-1}}\)
\citep{2003ApJ...586.1179D}.
\stopNEW
\citet{2008ApJ...688.1050G} propose an expanding \hii{} region or bubble blown by the winds of the massive progenitor as a mechanism that shapes the recent star formation in this region, in addition to the photoionization from the OB stars. 
This mechanism is similar to shell-like \hii{} regions, with a central cluster in a cavity, and with ongoing star formation triggered around their periphery.

\startNEW
\subsubsection{Observed structure function and model fit}
\label{sec:observ-struct-funct-346}
\stopNEW

Figure~\ref{fig:strucfunc-fit-MC}b shows the structure function of NGC 346.
\startNEW
The model-derived parameters are \(\sigma\pos^2 = \SI{33 \pm 3}{km^{2}.s^{-2}}\),
\(r_0 = \SI{2.4\pm 0.3}{pc} \)), and \(m = 0.95 \pm 0.07\).
These agree very well with the empirically derived parameters.
At large scales, an oscillatory behavior in \(B(r)\) is observed for \(r > \SI{10}{pc}\),
consistent with a periodic pattern with wavelength \(\lambda \approx \SI{15}{pc}\) and amplitude \(\SI{6}{km^{2}.s^{-2}}\).
Inspection of the velocity field (Figure~\ref{fig:velocity-maps}) shows that this pattern is predominantly oriented along the north-south direction and corresponds to the spacing between the filaments of molecular gas.
\stopNEW

\subsubsection{Comparison with previous studies}
\label{sec:comparison-346}

We have found no previous studies of NGC~346's structure function in the literature.

\subsection{Hubble X and Hubble V in NGC 6822}
\label{sec:6822-hubble}
The irregular dwarf galaxy, NGC 6822, is located at a distance of \SI{500}{kpc} \citetext{\SI{1}{\arcsecond} = \SI{2.42}{pc}; \citealp{2012A&A...540A.135S}} and hosts the \hii{} regions Hubble~X and Hubble~V which are the brightest regions in the galaxy
\NEW{\citep{Hodge:1989c}}.

Hubble V is located near the ``Hodge OB 8'' OB association with a number of \num{\sim 36} young stellar objects \citep{2021MNRAS.507.5106K} and its core is powered by an O3 V star \citep{1999PASP..111.1382O}.
It has an ionizing output \(Q_0\) of \SI{1.7e50}{photons.s^{-1}} (13 times higher compared to the Orion Nebula) and has a \ha{} luminosity of \SI{2e38}{erg.s^{-1}}\citep{2002MNRAS.329..481B}.
The mean non-thermal linewidth in \ha{} \(\siglosm = \SI{10.3 \pm 0.03}{km.s^{-1}}\) \citep{1986A&A...160..374H}.

Hubble X is located near the Hodge ``OB 13'' OB association with a number of \num{\sim 29} young stellar objects \citep{2021MNRAS.507.5106K} and its core is powered by an O4 V star \citep{1999PASP..111.1382O}.
It has an ionizing output \(Q_0\) of \SI{1.4e50}{photons.s^{-1}} (10 times higher compared to the Orion Nebula) and has a \ha{} luminosity of \SI{1.6e38}{erg.s^{-1}} \citep{2002MNRAS.329..481B}.
The mean non-thermal linewidth in \ha{} is \(\siglosm = \SI{10.5 \pm 0.02}{km.s^{-1}}\)  \citep{1986A&A...160..374H}.
\citet{1993PASJ...45..693T} studies the kinematics of both regions.

\subsubsection{Observed structure function and model fit}
\label{sec:observ-struct-funct-hubbles}

In Figure~\ref{fig:strucfunc-fit-ExtraGal}a we show the structure function of Hubble V.
\startNEW
The model-derived parameters are \(\sigma\pos^2 = \SI{10 \pm 3}{km^{2}.s^{-2}}\),
\(r_0 = \SI{3.6 \pm 1.0}{pc} \)), and \(m = 0.8 \pm 0.3\).
The velocity variance is slightly higher than the empirically derived value of \(\sigma\pos^2 = \SI{8}{km^{2}.s^{-2}}\) because the model-derived seeing is only 6 times smaller than the correlation length.

At the largest scales, the structure function turns downward,
as seen in Orion.
However, inspection of the velocity map suggests that the explanation
may be different in this source: the velocity varies more strongly across the short axis of the nebula than along its long axis.
There is also marginal evidence for weak periodic fluctuations
with wavelength \(\lambda \approx \SI{50}{pc}\).
\stopNEW

The Hubble X structure function is shown in Figure~\ref{fig:strucfunc-fit-ExtraGal}b.
\startNEW
The model-derived parameters are \(\sigma\pos^2 = \SI{15 \pm 3}{km^{2}.s^{-2}}\),
\(r_0 = \SI{4.0 \pm 0.5}{pc} \)), and \(m = 1.0 \pm 0.2\),
which again is slightly higher than the empirically derived value of \(\sigma\pos^2 = \SI{13}{km^{2}.s^{-2}}\) because the latter is reduced by the seeing.

At the largest scales the structure function is approximately flat,
consistent with homogeneous fluctuations. 
\stopNEW

\subsubsection{Comparison with previous studies}
\label{sec:comparison-carina}

We have found no previous studies of the structure functions
of Hubble~V or~X in the literature.

\startNEW
\subsection{NGC~595 and NGC~604 in M33}
\label{sec:m33-ngc}
\stopNEW

M33 is the third largest galaxy in the Local Group,
located at a distance of \SI{820}{kpc}
\citetext{\SI{1}{\arcsecond} = \SI{4.07}{pc}; \citealp{2015KamKinematics}}
and is of morphological type SA(s)cd.
The two brightest \hii{} regions in the galaxy are NGC~604 and NGC~595.

NGC 595 is the second most luminous giant \hii{} region in the M33 galaxy,
\startNEW
excited by a stellar population comprising
250 OB stars, 13 hot supergiants, and 11 Wolf-Rayet stars \citep{1993AJ....105.1400D, 1996AJ....111.1128M}.
\stopNEW
The ionizing output \(Q_0\) is \SI{7.6e50}{photons.s^{-1}} (60 times higher compared to the Orion Nebula) and
the \hii{} region has a luminosity in \ha{} of \SI{8.9e38}{erg.s^{-1}} \citep{2002MNRAS.329..481B}.
The mean non-thermal linewidth in \ha{} is
\(\siglosm = \SI{16.8 \pm 0.1}{km.s^{-1}}\) \citep{lagrois2009multi}.
\citet{lagrois2011} present a kinematic study of the regions using optical emission lines and and radio 21 cm line observations and find an \ha{} shell morphology that is related to the stellar winds of the several massive stars located in its interior.
They proposed champagne flows at the periphery of the molecular cloud leading to accelerated ionized material.


NGC 604 is the
\startNEW
most luminous \hii{} region in M33 and
\stopNEW
the second most luminous in the Local Group after 30~Dor.
\startNEW
The dominant stellar cluster is located in the center of the region and contains 200 OB stars and 4 Wolf-Rayet stars \citep{1996ApJ...456..174H, 2011MNRAS.411..235E}.
The optically visible stars come from a population
with an age of \SI{4}{Myr},
although infrared observations detect the presence of
ongoing massive star formation in the region
\citep{2012AJ....143...43F, 2012ApJ...761....3M}.
\stopNEW
The ionizing output is \(Q_0 = \SI{2.2e51}{photons.s^{-1}}\)
(175 times higher than the Orion Nebula) and
the region has an \ha{} luminosity of \SI{2.6e39}{erg.s^{-1}} \citep{2002MNRAS.329..481B}.
The mean non-thermal linewidth in \ha{} is
\(\siglosm = \SI{17.8 \pm 0.3}{km.s^{-1}}\) \citep{1986A&A...160..374H}.
\startNEW
The morphology of the region is strikingly similar to that of 30~Dor
(see Figure~\ref{fig:hii-regions})
\stopNEW
and is visually dominated by a predominant big loop, many expanding shells and different size filaments.


\startNEW
\subsubsection{Observed structure function and model fit}
\label{sec:observ-struct-funct-m33}
\stopNEW

In Figure~\ref{fig:strucfunc-fit-ExtraGal}c we show the structure function of NGC 595.
\startNEW
The model-derived parameters are \(\sigma\pos^2 = \SI{53 \pm 5}{km^{2}.s^{-2}}\),
\(r_0 = \SI{11 \pm 1.0}{pc} \), and \(m = 1.36 \pm 0.15\).
The velocity variance is slightly higher than the observed value of \SI{44}{km^{2}.s^{-2}},
indicative of a moderate finite box effect.
At the largest scales, the structure function trends slowly downwards,
which may be due to a radial decline in the amplitude of the turbulent velocity field,
but the effect is less marked than in other regions, such as Orion. 
\stopNEW

In  Figure~\ref{fig:strucfunc-fit-ExtraGal}d we show the structure function of NGC 604.
\startNEW
The model-derived parameters are \(\sigma\pos^2 = \SI{84 \pm 20}{km^{2}.s^{-2}}\),
\(r_0 = \SI{12 \pm 6}{pc} \)), and \(m = 0.8 \pm 0.2\).
Again the model-derived velocity variance is higher than the observed value of \SI{55}{km^{2}.s^{-2}},
which this time is due primarily to the fact that the
derived correlation length is only 6 times higher than the RMS seeing width,
so some of the small-scale fluctuation power is missing from the observations.
Of all our sources, NGC~604 shows the largest difference between the inferred
``true'' structure function and the observed one, so the derived parameters
should be treated with caution.
At the largest scales, the structure function shows a pronounced minimum
at \(r \approx \SI{150}{pc}\), which may be due to a periodic oscillation with \(\lambda \approx \SI{300}{pc}\),
which is slightly larger than the size of the map. 
On the other hand, if we fit the model for scales \(r \leq \SI{20}{pc}\), there is a decrease in the difference between the inferred ``true'' structure function and the observed one. With our current observations the fitting using less than half of the observational points cannot be justified; but with better quality observations and using our current model we would expect to derived more accurate parameters.
\stopNEW

\startNEW
\subsubsection{Comparison with previous studies}
\label{sec:comparison-m33}

\citet{lagrois2011} calculated the structure function for the \halpha\ emission line for NGC 595
over a region of diameter \SI{300}{kpc} with a resolution of \SI{6.6}{pc.pixel^{-1}} \citep{lagrois2009multi}, presenting results for separations \(r \ge \SI{16.5}{pc}\).
They found a total velocity variance of \(\sigma\pos^2 = \SI{35 \pm 2}{km^{2}.s^{-2}}\), which is smaller than our derived value of \(\sigma\pos^2 = \SI{53}{km^{2}.s^{-2}}\).
Our map covers a smaller region, so part of the reason for this discrepancy may be
a radial decline in the velocity variance, as mentioned above.
Another reason may be that the angular resolution of the \citet{lagrois2009multi} map
is of the same order as the correlation length, so some of the small-scale power is missing.

\citet{lagrois2011} employ a novel technique based on filtering the
bidimensional velocity field to remove large-scale non-turbulent motions
before calculating the structure function.
They find a total decorrelation length of \(\tau_0 = \SI{43}{pc}\) and a very steep
power-law index of \(m = \num{1.55 \pm 0.1}\).
As discussed in section~\ref{sec:comp-with-prev}, \(\tau_0\) is predicted to be 2--3 times larger than \(r_0\), given the way that we define the correlation length,
but even taking this into account, our derived value is still smaller than theirs
by at least 50\%.
Although this may be due to the difference in the areas covered,
it is more likely due to the lower resolution of the \citeauthor{lagrois2011} data.

The power-law index is somewhat steeper than our own value,
but in part this is due to differences in methodology:
the width of the low-pass filter that \citeauthor{lagrois2011} use to remove large-scale
motions is \SI{8}{pixels} or \SI{53}{pc}, which is comparable to the decorrelation length.
As a result, subtracting the filtered data causes a clear steepening of the structure function,
as can be seen in their Figure~6d.
The unfiltered data can be seen to show a shallower slope, more consistent
with our own results.
Which result is more reliable is difficult to say, but could potentially be checked
by experimenting on fake data in a similar fashion to our Appendix~\ref{sec:degr-struct-funct}. 

All previous studies of the structure function in NGC~604 are based
on the same TAURUS-II observational dataset that we use here
\citep{sabalisck1995supersonic}
and the results are unsurprisingly similar to our own,
although there are some differences in methodology and interpretation.
For instance, \citet{Medina-Tanco:1997a} identify an apparent ``knee''
in the structure function at \(r \approx \SI{10}{pc}\), which they interpret as
evidence for a broken power law.
However, in our own results the same feature is modeled as a combination
of two effects:
the natural flattening of
the structure function for \(r > r_0\),
plus a steepening at small scales due to the smoothing effects of the seeing. 
\citet{Melnick:2021x} also analysed the same structure function and concluded that
separations \(< \SI{2}{arcsec}\) should be discounted due to the seeing.
The results of our experiments with synthetic observations (Appendix~\ref{sec:effects-seeing-struc})
suggest that the situation is even worse than \citeauthor{Melnick:2021x} surmised
and that scales up to \SI{5}{arcsec} will be affected by the seeing.
On the other hand, if the effects of seeing are incorporated into the model structure function,
as we have done here, then the data at these separations do not need to be discarded entirely.

\stopNEW


\section{Additional covariance corner plots for model fits}
\label{sec:addit-covar-corn}
Supplementary online-only material.

\begin{figure*}
  \centering
  \sfcfigggg{OrionS}{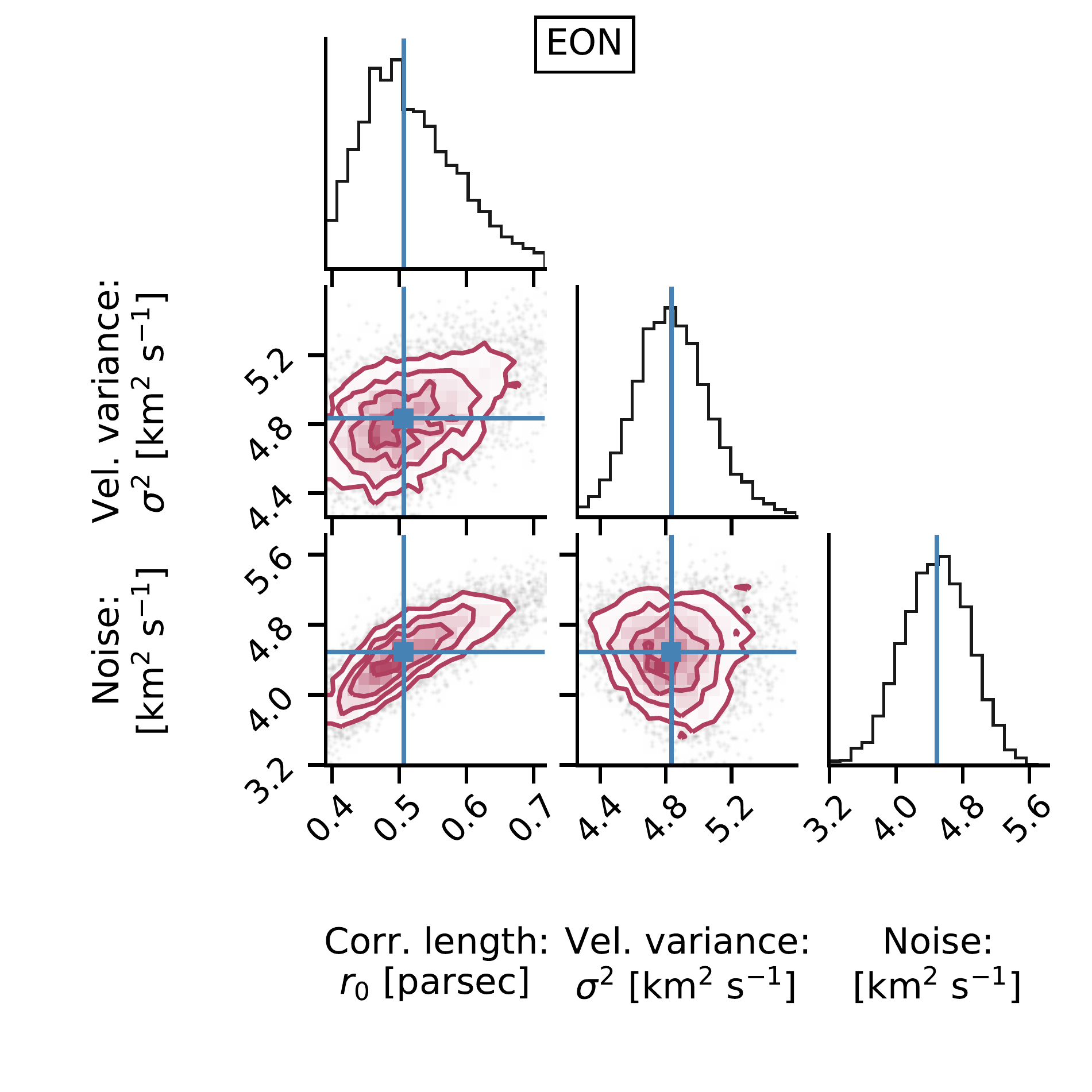}{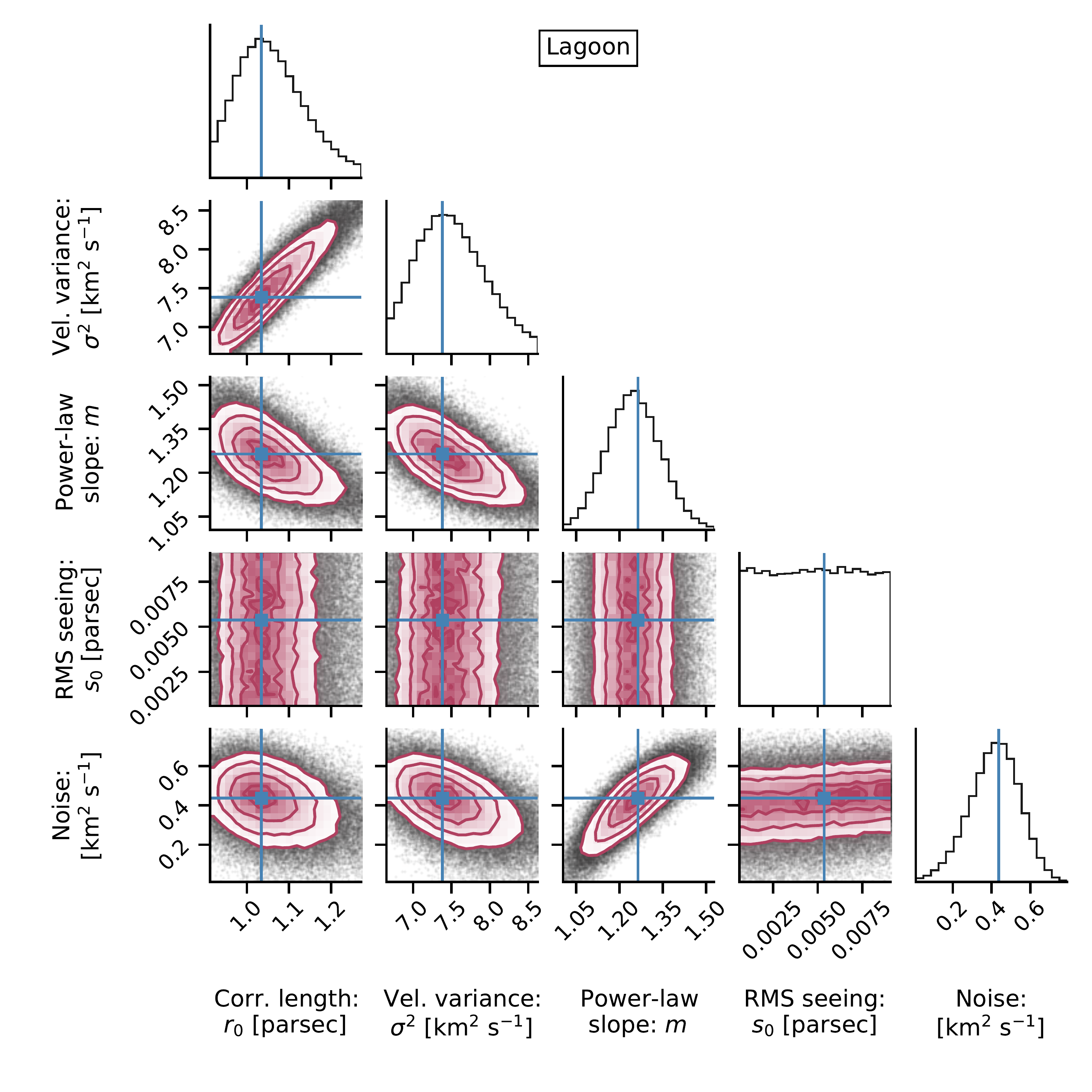}{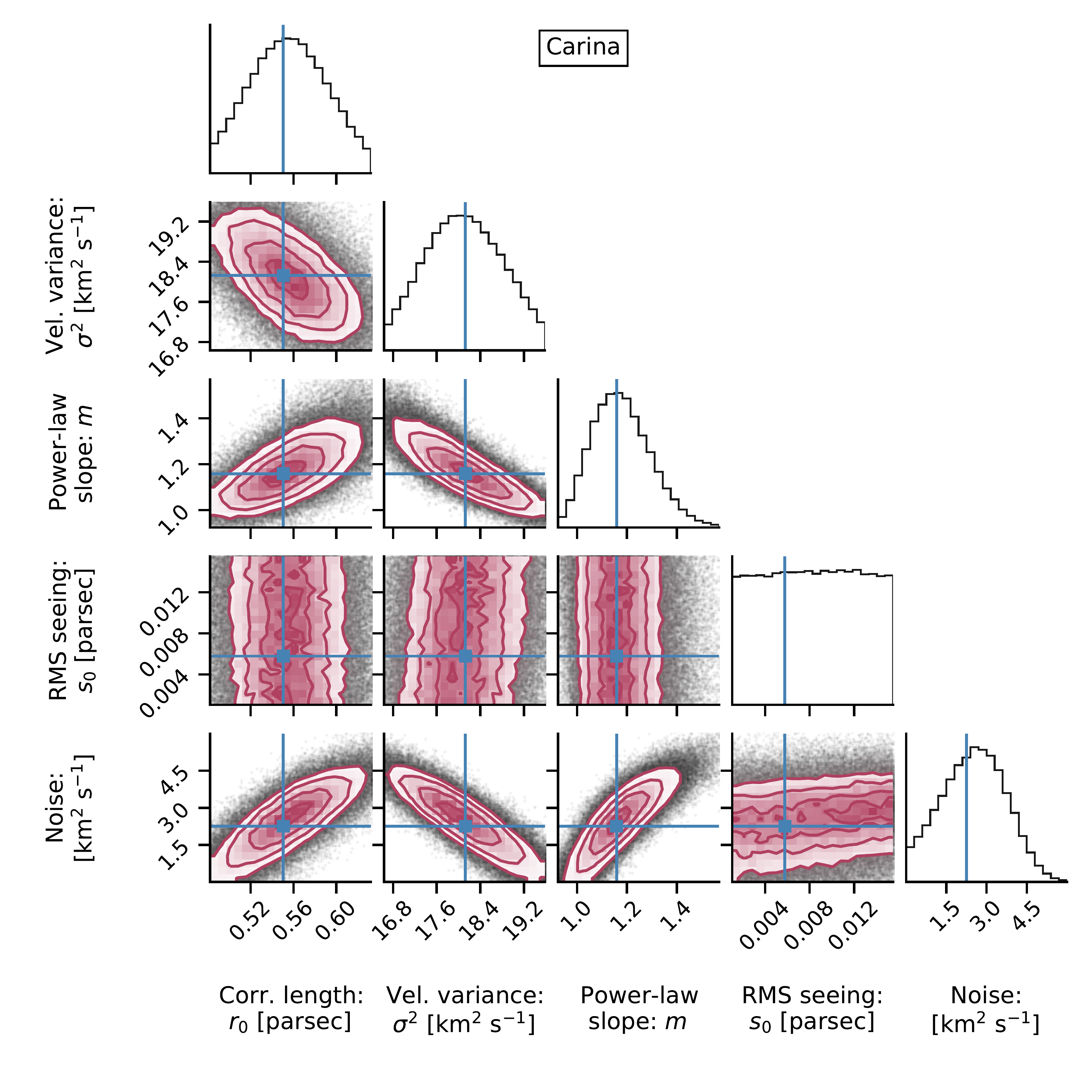}
  \caption{
    Corner plots of covariances between
    fitted model parameters
    of \ha{} structure function
    for Galactic \hii{} regions.
  }
  \label{fig:corner-Galactic}
\end{figure*}

\begin{figure*}
  \centering
  \sfcfigg{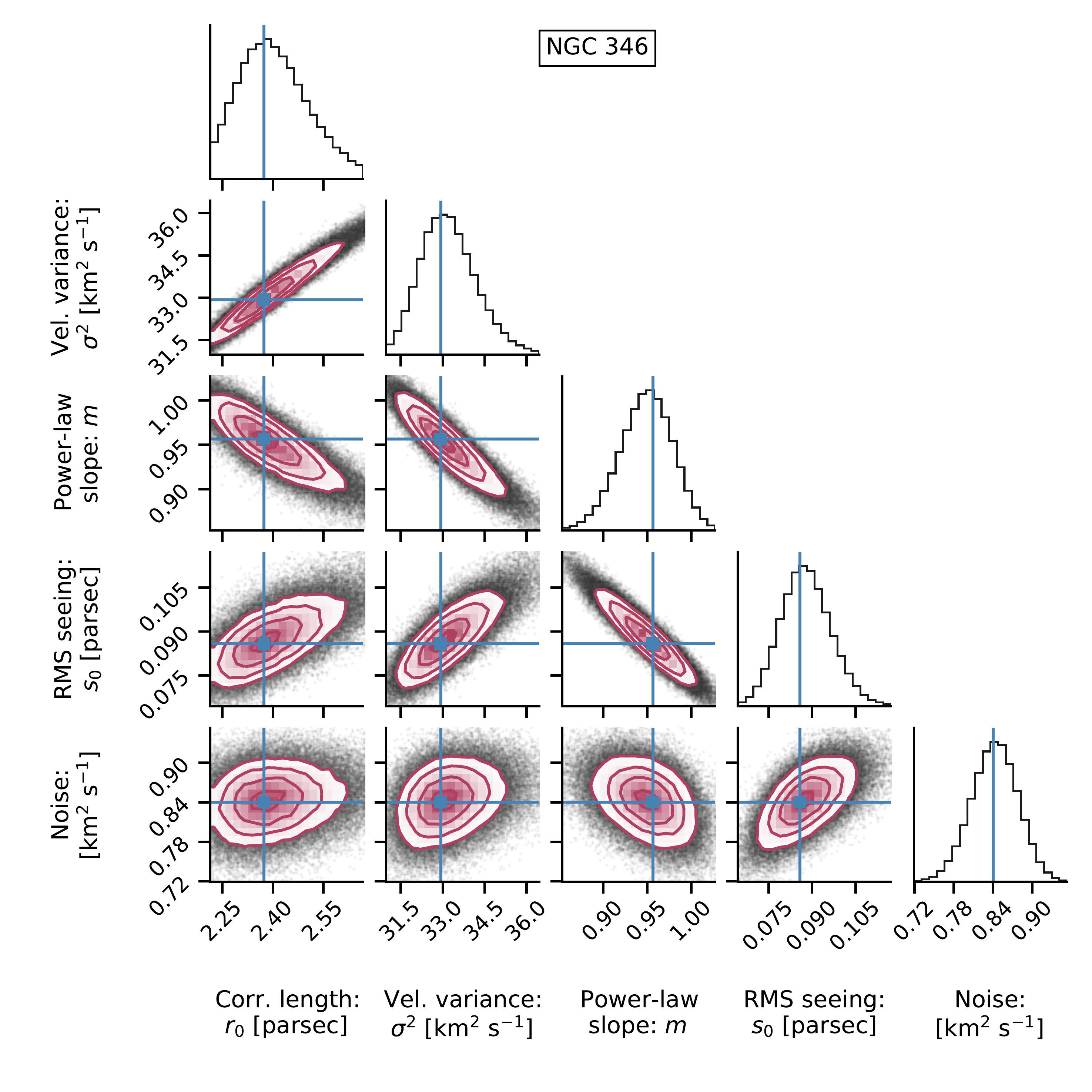}{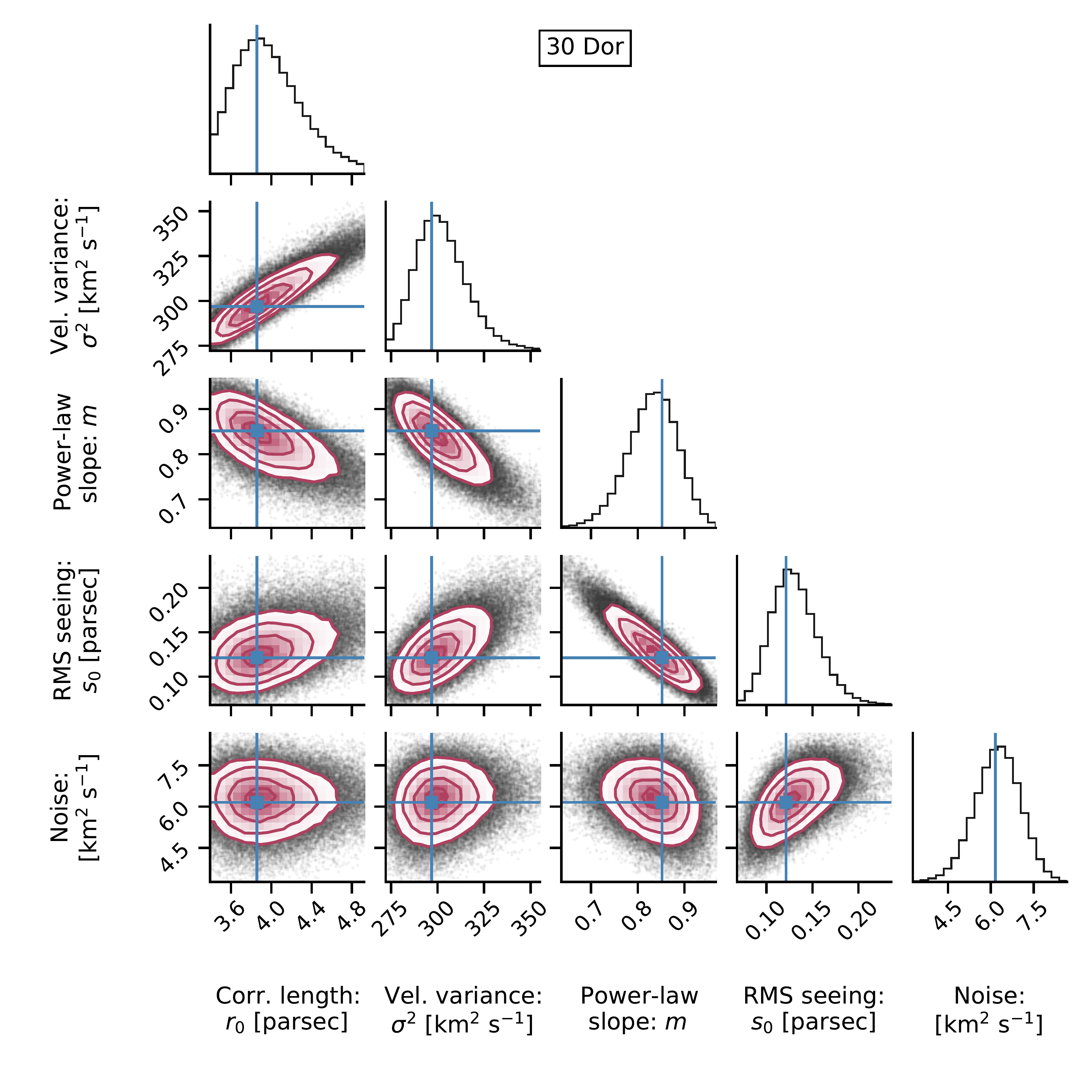}
  \caption{
    Same as figure~\ref{fig:corner-Galactic}
    except for Magellanic Cloud \hii{} regions. 
  }
  \label{fig:corner-MC}
\end{figure*}

\begin{figure*}
  \centering
  \sfcfigggg{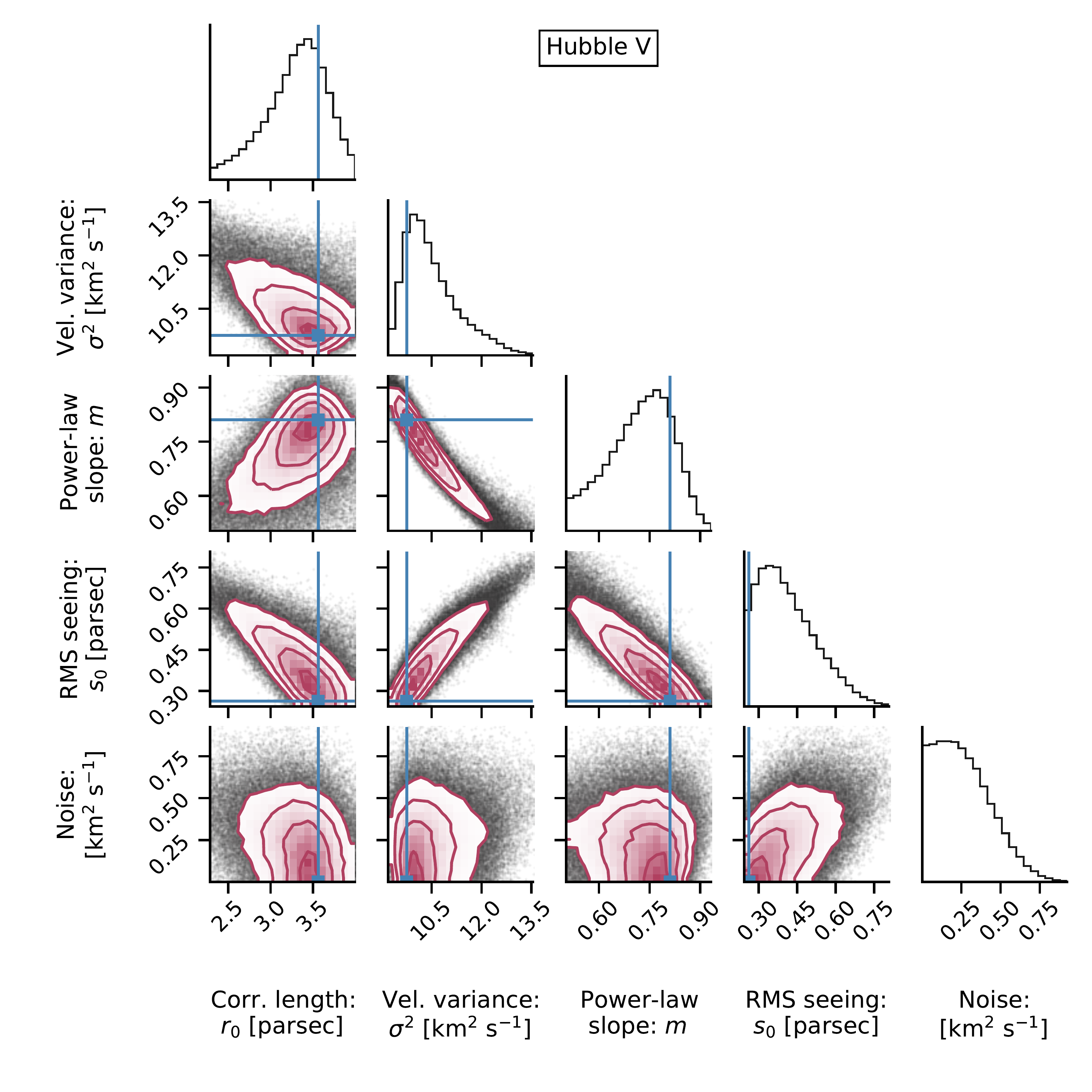}{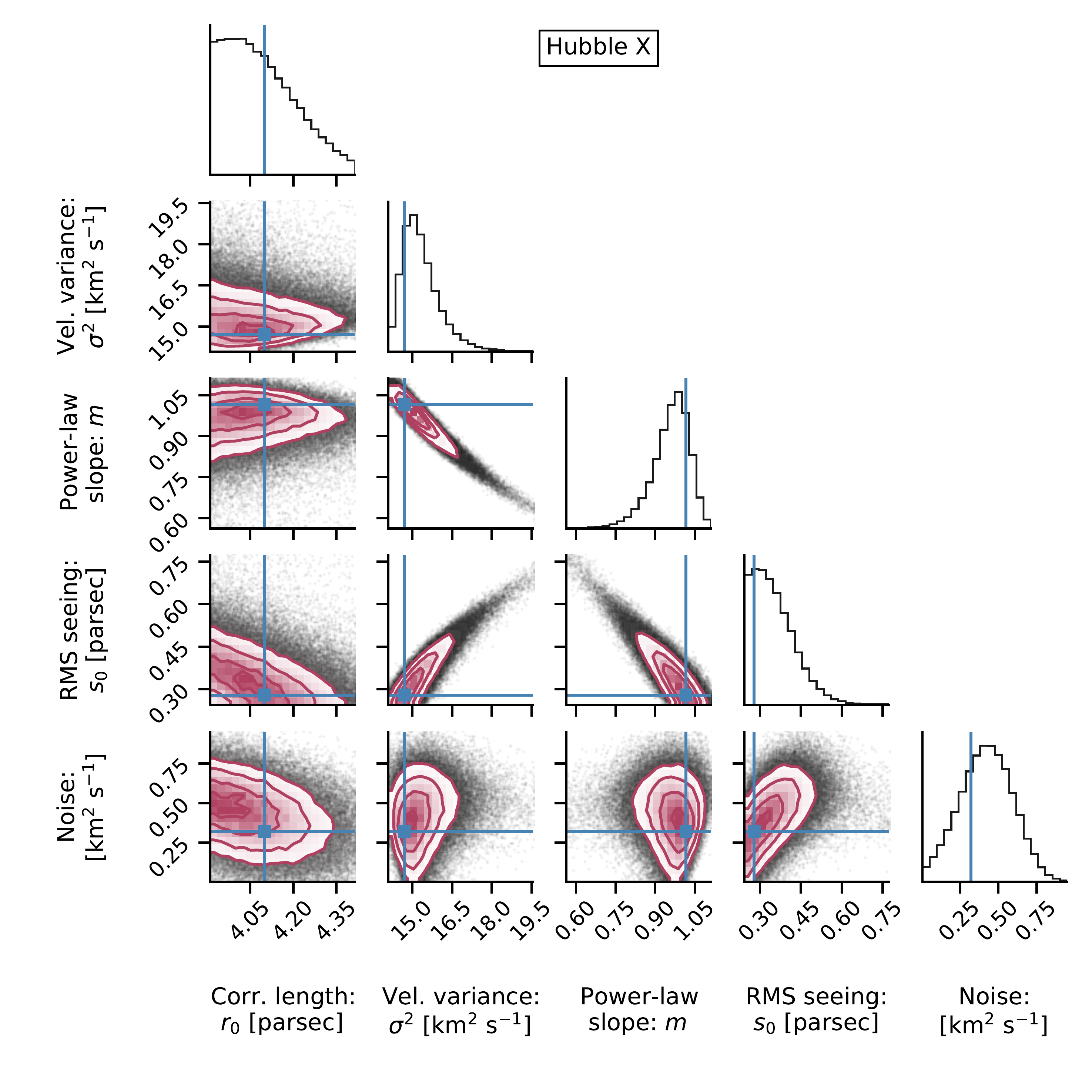}{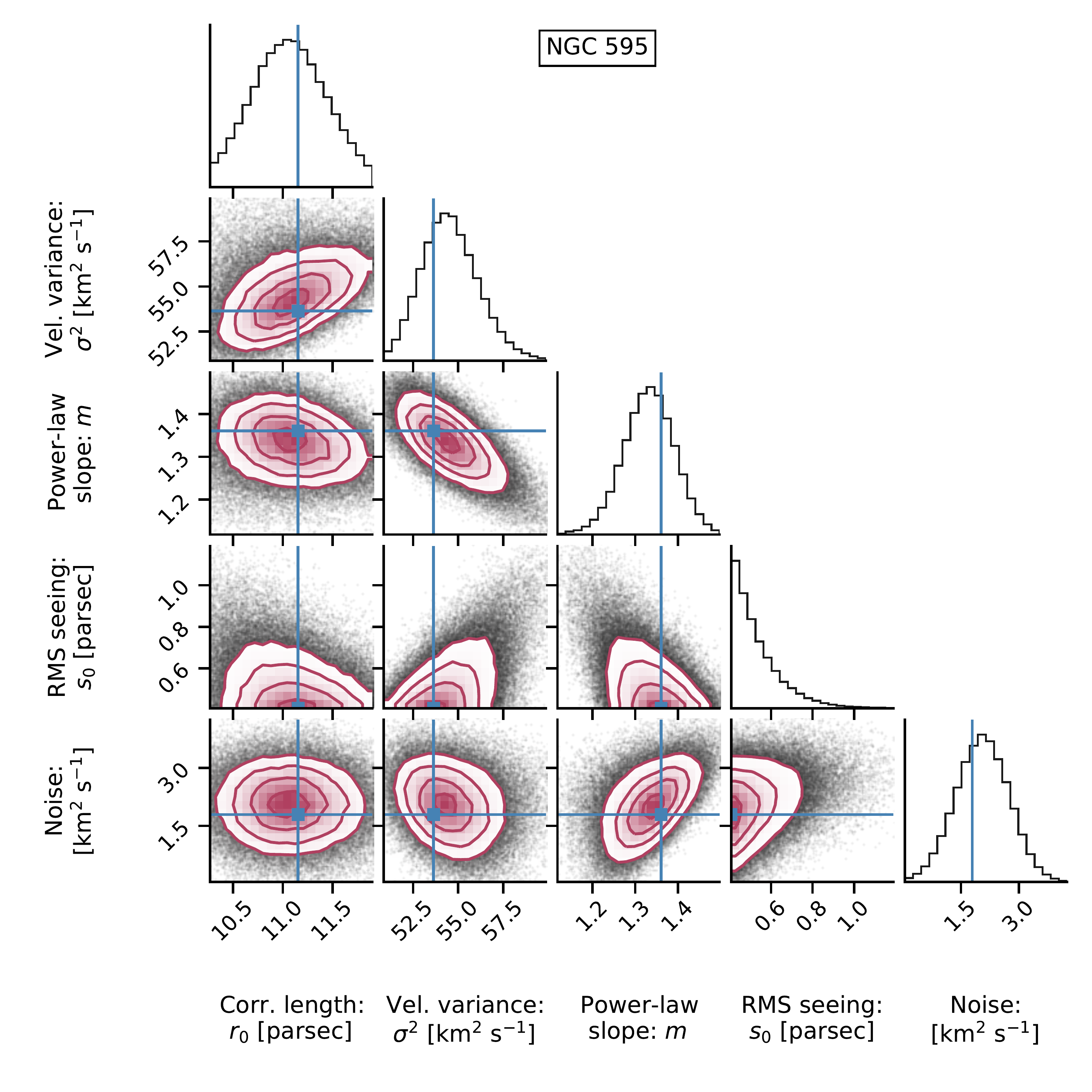}{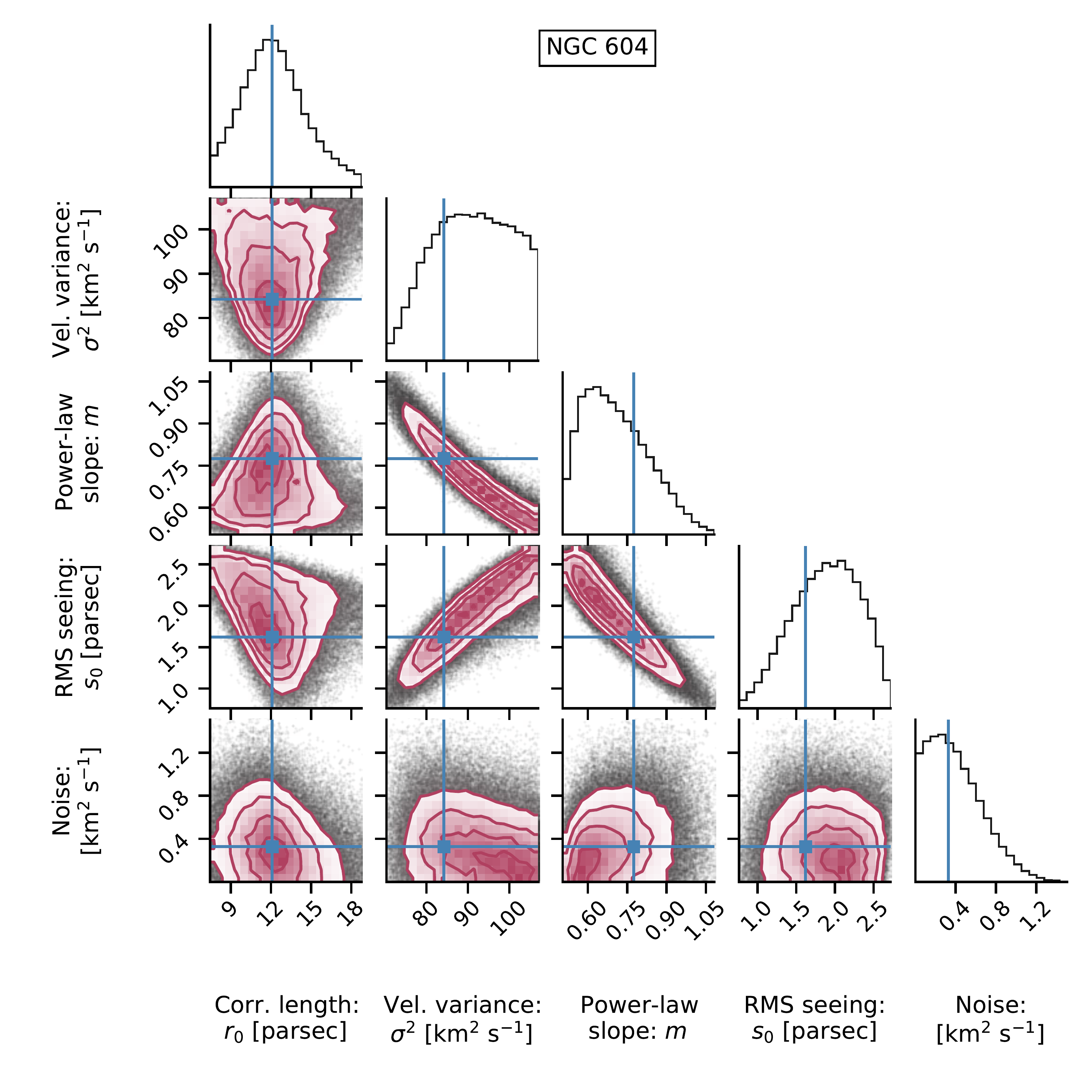}
  \caption{
    Same as figure~\ref{fig:corner-Galactic}
    except for \hii{} regions in more distant
    Local Group Galaxies.
  }
  \label{fig:corner-ExtraGal}
\end{figure*}

\begin{table*}
\begin{center}\caption{Fitting parameters.}
\begin{tabular}{cCCCCCCCC}\toprule
\hii{}    &  \text{No. points} & \text{Relatively}  & \text{Weight / No. points}     &  \text{Weight / No. points}     & \text{Max.}  & \text{Reduced} & \text{Autocorrelation } & \text{Acceptance}\\
Region    &  \text{merged} & \text{uncertainty} & \text{Small scales} &  \text{Large scales} & \text{separation}  & \text{chi-square} & \text{time} & \text{fraction}\\
\midrule
Orion     & 3 & 0.02    &  -           &   -        & 0.5L  & 0.82 & 60  & 0.55\\
EON       & - & 0.03   &  2.0 / 3     &   3.0 / 3        & 0.7L  & 0.62 & 40 & 0.64\\
Lagoon    & 2 & 0.08    &  2.5 / 16    &   2.5 / 8  & 0.5L  & 0.94 & 72  & 0.50\\
Carina    & 2 & 0.055   &  3.0 / 14    &   3.0 / 4  & 0.5L  & 0.94 & 72  & 0.50\\
30 Dor    & 5 &  0.07   &  -           &   -        & 0.9L  & 0.98 & 84  & 0.54\\
NGC 346   & 3 & 0.02    &  -           &   -        & 0.5L  & 0.83 & 64  & 0.54\\
Hubble V  & - & 0.05    &  2.0 / 3     &   -        & 0.6L  & 0.86 & 340 & 0.44\\
Hubble X  & - & 0.032   &  2.0 / 4     &   2.0 / 6  & 0.5L  & 0.73 & 79  & 0.49\\
NGC 595   & - &  0.055  &  4.5 / 3     &   3.0 / 5  & 0.5L  & 0.83 & 67  & 0.51\\
NGC 604   & - &  0.08  &   2.0 / 3    &   1.5 / 2 & 0.5L  & 0.92 & 112 & 0.45\\
\bottomrule
\end{tabular}\label{tab:fitting-parameters}
\end{center}
\end{table*} 




%
\bsp	
\label{lastpage}
\end{document}